\documentclass[twocolumn,twocolappendix]{aastex63}

\usepackage{color} 
\usepackage{float}
\usepackage{amsmath}
\usepackage{array}
\usepackage{multirow}

\newcolumntype{P}[1]{>{\centering\arraybackslash}p{#1}}
\def\arcsec{\hbox{$^{\prime\prime}$}}

\definecolor{emerald}{rgb}{0.1,0.5,0.3}

\definecolor{purple}{HTML}{4B0082}

\definecolor{brown}{HTML}{8B4513}

\shorttitle{A Multiwavelength view of IC 860}
\shortauthors{Luo et al.}

\begin{document}

\title{A Multiwavelength view of IC 860: What Is in Action inside Quenching Galaxies\footnote{{\it Herschel} is an ESA space observatory with science instruments provided by European-led Principal Investigator consortia and with important participation from NASA.}} 

\correspondingauthor{Yuanze Luo}
\email{yluo37@jhu.edu}

\author[0000-0002-0696-6952]{Yuanze Luo}
\affiliation{William H. Miller III Department of Physics and Astronomy, Johns Hopkins University, Baltimore, MD 21218, USA}

\author[0000-0001-7883-8434]{Kate Rowlands}
\affiliation{AURA for ESA, Space Telescope Science Institute,
3700 San Martin Drive, Baltimore, MD 21218, USA}
\affiliation{William H. Miller III Department of Physics and Astronomy, Johns Hopkins University, Baltimore, MD 21218, USA}

\author[0000-0002-4261-2326]{Katherine Alatalo}
\affiliation{Space Telescope Science Institute, 3700 San Martin Dr, Baltimore, MD 21218, USA}
\affiliation{William H. Miller III Department of Physics and Astronomy, Johns Hopkins University, Baltimore, MD 21218, USA}

\author[0000-0001-6245-5121]{Elizaveta Sazonova}
\affiliation{William H. Miller III Department of Physics and Astronomy, Johns Hopkins University, Baltimore, MD 21218, USA}

\author[0000-0002-5258-8761]{Abdurro'uf}
\affiliation{Institute of Astronomy and Astrophysics, Academia Sinica, 11F of AS/NTU Astronomy-Mathematics Building, No.1, Sec. 4, Roosevelt Rd, Taipei 10617, Taiwan, R.O.C.}

\author[0000-0001-6670-6370]{Timothy Heckman}
\affiliation{William H. Miller III Department of Physics and Astronomy, Johns Hopkins University, Baltimore, MD 21218, USA}

\author[0000-0001-7421-2944]{Anne M. Medling}
\affiliation{Ritter Astrophysical Research Center, University of Toledo, Toledo, OH 43606, USA}
\affiliation{ARC Centre of Excellence for All Sky Astrophysics in 3 Dimensions (ASTRO 3D)}

\author[0000-0003-2823-360X]{Susana E. Deustua}
\affiliation{National Institute of Standards and Technology, 100 Bureau Drive, Gaithersburg, MD 20899, USA}

\author[0000-0003-1991-370X]{Kristina Nyland}
\affiliation{National Research Council, resident at the U.S. Naval Research Laboratory, 4555 Overlook Ave SW, Washington, DC 20375, USA}

\author[0000-0002-3249-8224]{Lauranne Lanz}
\affiliation{The College of New Jersey, Ewing, NJ 08618, USA}

\author[0000-0003-4030-3455]{Andreea O. Petric}
\affiliation{Space Telescope Science Institute, 3700 San Martin Dr, Baltimore, MD 21218, USA}

\author[0000-0003-3191-9039]{Justin A. Otter}
\affiliation{William H. Miller III Department of Physics and Astronomy, Johns Hopkins University, Baltimore, MD 21218, USA}

\author{Susanne Aalto}
\affiliation{Department of Space, Earth and Environment, Onsala Space Observatory, Chalmers University of Technology, 439 92 Onsala, Sweden}

\author{Sabrina Dimassimo}
\affiliation{Institute for Defense Analyses, 730 East Glebe Road, Alexandria, VA 22305, USA}

\author[0000-0002-4235-7337]{K. Decker French}
\affiliation{Department of Astronomy, University of Illinois Urbana-Champaign, Urbana, IL 61801, USA}

\author[0000-0001-8608-0408]{John S. Gallagher III}
\affiliation{Department of Astronomy, University of Wisconsin, 475 N. Charter Street, Madison, WI, 53706, USA}

\author[0000-0002-0363-4266]{Joel C. Roediger}
\affiliation{National Research Council of Canada, Herzberg Astronomy and Astrophysics, 5071 West Saanich Road, Victoria, BC V9E 2E7, Canada}

\author[0000-0002-7850-7093]{Sofia Stepanoff}
\affiliation{The College of New Jersey, Ewing, NJ 08618, USA}

\begin{abstract}
We present a multiwavelength study of IC 860, a nearby post-starburst galaxy at the early stage of transitioning from blue and star-forming to red and quiescent. Optical images reveal a galaxy-wide, dusty outflow originating from a compact core. We find evidence for a multiphase outflow in the molecular and neutral gas phase from the CO position-velocity diagram and NaD absorption features. We constrain the neutral mass outflow rate to be $\sim$0.5 M$_{\odot}/$yr, and the total hydrogen mass outflow rate to be $\sim$12 M$_{\odot}$/yr. Neither outflow component seems able to escape the galaxy. We also find evidence for a recent merger in the optical images, CO spatial distribution, and kinematics, and evidence for a buried AGN in the optical emission line ratios, mid-IR properties, and radio spectral shape. The depletion time of the molecular gas reservoir under the current star formation rate is $\sim$7 Gyr, indicating that the galaxy could stay at the intermediate stage between the blue and red sequence for a long time. Thus the timescales for a significant decline in star formation rate (``quenching") and gas depletion are not necessarily the same. Our analysis supports the quenching picture where outflows help suppress star formation by disturbing rather than expelling the gas and shed light on possible ongoing activities in similar quenching galaxies.
\end{abstract}

\section{Introduction} \label{sec:intro}
If we take a snapshot of all galaxies and sort them based on their colors and morphologies, we find that present-day galaxies fall predominantly in two distinct populations: either blue, star-forming spirals (late-type galaxies, LTG) or red, quiescent ellipticals/lenticulars (early-type galaxies, ETG, \citealt{Hubble_1926,Baldry_2004}). This bimodality is often demonstrated on the stellar mass vs. star formation rate (or related quantities like color) diagrams. We refer to the two galaxy populations as the ``blue/red sequence" and the intermediate stage as the ``green valley"\footnote{Note that color-mass based classification does not contain information about the gas in the galaxies. When we refer to the blue/red sequence galaxies in this paper we implicitly assume that they are gas rich/deprived as in the traditional picture.}. Through cosmic time, the galaxy number density increases in the red sequence whilst it decreases in the blue sequence \citep[e.g.,][]{Bell_2007,Bell_2012}. Such findings prompt a galaxy evolution scenario where blue galaxies evolve onto the red sequence as star formation (SF) diminishes because gas is consumed in stars or becomes unavailable for stellar synthesis. 

Many mechanisms can cause the reduction in star formation rate (``quenching"). External processes such as galaxy mergers can result in energetic starbursts that exhaust the available gas, leaving the merger remnant quenched \citep[e.g.,][]{Hopkins_2006,Sparre_2016}. Mergers may as well trigger stellar and/or active galactic nuclei (AGN) feedback that could violently heat and expel the remaining gas \citep[e.g.,][]{Naab_2003,Bournaud_2005}. Ram pressure stripping can also remove gas as a galaxy falls into a cluster \citep{Gunn_1972}. Examples of internal quenching mechanisms include morphological quenching and AGN feedback. Morphological quenching refers to the scenario where a galaxy builds up its central bulge over time to a point that the gas in the galaxy is stable against gravitational collapse. Morphological quenching does not require removal or termination of gas supply and acts on a long timescale of several Gyrs \citep[e.g.,][]{Martig_2009,Colombo_2018,Lin_2019,Mendez-Abreu_2019}. AGN feedback refers to the scenario where the AGN re-deposits energy and momentum into its host galaxy via outflows and/or radiation, and is often invoked to explain the rapidity of gas heating or expulsion necessary for short quenching timescales. AGN feedback is thought to be responsible for keeping a galaxy quenched \citep[e.g.,][]{Hopkins_2008,Terrazas_2016,Terrazas_2017}. However, the nature of AGN feedback is still uncertain, as is its efficacy at rapidly removing the entire reservoir of star-forming fuel and globally halting SF. Observational evidence of galaxy quenching from AGN feedback remains circumstantial \citep[e.g.,][]{Smethurst_2017,French_2018}, although in simulations AGN feedback plays an important role in quenching \citep[e.g.,][]{Zheng_2020}.

It is unlikely that a single mechanism is responsible for galaxy quenching, and multiple quenching mechanisms will affect galaxies across their lifetime, working in collaboration to ensure galaxies stay quenched \citep{Smethurst_2017}. The goal of studying this transition stage in galaxy evolution is to disentangle the relative importance of different quenching mechanisms. 
In addition to the mysterious interplay between different quenching mechanisms, the blue to red evolutionary path has its own complications. \citet{Schawinski_2014} showed that blue early-type galaxies and blue late-type galaxies transit quite differently to the red sequence. Early-type galaxies show a rapid cessation of SF and transit the green valley rapidly (``the fast quenching track"), while late-type galaxies experience a more gradual decrease in star formation rate. Thus regarding the green valley as a transition phase for all galaxies might be overly simplistic. Identifying the initial properties and pathways taken by these ``dying galaxies" is essential to build a complete understanding of galaxy evolution.

In this work we examine the fast quenching track by studying one of the post-starburst galaxies (PSBs): IC 860. PSBs have quenched their SF both rapidly and recently ($\lesssim$1 Gyr), so it is promising that they will reveal visible evidence for the quenching triggers (see \citealt{French_2021} for a review on PSBs). PSBs consist of $<$1\% of the galaxy population by $z\sim$ 0.5 and are more common at higher redshifts \citep[e.g.,][]{Wild_2016}, where high quality and spatially resolved data is harder to obtain. Despite being rare, PSBs might be an important phase in galaxy evolution, as $\sim$40\%--100\% of galaxies are expected to pass through this phase \citep{Zabludoff_1996,Snyder_2011,Wild_2016}.
Recent studies \citep{Rowlands_2015,French_2015,Alatalo_2016a} showing that local PSBs have significant cold gas reservoirs highlights the unique timescale they represent: we are catching them in the act of depleting their remaining star-forming fuel. 

IC 860 is a nearby PSB for which we have obtained exquisite multiwavelength data to probe different activities such as outflows and AGN. Parent PSB samples containing IC 860 have been statistically studied in several works \citep[e.g.,][]{Alatalo_2016a,French_2018}. Particularly, \citet{Aalto_2019} did a case study on IC 860 with millimeter wavelength data focusing on the galactic center, and found that IC 860 hosts a nuclear outflow of dense molecular gas. It is also probable that the infrared emission within the galaxy originates from a compact core of $\sim$10 pc in size \citep{Aalto_2019}, which makes IC 860 a particularly interesting example to study the impact of nuclear activity on the cessation of SF. IC 860 can serve as a local laboratory of galaxy evolution to reveal the multiple events going on inside these quenching galaxies.  We present the data used in this study in Section 2, and describe our analysis in Section 3. Results and discussion are provided in Section 4 and Section 5, respectively. We summarize our findings in Section 6. Throughout this paper, we assume a flat cosmological model with $H_0=70\ \rm{km\ s^{-1}\ Mpc^{-1}}$, $\Omega_m=0.3$, and $\Omega_{\Lambda}=0.7$.

\section{Data}
IC 860 is a nearby ($z$ = 0.0129 from SDSS DR7, \citealt{Abazajian_2009}) luminous infrared (IR) galaxy (LIRG) with $L_{\rm{IR}}=10^{11.14}\ L_{\odot}$ \citep{Chu_2017}. It is a late-type spiral galaxy with a bar at the center and appears to be heavily obscured by dust. We first briefly describe the parent sample IC 860 belongs to in \S2.1, and present other data on this particular galaxy in the rest of \S2.

\subsection{Shocked Post-Starburst Galaxies}
PSBs are defined as galaxies that have rapidly quenched their SF in the last $\sim$1 Gyr and are traditionally found by requiring simultaneously the absence of prominent ionized emission lines (associated with young energetic stars) and the presence of strong Balmer absorption features (associated with intermediate age A-type stars). While these criteria can identify transitioning galaxies, they ignore other energetic processes like AGN and shocks that can also produce strong emission lines, and thus catch galaxies only after they have already quenched and transformed. 

To account for this selection bias, \citet{Alatalo_2016} created the Shocked POststarbust Galaxy Survey (SPOGS\footnote{\url{http://www.spogs.org}}), which is a catalog of PSBs hosting emission line ratios consistent with shocks. This catalog is based on Sloan Digital Sky Survey (SDSS) DR7 \citep{Abazajian_2009} and the Oh-Sarzi-Schawinski-Yi absorption and emission line catalog \citep[OSSY,][]{Oh_2011}. First, a sub-sample of emission line galaxies are selected based on their high-quality continuum and emission line fitting \citep[see][\S2.1 for more details]{Alatalo_2016}. Gas with emission lines consistent with shock excitation are then selected utilizing the [\ion{O}{3}]/H$\beta$ vs [\ion{N}{2}]/H$\alpha$, [\ion{S}{2}]/H$\alpha$, and [\ion{O}{1}]/H$\alpha$ line diagnostic diagrams \citep[the BPT diagrams,][]{Baldwin_1981,Veilleux_1987}. Shocked Post-starburst Galaxies (SPOGs) are galaxies that satisfy:
\begin{enumerate}
\item Strong Balmer absorption characterized by EW(H$\sigma$) $>$ 5 \AA;
\item Lie within the polygon ``shock region" defined by Equations (1)--(9) in \citet{Alatalo_2016} of all three diagnostic diagrams;
\item Not falling consistently inside the SF and composite regions of all three diagnostic diagrams.
\end{enumerate}

SPOGs selected with the above criteria trace a younger population \citep[e.g.,][]{French_2018} than traditional PSBs selected by requiring no prominent emission lines \citep[e.g.,][]{Goto_2007}, and can catch a galaxy when its SF is abruptly diminishing. IC 860 is one of the SPOGs that lies on the blue end of the green valley, and thus provides a window into the very early stages of galaxy transition.

\subsection{Hubble Space Telescope Imaging}
\subsubsection{Multi-band Imaging}
Near-UV (NUV), optical, and near-infrared (NIR) images of IC 860 were obtained with the Wide Field Camera 3 (WFC3) across seven broadband filters on the \emph{Hubble Space Telescope (HST)} in March and December 2017 (See Table~\ref{tab:hst} for details). The final data products are obtained from the Mikulski Archive for Space Telescopes (MAST)\footnote{\url{https://mast.stsci.edu/portal/Mashup/Clients/Mast/Portal.html}} at the Space Telescope Science Institute and can be accessed via\dataset[10.17909/tvgs-sa69]{https://doi.org/10.17909/tvgs-sa69}. All images are full frame and were processed with the standard reduction pipeline \texttt{CALWF3} \citep{Ryan_2016}. Cleaned images were coadded, registered, and scaled to a common pixel scale with the \texttt{AstroDrizzle} task of the \textit{HST} DrizzlePac software package \citep{Gonzaga_2012}. There are two independent observations with filters F140W (IR) and F275W (NUV), and we again use \texttt{AstroDrizzle} to combine them into one final image for each filter. We use the \texttt{lacosmic}\footnote{\url{https://lacosmic.readthedocs.io/en/latest/}} package \citep{vanDokkum_2001} to remove cosmic rays in the final image products.

\subsubsection{Spatially Resolved Spectral Energy Distributions}
We also construct spatially resolved spectral energy distributions (SEDs) using the seven \textit{HST} images. The five UV-optical \textit{HST} images have a resolution of $\sim$0.0396 arcsec per pixel, while the two IR images have a resolution of $\sim$0.128 arcsec per pixel. We first align the images and match them in resolution using \texttt{AstroDrizzle}. With the flat field
calibrated products from MAST, we drizzle all the images to align with north, and set the final resolution of the UV-optical images to that of the IR images. The flat field calibrated products also contain the estimated error in each pixel, and we drizzle the error images in the same way to get the final error images matched in orientation and resolution. With the matched \textit{HST} images, we have a 7-point SED with its associated flux error for each pixel at \textit{HST} IR pixel resolution.

\begin{table}[t]
\caption{\textit{HST} WFC3 Observations (Program 14715, 4 orbits)}
\centering
\begin{tabular*}{\columnwidth}{p{2.2cm} p{1.7cm} p{1.7cm} p{1.7cm}}
\hline \hline
Obs ID & Channel & Filter & Exp. Time \\ 
& & & (seconds) \\
\hline
ID8N02010 & IR-FIX  & F160W  & 406                 \\
ID8N02020 & UVIS2   & F606W  & 800                 \\
ID8N02WPQ & IR-FIX  & F140W  & 203                 \\
ID8N02030 & UVIS2   & F814W  & 750                 \\
ID8N01010 & UVIS2   & F438W  & 600                 \\
ID8N01020 & UVIS2   & F275W  & 2500                \\
ID8N01030 & UVIS2   & F336W  & 2000                \\
ID8N01040 & UVIS2   & F275W  & 2400                \\
ID8N01Q9Q & IR-FIX  & F140W  & 203                \\
\hline \hline
\label{tab:hst}
\end{tabular*}
\end{table}

\subsection{Integral Field Spectroscopy (IFS)}
Optical IFS data of IC 860 is from the Wide Field Spectrograph \citep[WiFeS,][]{Dopita_2007, Dopita_2010}. WiFeS is an integral field spectrograph on the Australian National University 2.3m telescope at the Siding Spring Observatory. The data were obtained in May 2015 with the spectral resolution configuration R$_{\rm{blue}}$=3000 and R$_{\rm{red}}$=7000. The standard \texttt{pyWiFeS} data reduction pipeline \citep{Childress_2014} was applied to all collected image slices, which were then synthesized into calibrated spectral cubes. The final data products are one blue and one red spectral cube, each with pixel resolution of 1\arcsec, covering air wavelength ranges of $\sim$3500$-$5700\AA\ and $\sim$5400$-$7000\AA, respectively.

In order to better fit the stellar continuum and determine the stellar kinematics, we combine the spectral cubes into one final cube. The blue cube and the red cube overlap in the wavelength range $\sim$5400$-$5700\AA, and have slightly different overall flux level. To keep the information from both cubes, we first scale the red cube flux to the same level as the blue cube flux for each spaxel using fluxes near the overlapping region, but exclude the overlapping region because the red spectra are very noisy. We then take the weighted average of the blue and red spectra for the overlapping region\footnote{The blue and red IFS data have different wavelength spacing. We thus first resample the red spectra (flux and error) to match the blue spectra in a flux-conserving way with \texttt{SpectRes} by \citet{Carnall_2017}.} and propagate the flux error accordingly. We set the weights of the blue spectra to be slowly declining from 1 to 0.58 and the weights of the red spectra to be one minus those of the blue by trial and error, so as to minimize the influence of noisy red spectra on the spectral shape in the overlapping region while still keeping the red data. To further check the quality of this combination, we sum all spectra in the combined cube and convolve the summed spectrum with SDSS $g$ and $r$ filters to calculate fluxes as observed through these two filters. The resulting $g$-band and $r$-band fluxes agree well with the SDSS values. Finally, we resample the combined data cube to a logarithmically uniform wavelength grid covering $\sim$3500$-$7000\AA\ (the same wavelength range of the combined spectra) with $\Delta$log($\lambda$) = 1.53$\times$10$^{-4}$\AA.

The original IFS data also lack a world coordinate system (WCS). We manually add a WCS to the combined cube by matching the blue cube image to the SDSS $g$-band image of the galaxy. We check that under this WCS the red cube image also matches the SDSS $r$-band image.

\begin{figure*}
        \centering \includegraphics[width=2\columnwidth]{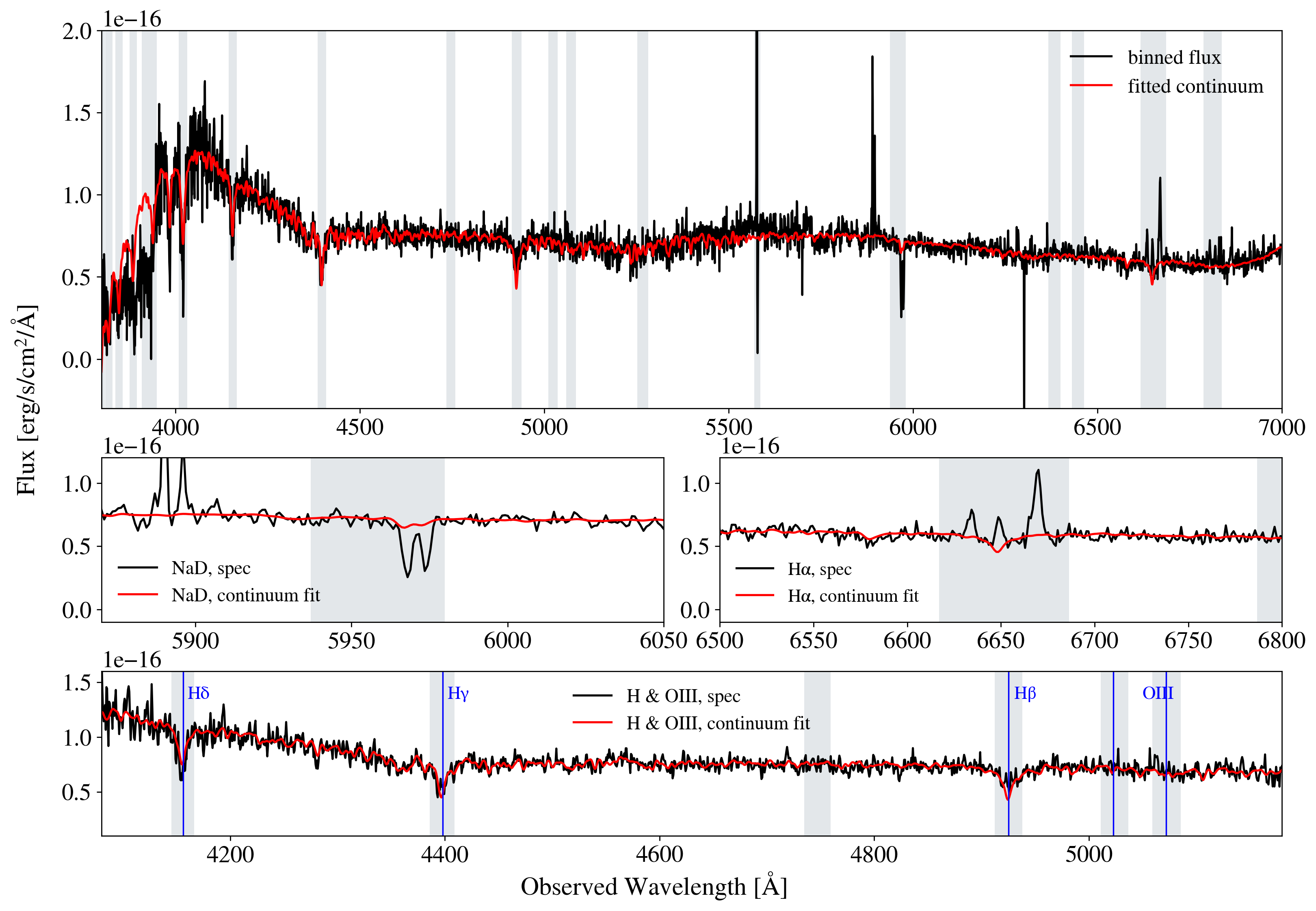}
        \caption{
                \label{fig:comb} 
                One of the spectra from the combined IFS data cube after binning, overplotted with its best fit continuum (\S3.1). This spectrum is from the bin at the center of the galaxy. The grey shaded areas are masked during the continuum fitting and the zoomed-in views for some important spectral features are shown in the smaller panels. The overall weak emission lines and strong Balmer absorption lines reflect the post-starburst nature of IC 860. There appears to be a small mismatch between the spectrum and the fitted continuum between 5400--5700\AA, where the original blue and red data cubes overlap, which could be caused by the noisy data in the original red cube. This small mismatch should not affect our results since there are no prominent spectral features within this region, and we test that our data combination is reliable in \S2.3.
        }
\end{figure*}

\subsection{CO Molecular Gas and Radio Continuum}
IC 860 was observed in $^{12}$CO(1–0) with the Combined Array for Research in Millimeter-wave Astronomy (CARMA)\footnote{\url{ http://www.mmarray.org}} in semesters 2012a and 2014a for a total of 5 hours. CARMA is an interferometric array of 15 radio dishes (6$\times$10.4m and 9$\times$6.1m) located in the Eastern Sierras in California \citep{Bock_2006}. The data reduction was done using the Multichannel Image Reconstruction, Interactive Analysis and Display \citep[\texttt{MIRIAD,}][]{Sault_1995} package. The integrated intensity (moment0) and mean velocity (moment1) maps were then constructed from the  data cube, following the procedure in \citet{Alatalo_2013}. The pixel resolution of the final data products are 0.42\arcsec, with a beam size $2.3\arcsec \times1.3$\arcsec and an rms of 7.1\,mJy in a 20 km/s channel. The CO velocity in this data is in radio convention. We convert it to optical convention in later analysis for comparison with other velocity data. 

The CARMA data also measured the 113 GHz radio continuum emission in IC 860. The continuum map was obtained by first isolating the line-free channels using the \texttt{MIRIAD} task \texttt{uvlin}\footnote{\url{http://www.atnf.csiro.au/computing/software/miriad/doc/uvlin.html}}, and then inverting the line-free visibilities using the multi-frequency synthesis options with \texttt{MIRIAD}. We measure the continuum flux to be 6.15 $\pm$0.06 mJy and include the flux in Table \ref{tab:full_sed}.

We use the total CO luminosity of IC 860 (see also \citealt{McBride_2014}) and a Milky Way CO-to-H$_2$ conversion factor $\alpha_{CO} = 4.3$\ M$_{\odot}$ (K\ km\ s$^{-1}$\ pc$^2)^{-1}$ \citep{Bolatto_2013} to derive the total molecular gas mass in the galaxy, M$(H_2)_{tot} = (1.79 \pm 0.54)\times 10^9$ M$_{\odot}$, where the error represents the 30\% calibration uncertainty.

\subsection{Spectral Energy Distribution of the Entire Galaxy}
Photometric data of the entire galaxy has been measured from far UV (FUV) to far infrared (FIR), allowing us to construct and fit the SED. IC 860 is observed with the Galaxy Evolution Explorer \citep[\textit{GALEX},][]{Martin_2005}, the 2 Micron All Sky Survey \citep[\textit{2MASS},][]{Skrutskie_2006}, and \textit{Spitzer} \citep{Werner_2004}. We measure the fluxes following Alatalo et al. (in preparation), similar to the way in \citet{U_2012}. IC 860 was also observed with the \textit{Herschel} Space Observatory \citep{Pilbratt_2010}, and the photometry was published in \citet{Chu_2017}. The \textit{Spitzer} and \textit{Herschel} observations are part of the Great Observatories All-sky LIRG Survey (GOALS) with data accessible via \citet{goals}. In addition, we include optical photometry from SDSS DR7 $ugriz$ bands, mid-infrared (MIR) photometry from the \textit{Wide-Field Infrared Survey Explorer} \citep[\textit{WISE},][]{Wright_2010,allwise}, NUV to NIR photometry from \textit{HST}/WFC3, and FIR photometry from \textit{AKARI} \citep{akari_mission}. For \textit{WISE} data, we take the aperture magnitudes (e.g., \textit{w3gmag}) for extended sources of the \textit{W3} and \textit{W4} filters, then add zero-point corrections of 0.03 and -0.03 following \citet{Jarrett_2012}. We convert the \textit{WISE} magnitudes to fluxes including a color correction dependent on the spectral slope\footnote{\url{http://wise2.ipac.caltech.edu/docs/release/allsky/expsup/sec4_4h.html}} with a further color correction to the \textit{W4} band due to rising spectral slopes in this band \citep{Brown_2014}. The \textit{HST} fluxes are measured 
from the imaging described in \S2.2.1 with the  \texttt{photutils}\footnote{\url{https://photutils.readthedocs.io/en/stable/}} package. The \textit{AKARI} data is from the Far-Infrared Surveyor \citep[FIS,][]{akari_fis} Bright Source Catalogue \citep[version 1,][]{akari-doi}.

Every flux with wavelength $<2\mu$m is corrected for Milky Way extinction using the Galactic extinction values from \citet{Schlafly_2011} provided in the NASA Extragalactic Database (NED\footnote{\url{http://ned.ipac.caltech.edu}}), and converted to Jy before the SED fitting. Finally, we convolve the catalogue errors in quadrature with a calibration error of 10\% of the flux for SDSS, \textit{HST}, \textit{Herschel}, \textit{Spitzer}, \textit{WISE W3}, \textit{AKARI} data, and 20\% of the flux for \textit{GALEX}, \textit{2MASS}, \textit{WISE W4} data, to account for differences in the methods used to measure the total photometry across different surveys and uncertainties in the spectral synthesis models used to fit the underlying stellar populations. The flux densities are summarized in Table \ref{tab:full_sed} in the Appendix.

\subsection{X-ray Observations}
IC 860 has been observed both by the \textit{Nuclear Spectroscopic Array Telescope} (\textit{NuSTAR}; \citealt{Harrison_2013}) and the Advanced CCD Imaging System (ACIS; \citealt{Weisskopf_2000}) on \textit{Chandra}. \textit{Chandra} observed IC 860 twice, once in Cycle 10 on 2009-03-24 for 19.15\,ks (PI Alexander; ObsID 10400) and again serendipitously in Cycle 22 on 2020-05-22 for 19.8\,ks (PI Lanz, ObsID 22622). {\em NuSTAR} observed IC 860 on 2018-02-01 for 72.5\,ks (PI Lanz, ObsID 6031024002). 

While this would seem like a wealth of data, IC 860 is not significantly detected by \textit{NuSTAR} and only weakly, if significantly, detected by \textit{Chandra}. In a 30\arcsec\, aperture placed around IC 860 in the \textit{NuSTAR} observation, the count rate is consistent with the background level. By contrast, a 3\arcsec\, aperture on the \textit{Chandra} observations yields 9 and 6 total counts in the Cycle 10 and Cycle 22 observations, respectively\footnote{The decrease in counts is probably due to a combination of low number statistics and the decreased sensitivity of ACIS due to the build up of contaminant \citep{Marshall_2004}.}. We use these observations to constrain the intrinsic X-ray luminosity of the galaxy's core in \S3.5.

\begin{table*}[t]
\caption{Radio Survey Data for IC 860}
\centering
\begin{tabular}{c|c|c|c|c|c}
\hline \hline
\textbf{Survey} & \textbf{Freq [MHz]} & \textbf{$\theta_{\rm FWHM}$ [arcsec]} & \textbf{S$_{\rm total}$ [mJy]} & \textbf{S$_{\rm peak}$ [mJy]} & \textbf{References}\\ 
\hline
TGSS ADR1 & 150  & 25  & 48.4 $\pm$ 6.9 & 39.3 $\pm$ 4.9  & \citet{Intema_2017} \\
VCSS$^{*}$  & 340  & 15  & 44.4 $\pm$ 8.8 & 44.0 $\pm$ 8.8 & \citet{Peters_2021} \\
RACS & 887  & 15  & 44.2 $\pm$ 8.8 & 42.0 $\pm$ 8.4 & \citet{McConnell_2020} \\
NVSS  & 1400 & 45  &  30.8 $\pm$ 6.2 & no data & \citet{Condon_1998} \\
FIRST & 1400 & 5   & 32.3 $\pm$ 6.5 & 31.2 $\pm$ 6.2 & \citet{Becker_1995} \\
VLASS & 3000 & 2.5 & 34.2 $\pm$ 0.2  & 32.9 $\pm$ 0.1  & \citet{Lacy_2020, Gordon_2020} \\
\hline \hline
\multicolumn{6}{p{18cm}}{\textbf{Notes: }$^{*}$Data from the VLITE$^{a}$ Commensal Sky Survey (VCSS$^{b}$) are preliminary. Flux bias corrections of 38\% and 48\% have been added to the peak and total flux values, respectively. 

$^{a}$The VLA Low-band Ionosphere and Transient Experiment (VLITE; \citealt{Clarke_2016,Polisensky_2016}) is a commensal instrument on the VLA that records data simultaneously at 340 MHz during regular VLA observations.

$^{b}$ VCSS data are taken simultaneously during VLASS.}
\label{tab:radio}
\end{tabular}
\end{table*}

\subsection{Radio Observations}
The basic radio properties of IC 860 have been reported previously in the literature (e.g., \citealt{Baan_2006, Parra_2010}).  IC 860 harbors a compact (sub-arcsecond), nuclear radio source with a luminosity of $1.15 \times10^{22}$~W~Hz$^{-1}$ ($1.61 \times10^{38}$ erg/s) at 1.4 GHz based on the NRAO Very Large Array (VLA) Sky Survey (NVSS; \citealt{Condon_1998}).

In addition to NVSS, IC 860 has been observed and detected in several other wide-field, public radio surveys spanning a broad range of frequencies. We summarize these survey data in Table \ref{tab:radio} and use them to explore the radio spectral properties of IC 860 in \S5.3.2.

\begin{figure*}
        \centering \includegraphics[width=2\columnwidth]{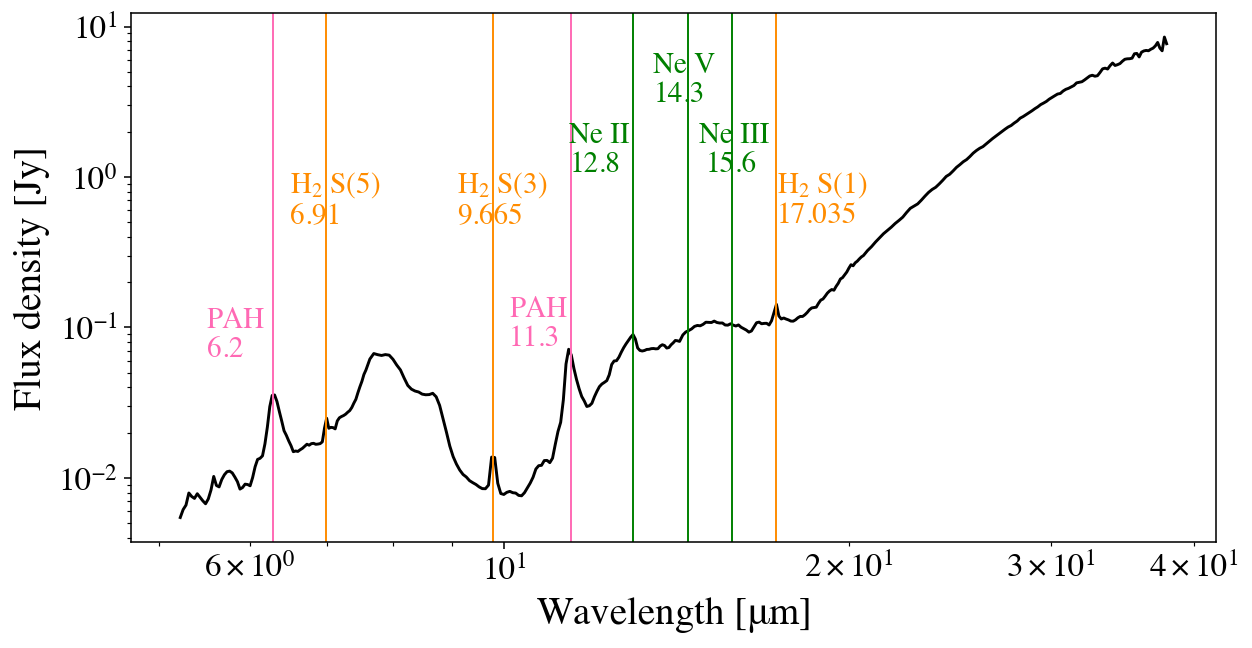}
        \caption{
                \label{fig:irs} 
                Low resolution \textit{Spitzer} IRS spectrum of IC 860. Some important line features are labelled with vertical lines and the numbers near the lines indicate the wavelengths. IC 860 exhibits deep silicate absorption feature at 9.7$\mu$m, emission from PAH, nebular emission lines, and H$_2$ rotational lines. The steep slope $F_{30\mu m}/F_{15\mu m}$ is indicative of hot dust possibly heated by an AGN.
        }
\end{figure*}

\subsection{Spitzer Infrared Spectrograph spectrum}
IC 860 was observed with the \textit{Spitzer} Infrared Spectrograph (IRS). The low resolution IRS spectrum covers 5--38$\mu$m and has been analyzed in previous work \citep[e.g.,][]{petric2011,Stierwalt_2013,inami2013,Stierwalt_2014,petric2018,Lambrides_2019}. IC 860's MIR emission is relatively compact and is completely covered by the IRS slit (\citealt{Stierwalt_2014} calculated the total-to-slit flux ratio at 8$\mu$m to be 1). We show the spectrum with some important features labelled in Figure \ref{fig:irs}.

The MIR spectrum contains important information about the galaxy's dust properties and any obscured AGN. IC 860 shows deep silicate absorption at 9.7$\mu$m, indicating the existence of a large amount of dust along the line of sight. The MIR slope characterized by the flux ratio $F_{30\mu m}/F_{15\mu m}$ is particularly steep compared to other LIRGs analyzed in \citet{Stierwalt_2013}, where they proposed that the steep slope could be caused by both high dust obscuration and high dust temperature.

The spectrum also contains emission from polycyclic aromatic hydrocarbons (PAH), nebular emission lines, and H$_2$ rotational lines. PAH emission is a good tracer of SF. \citet{Stierwalt_2013} measured the equivalent width of the 6.2$\mu$m PAH emission of IC 860 to be 0.43$\mu$m. Compared to that of typical SF galaxies ($>0.54\mu$m) and AGN host galaxies ($<0.27\mu$m), they suggested that the MIR emission in IC 860 is likely from a SF-AGN composite origin. The lack of robust detection of [\ion{Ne}{3}] and [\ion{Ne}{5}], which are tracers of SF and AGN respectively, also indicates neither SF nor AGN emisision is particularly strong or is heavily obscured in this galaxy.
The detection of H$_2$ rotational lines points to the existence of warm molecular gas in the galaxy. Using the IRS spectrum, \citet{petric2018} estimated for IC 860 the mass and temperature of the warm molecular gas to be 1.43 $\times$10$^{7}$M$_{\odot}$ and 404K, which suggests a non-SF heating source. We discuss more about the star formation rates and existence of an AGN later in \S4.6, \S5.1, and \S5.3.

\section{Analysis}
\subsection{Stellar Kinematics with pPXF}
Before fitting the kinematics, we first apply the Voronoi binning method \citep{Cappellari_2003} to the combined IFS cube based on data near the H$\alpha$ wavelength to increase the signal to noise ratio (SNR) of the spectra. We choose a wavelength range that encompasses the entire H$\alpha$ line, and use the mean SNR of data slices within this wavelength range as the initial SNR for the binning process. We exclude spaxels with SNR$_{\rm{H\alpha}}<$1 and set the target SNR of H$\alpha$ to be 8, as a high emission line SNR is also desired when fitting the ionized gas kinematics (\S3.2). The spectra in each bin are combined by taking the average. Under this binning scheme, the final SNR at 4000\AA, where important information about the stellar populations lie, is $\sim$5. This binning scheme thus should allow us to adequately measure the stellar kinematics. 
We test that Voronoi binning at 6000\AA\ (a line-free continuum region) to a target SNR of 10 produces almost identical fitting results. 

\begin{figure}
        \centering \includegraphics[width=\columnwidth]{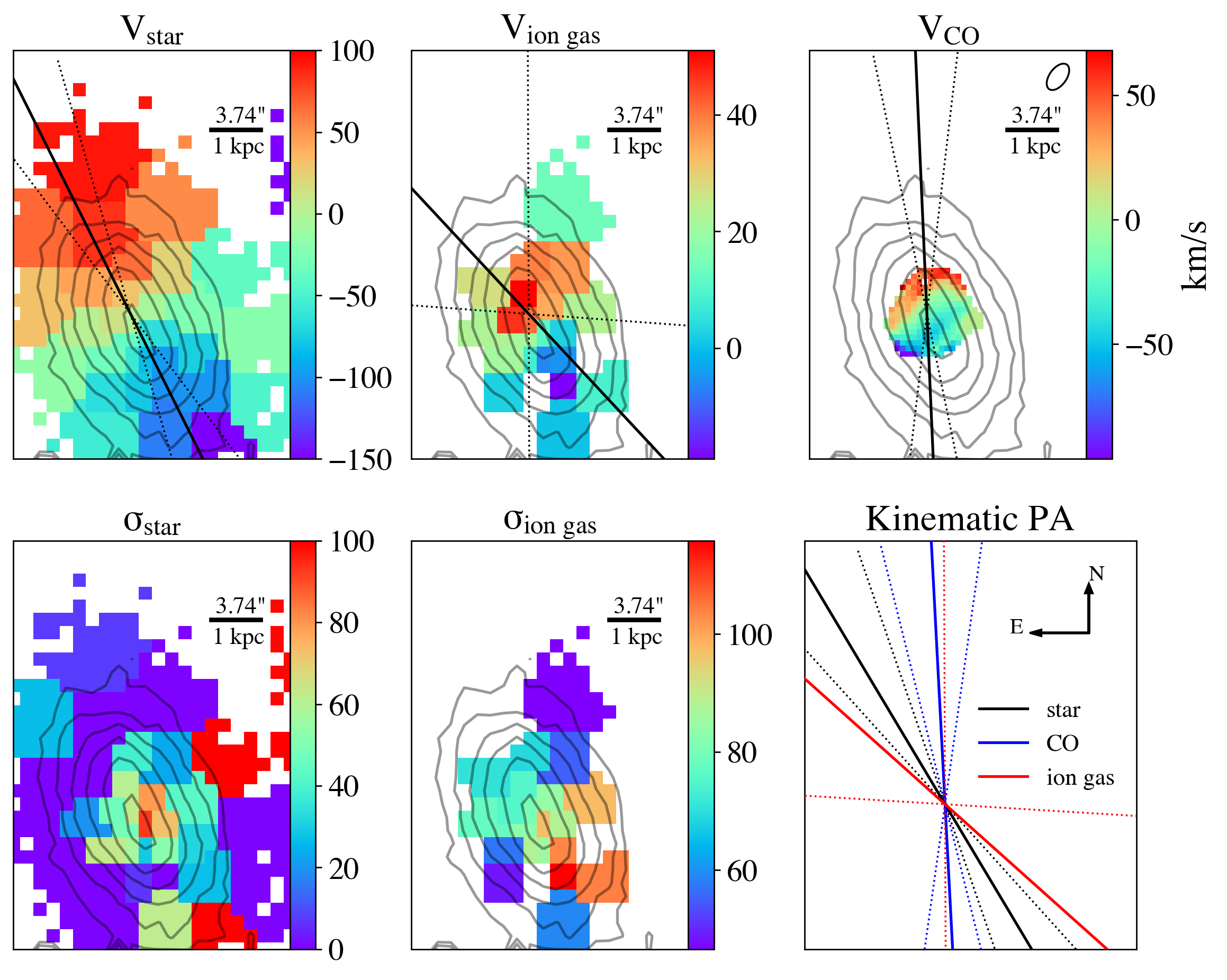}
        \caption{
                \label{fig:vmaps} 
                Fitted kinematic position angles (black solid lines, with black dotted lines representing uncertainties) overplotted on the corresponding velocity fields (centered on the galaxy systemic velocity). The axes are in spatial units (1 kpc indicated by a bar) and the colorbars are in km/s. The gray lines are logarithmically spaced contours from 30\% to 95\% of the integrated IFS flux peak. North is up in all plots, and the small ellipse in the CO velocity panel shows the beam. The position angles indicate that the stars and molecular gas are kinematically misaligned.
        }
\end{figure}

We fit the stellar continuum and obtain the stellar velocity and velocity dispersion using the penalized pixel-fitting method (\texttt{pPXF}\footnote{We use the python package ppxf version 7.4.2: \url{https://pypi.org/project/ppxf/}.}) by \citet{Cappellari_2017}, which is an upgrade to the original algorithm by \citet{Cappellari_2004}. We use the theoretical model-based stellar templates from \citet{Delgado_2005}, computed with the stellar isochrones from the Padova group \citep{Bertelli_1994,Girardi_2000,Girardi_2002}, covering ages of 4 Myr to 17 Gyr and metallicities Z = 0.019 (solar), 0.008 and 0.004. We first match (degrade) the spectral resolution of the templates to that of our data using the \texttt{match\_spectral\_resolution} function from the data analysis pipeline (DAP\footnote{\url{https://sdss-mangadap.readthedocs.io/en/latest/}}) of the Mapping Nearby Galaxies at Apache Point Observatory (MaNGA) survey \citep{Westfall_2019}. Each template spectrum is then resampled to log-linear wavelengths and normalized by its median flux. We fit the continuum with major emission lines, the NaD absorption feature, and the 5577\AA\ skyline masked, plus an additive Legendre polynomial of degree 10 to account for template mismatch and systematics in the flux calibration. An example of the binned spectrum and its fitted continuum is shown in Figure \ref{fig:comb}. The final stellar velocity and stellar velocity dispersion maps are shown in Figure \ref{fig:vmaps} and further discussed in \S4.3.

\begin{figure*}
        \centering \includegraphics[width=2\columnwidth]{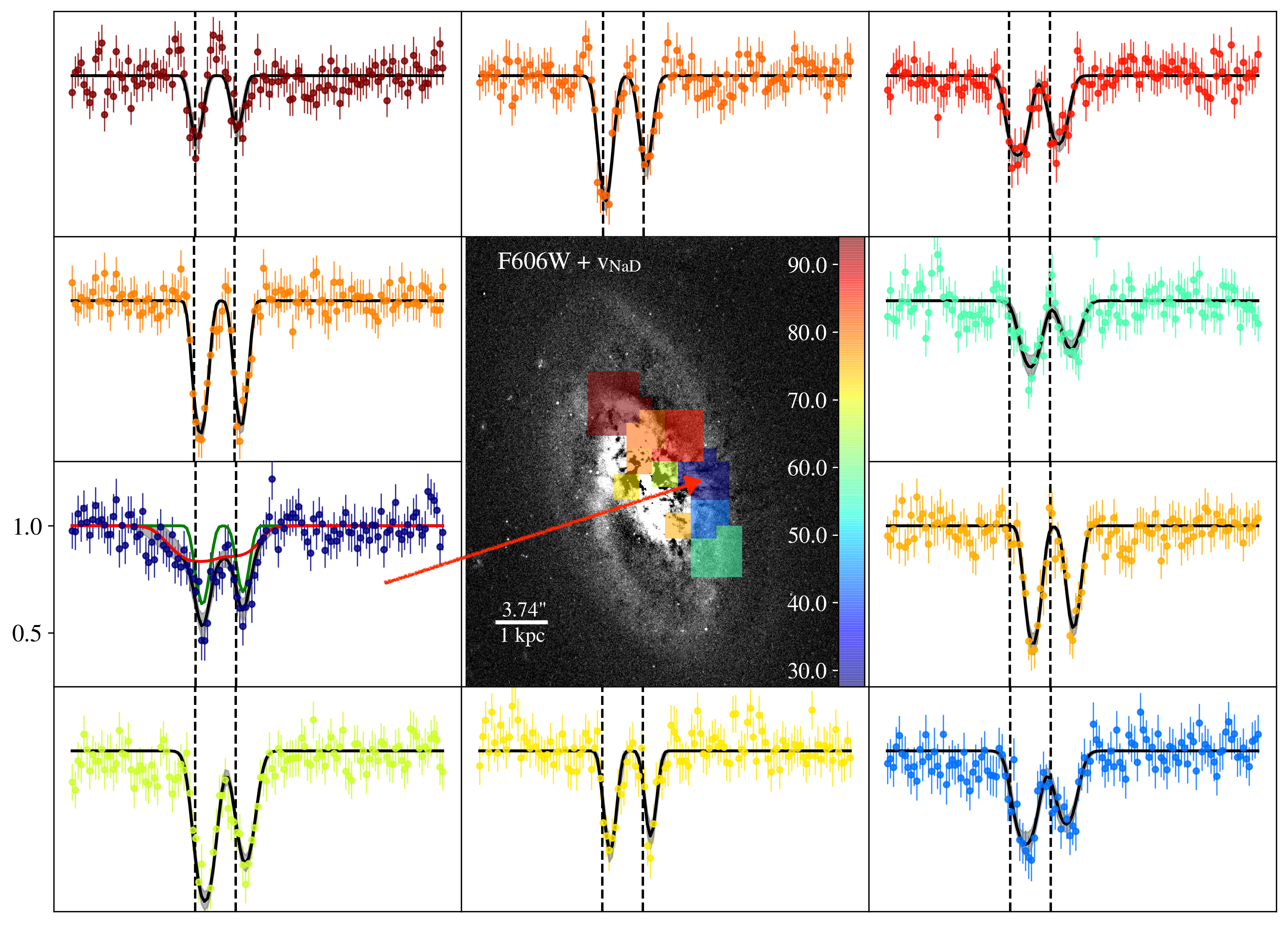}
        \caption{
                \label{fig:nad} 
                The middle panel shows the $v_{\rm{NaD}}$ map (in km/s, with respect to the galaxy systemic velocity) on top of the \textit{HST} V-band image, while the surrounding panels show the normalized NaD ISM profiles overplotted with the best fit models (black lines with shaded areas representing uncertainties), in the same color as the corresponding bin on the map. The outflow bin is marked with an arrow. The y-axes of all panels are set to the same scale. The blueshifted wing (red line) around the systemic component (green line) of the NaD absorption in the bottom left panel indicates the presence of an outflow, and the corresponding bin on the map overlaps with the dust cone. 
                The vertical dashed lines in the small panels mark the stellar restframe vacuum wavelengths of the NaD doublet.
        }
\end{figure*}

\begin{figure*}
        \centering \includegraphics[width=2\columnwidth]{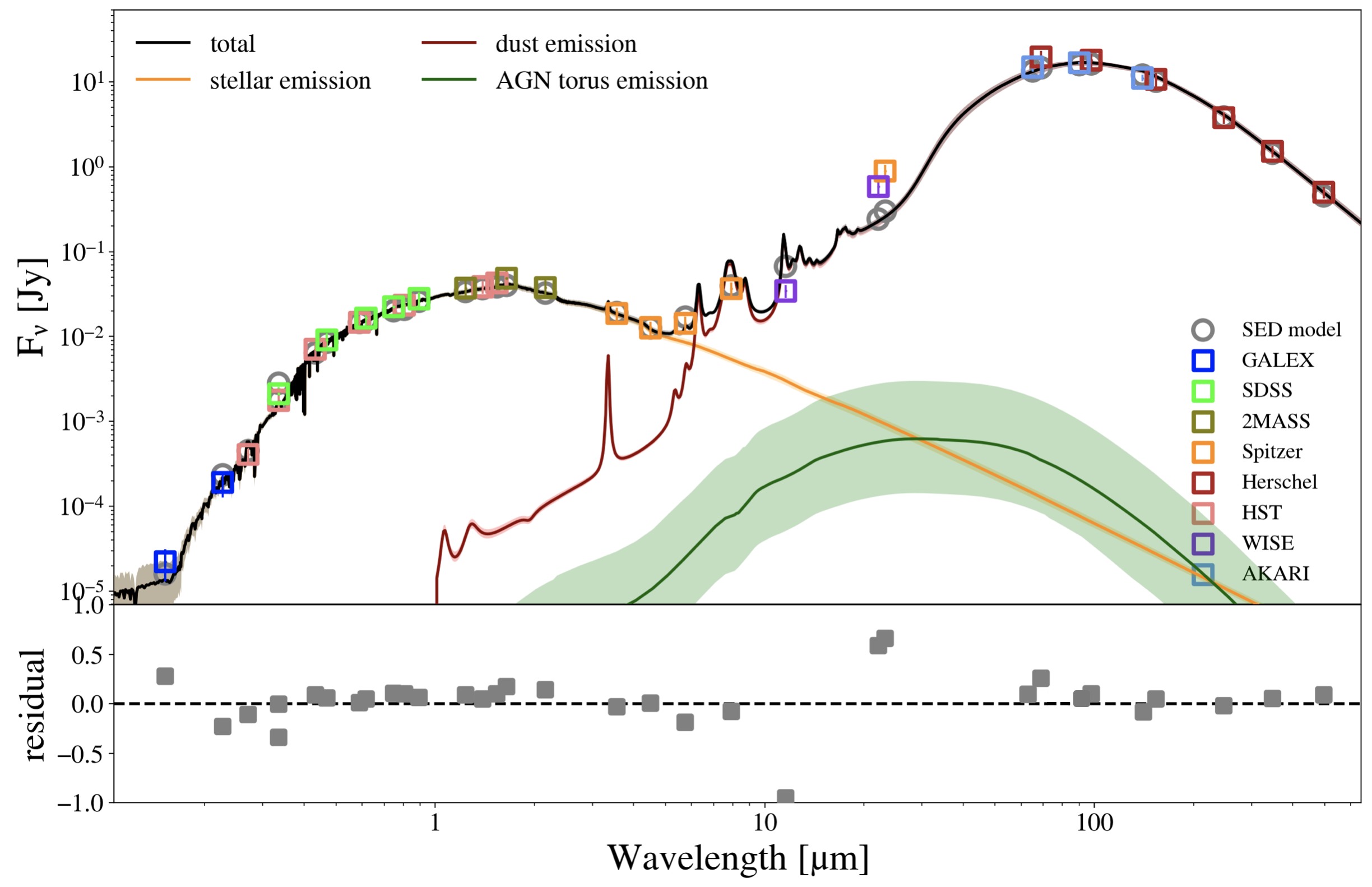}
        \caption{
                \label{fig:sed} 
                FUV to FIR SED of the entire galaxy plotted with the median posterior model spectrum from the MCMC SED fitting technique (\S3.4.1), with shaded regions showing the 16th--84th percentile uncertainties. The colored squares represent the actual data while the gray circles represent the model predictions. Decomposition of the model is plotted with different colors and the residual is calculated as $(F_{obs}-F_{model})/F_{obs}$.
        }
\end{figure*}

\begin{figure*}
        \centering \includegraphics[width=2\columnwidth]{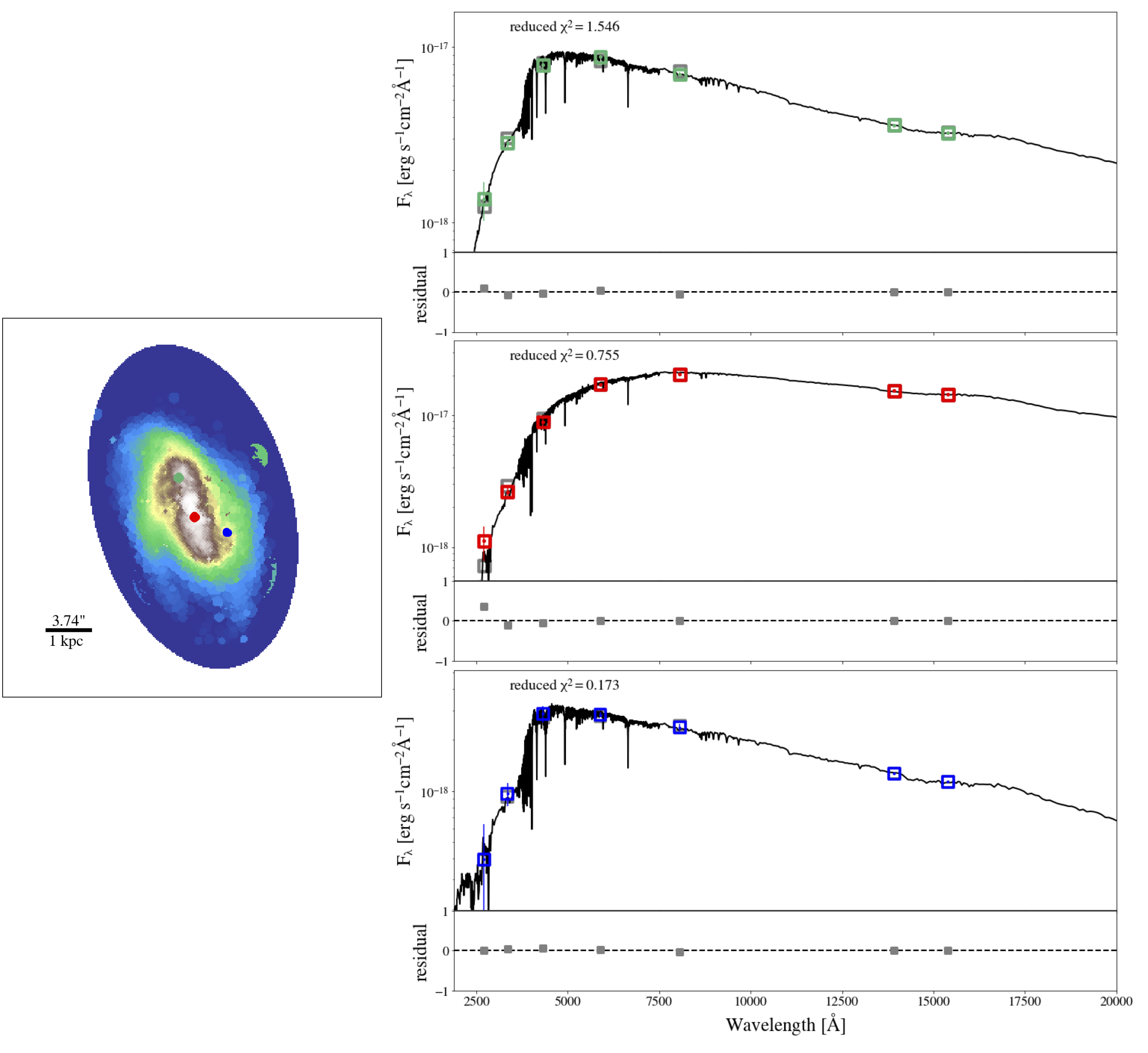}
        \caption{
                \label{fig:hst_sed} 
               Examples of spatially resolved SEDs constructed from \textit{HST} imaging data plotted with their best fit models (\S3.4.2). The left panel is a colormap showing the binning scheme and the right panels are SEDs from different galaxy regions marked by colored points on the bin map. The color of the data points on the SED matches the color of its location markers on the bin map.
        }
\end{figure*}

\subsection{Ionized Gas Properties with LZIFU}
We use \texttt{LZIFU} \citep{Ho_2016}, an \texttt{IDL} toolkit for fitting multiple Gaussian components to emission lines to determine the properties of ionized gas in IC 860. Considering that the two IFS data cubes have different spectral resolutions, we run \texttt{LZIFU} in its two-sided data mode on the blue and red data cubes to best preserve the information. \texttt{LZIFU} accepts an external continuum cube with linear wavelength grid, so we interpolate the log-linear stellar continuum cube obtained with \texttt{pPXF} (\S3.1) to the blue and red data cube wavelength grids, and have \texttt{LZIFU} skip its continuum fitting step. 

\texttt{LZIFU} subtracts the given continuum before the emission line fitting process. To account for the possibility of multiple kinematic components, we run the program three times, using 1--3 Gaussian components to fit all emission lines. 
Visual inspection of the model fits indicates that the one-component case best fits the data. The flux errors in the IFS cubes are propagated through the emission line fitting process to produce the error on the flux measured for each emission line. We use the results from the one Gaussian component fitting hereafter. 

\texttt{LZIFU} provides maps of emission line fluxes, velocity and velocity dispersion. The program was designed to fit all emission lines simultaneously with each kinematic component constrained to share the same velocity and velocity dispersion. In our case where we only fit one Gaussian (one kinematic component), we get one velocity map and one velocity dispersion map representative of all emission lines. The maps are shown in Figure \ref{fig:vmaps}. Given that the ionized gas velocity information is mainly constrained by the strongest emission line H$\alpha$, the velocity maps of ionized gas are masked such that only bins with H$\alpha$ flux SNR $>$ 3 are used.

\subsection{NaD kinematics fitting}
We detect absorption features of the resonant \ion{Na}{1} doublet at 5891.583 and 5897.558\AA\ (vacuum wavelengths, also referred to as NaD) in our binned IFS data. Stars and interstellar medium (ISM) can both contribute to the NaD absorption in galaxy spectra. With our primary interest in finding the evidence of an outflow, we focus on the ISM contribution to the NaD doublet, which traces cool (T $\lesssim$ 10$^4$ K), metal-enriched gas. We convert our IFS data from air to vacuum wavelengths then fit the velocity of NaD \textit{with respect to the stellar velocity} using a Bayesian inference method utilizing the \texttt{EMCEE}\footnote{\url{https://emcee.readthedocs.io/en/stable/}} package \citep{mcmc}. Our approach closely follows that in \citet{Roberts-Borsani_2019} and is described briefly below. We refer the readers to their work (and references therein) for more details.

We exclude the stellar contribution to the NaD absorption by removing from the spectra the stellar continuum fit in \S3.1. The NaD wavelength range is masked when fitting the continuum, since the stellar models only account for NaD absorption from the stars, not the ISM. 
The \ion{He}{1} emission line at 5875.67\AA\ is also masked in continuum fitting since it is close enough to the NaD line that it could affect the residual profile. The choice of stellar model used in the continuum fitting can affect the estimated NaD due to stars. \citet{Roy_2021} found that different models (theoretical and empirical) estimated the stellar component of NaD absorption generally within a variation of $\sim$10$-$30\%. This variation is relatively small and mostly affects the fitted line depths rather than line centers. The quality of the continuum fitting is verified by visual inspection. We remove the stellar continuum by dividing each spectrum by its continuum fit. The \ion{He}{1} line is separately fitted and subtracted if it is present. Additionally, we fit a first-order polynomial to the continuum-normalized spectrum in the wavelength range immediately ($\sim$20\AA) blueward and redward of the NaD profile and divide the residual by the polynomial fit. This is to account for any systematic errors in continuum fitting. Before fitting the NaD feature, each spectrum is shifted to the stellar restframe using both the redshift and the corresponding stellar velocity, so that the final fitted NaD velocity is with respect to the stars (which we call $\Delta v_{\rm{NaD}}$). We use the velocity of the fitted NaD line center throughout this work. 

We fit the NaD equivalent width (EW) with and without removing the stellar continuum (EW$_{\rm{ISM}}$ and EW$_{\rm{ISM+star}}$, respectively) and proceed to fit the NaD velocity only when the spectrum shows significant NaD absorption after removing the stellar continuum (SNR$_{\rm{NaD}}>$ 10, EW$_{\rm{ISM}}>$ 0.5\AA, and EW$_{\rm{ISM}}>$ 0.1EW$_{\rm{ISM+star}}$). The absorption profile is modelled by the analytical function from \citet{Rupke_2005a}:
\begin{equation}
    I(\lambda) = 1 - C_f + C_f \times e^{-\tau_{\rm{B}}(\lambda)-\tau_{\rm{R}}(\lambda)}
\end{equation}
where $C_f$ is the velocity-independent covering factor, related to the clumpiness of the gas along the line of sight, and $\tau_{\rm{B}}(\lambda)$ and $\tau_{\rm{R}}(\lambda)$ are the optical depths of the \ion{Na}{1} line at 5891.583 and 5897.558\AA, respectively. Specifically, the optical depth is expressed as following:
\begin{equation}
    \tau(\lambda) = \tau_0 \times e^{-(\lambda-(\lambda_0+\Delta \lambda_{\rm{offset}}))^2/((\lambda_0+\Delta \lambda_{\rm{offset}})b_{\rm{D}})/c)^2}
\end{equation}
where $\tau_0$, $\lambda_0$, $b_{\rm{D}}$, and $c$ are the central optical depth of each line component, the central wavelength of each line component, the Doppler line width, and the speed of light. The wavelength offset is converted from the velocity offset by $\Delta \lambda_{\rm{offset}} = \Delta v \lambda_0/c$. The \ion{Na}{1} 5891.583\AA\ line has twice the depth of the \ion{Na}{1} 5897.558\AA\ line \citep[$\tau_{\rm{0,B}}/\tau_{\rm{0,R}}=$ 2,][]{Morton_1991}, and the optical depth parameter can be derived from the column density of Na:
\begin{equation}
    N(\rm{Na\ I}) = \frac{\tau_0 b}{1.497\times10^{-15}\lambda_0 f}\  \rm{cm}^{-2}
\end{equation}
where $\lambda_0$ and $f$ are the restframe vaccuum wavelength and oscillator strength, respectively. Throughout this study we assume $\lambda_0$ = 5897.55\AA\ and $f$ = 0.318 following \citet{Morton_1991}. The parameters left free in the fitting process are: $b_{\rm{D}}$, $C_f$, N(\rm{Na\ I}), $\Delta v$.

When the absorption line profile is well fitted by only one kinematic component, no outflow is detected and the fitted velocity is the systemic velocity of the neutral gas traced by NaD. We fit all our qualified spectra (10 in total) with one (systemic) and then two (systemic + outflow) kinematic component(s) and inspect the results to determine which scheme best fits each spectrum. When fitting with two components, the systemic component velocity is fixed to the one-component fitted velocity, since any inflow or outflow feature would manifest itself as a shifted component with respect to the systemic line center. 9 out of 10 spectra are best fit by one component, and one spectrum is best fit by two components. The second component in this bin traces a blueshifted wing compared to the systemic line center, indicating the presence of an outflow. The outflow velocity is taken to be the velocity difference between the systemic and outflow components (58 km/s). We also obtain from this fitting other parameters such as N(\ion{Na}{1}) and $C_f$ which are later used to calculate the mass outflow rate. To make the derived NaD velocity field with respect to the stellar velocity ($\Delta v_{\rm{NaD}}$) comparable to the velocity fields of other components in Figure \ref{fig:vmaps}, we obtain the NaD velocity field with respect to the galactic systemic velocity ($v_{\rm{NaD}}$) from $v_{\rm{NaD}} = \Delta v_{\rm{NaD}} + v_* - v_{\rm{sys}}$, where $v_{\rm{sys}}$ is defined as the stellar velocity at the galactic center. We show this $v_{\rm{NaD}}$ map and NaD profiles in Figure \ref{fig:nad} and further discuss the results in \S4.4. 

We observe in Figure \ref{fig:nad} that the line center of NaD is redshifted compared to the vacuum restframe wavelength in all bins. This could imply that there might be cool neutral gas infalling towards the galaxy, but further analysis with higher quality data is required to confirm this.

\begin{table*}[t]
\caption{IC 860 Derived Characteristic Properties}
\centering
\begin{tabular}{c c p{11cm}}
\hline \hline
\textbf{Derived Property} & \textbf{Value} & \textbf{Descriptions}\\
\hline
log(M$_*$[M$_{\odot}$]) & 10.96$^{+0.06}_{-0.05}$ & Stellar mass \\
$\tau_1$ & 1.27$^{+0.35}_{-0.19}$ & Dust optical depth of the birth cloud in the \citet{Charlot_2000} dust attenuation law \\
$\tau_2$ & 2.11$^{+0.11}_{-0.12}$ & Dust optical depth of the diffuse ISM in the \citet{Charlot_2000} dust attenuation laws\\
log(M$_{dust}$[M$_{\odot}$]) & 7.42$^{+0.09}_{-0.05}$ & Dust mass \\
log(Age$_{*}$[Gyr]) & 0.80$^{+0.11}_{-0.11}$ & Mass-weighted age of the stellar population\\
log(f$_{\rm{AGN}}$) & -3.97$^{+0.69}_{-0.87}$ & AGN luminosity as a fraction of the galaxy bolometric luminosity\\
\hline
$i$ [deg] & 54 & Galaxy inclination angle derived from axial ratio \\
M(H$_2$) [M$_{\odot}$] & (1.79 $\pm$ 0.54)$\times$10$^9$ & Total molecular gas mass derived from CO luminosity \\
$v_{\rm{neutral}}$ [km/s] & 58--100 & Neutral outflow velocity from NaD absorption fitting\\
$v_{\rm{molecular}}$ [km/s] & 200--340 & Molecular outflow velocity from the CO PV diagram\\
$\dot{M}_{\rm{neutral}}$ [M$_{\odot}$/yr] & 0.5 & Neutral mass outflow rate from NaD absorption fitting\\
M(H)$_{\rm{outflow}}$ [M$_{\odot}$] & 2$\times$10$^8$ & Total hydrogen outflow mass derived from extinction maps\\
$\dot{M}(H)$ [M$_{\odot}$/yr] & 12 & Total hydrogen mass outflow rate \\
M(H$_2$)$_{\rm{outflow}}$ [M$_{\odot}$] & 10$^6$ & Molecular outflow mass derived from redshifted emission on the CO PV diagram\\
$\dot{M}$(H$_2$) [M$_{\odot}$/yr] & 0.4 & Molecular mass outflow rate from outflow mass and velocity\\
\hline \hline
\multicolumn{3}{p{18cm}}{\textbf{Notes: }Parameters in the upper half of the table are obtained from the SED fitting (\S3.4.1), with superscripts and subscripts indicating the 84th and 16th percentile errors from the posterior distributions. The outflow velocities and mass outflow rate are represented as ranges here to account for the effect of galaxy inclination.}
\label{tab:property}
\end{tabular}
\end{table*}

\begin{figure*}
        \centering \includegraphics[width=2\columnwidth]{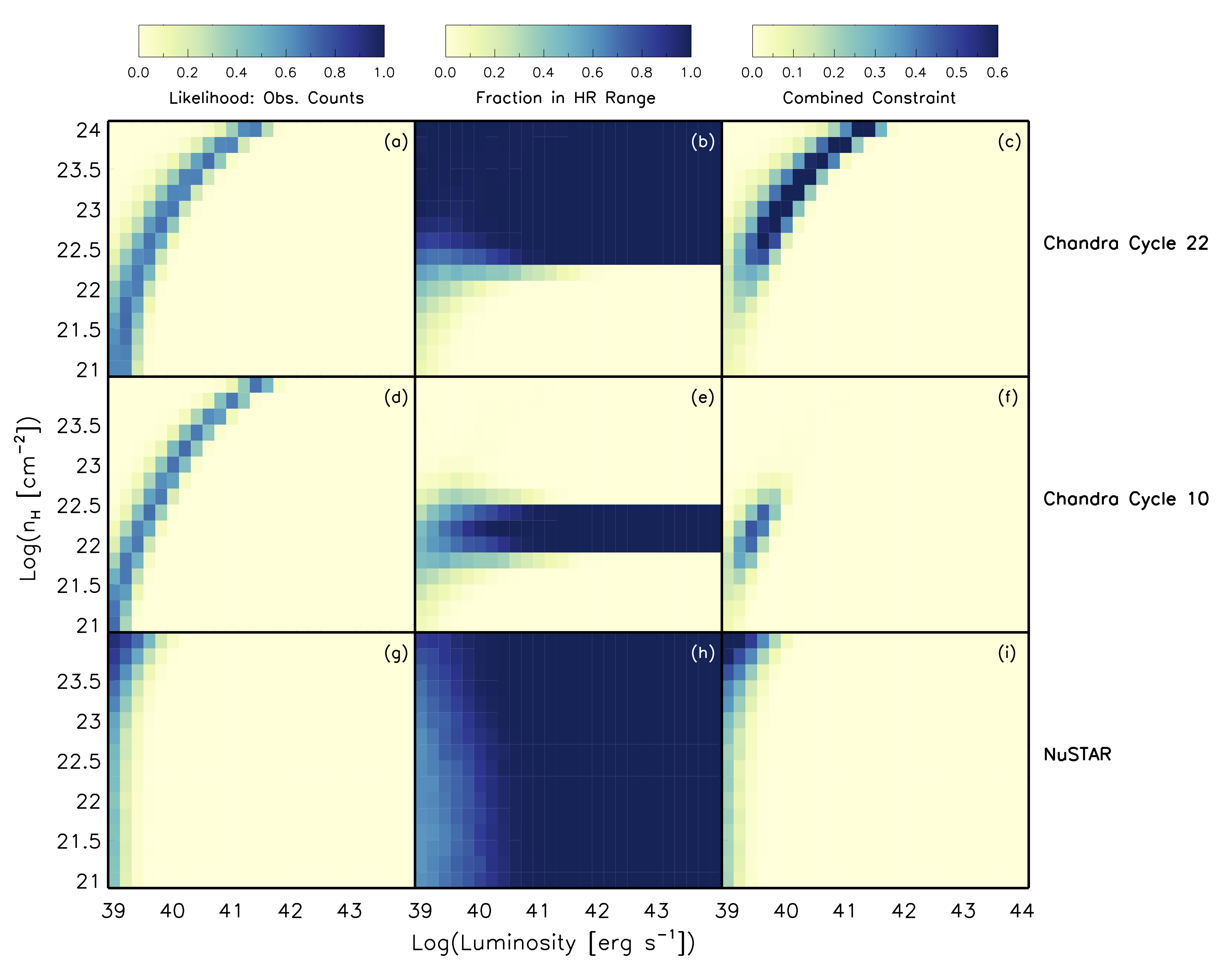}
        \caption{
                \label{fig:xray} 
               Agreement of mock spectra created during forward modeling analysis (see \S3.5 for more detail) with observed constraints from three X-ray observations of IC 860. The x-axes are the intrinsic 2--10 keV luminosity of the AGN, while the y-axes are the local obscuration column density. Each row shows the comparison to a different dataset, labeled on the right. \textbf{Left:} the likelihood that each model (defined by an intrinsic luminosity and its obscuration) could create the number of observed 0.5--8 keV (\textit{Chandra}) or 3--24 keV (\textit{NuSTAR}) photons, effectively acting as a constraint on normalization of the model. \textbf{Middle:} the fraction of the 1000 spectra generated for each model that had a hardness ratio consistent with the range determined with BEHR based on the observation, effectively acting as a constraint on the spectral shape of the model. \textbf{Right:} the combined constraint to determine the models most likely to have both the normalization and spectral shape of the AGN needed to reproduce the sparse observed data. While the three datasets do not tell a completely consistent story, they firmly rule out bright AGN with intrinsic luminosity above $10^{42}$ erg/s.
        }
\end{figure*}

\subsection{SED Fitting with piXedfit}
We perform the SED fitting with the python package \texttt{piXedfit}\footnote{\url{https://github.com/aabdurrouf/piXedfit}} \citep{Abdurrouf_2021}, which is a powerful tool in fitting both spatially resolved SEDs and a single SED for the entire galaxy. In our fitting of both the entire galaxy's SED and the spatially resolved SEDs, we use models that assume a double power law star formation history (because of its flexibility in modeling the rising and falling phases), the \citet{Chabrier_2003} initial mass function (IMF), and the two-component dust model of \citet{Charlot_2000}. The model spectra are generated using the Flexible Stellar Population Synthesis \citep[FSPS,][]{Conroy_2009}. We also add in the dust emission and AGN dusty torus emission when generating the models, but do not include the intergalactic medium absorption and the nebular emission because IC 860 is relatively nearby and shows weak emission lines in the IFS data. The dust emission modeling follows the energy balance principle, where the amount of energy attenuated by the dust in the UV is equal to the amount of energy re-emitted in the infrared \citep{daCunha_2008}. \texttt{piXedfit} uses the \citet{Draine_2007} dust emission templates to describe the shape of the infrared SED, and the AGN templates from the \citet{Nenkova_2008a,Nenkova_2008b} \texttt{CLUMPY} models to model the AGN-heated dusty torus emission. When generating random models for the fitting, we follow the default settings on the prior ranges in \texttt{piXedfit} for most of the parameters, while changing the ranges for some parameters to better fit the situation of IC 860. We provide the list of parameter prior ranges that we modify as online material. The final prior ranges of the parameters are wide enough to cover a variety of physical situations.

\subsubsection{The Entire Galaxy}
We use \texttt{piXedfit}'s MCMC posterior sampling method when fitting the SED of the entire galaxy, using the total photometry measurements described in \S2.5. The model library contains 800,000 random SEDs. We use 100 MCMC walkers, and 600 steps per walker. We fit the SED using sets of models with and without including the AGN dusty torus emission and find models with this component produce a better fit to the photometry. We thus adopt the results from fitting with models including the AGN component. The best fit SED model is shown in Figure \ref{fig:sed} and relevant fitted parameters are listed in Table \ref{tab:property}. Our best fit model seems to underestimate the flux in the MIR, while fitting the rest of the data points well. Strong MIR emission has been widely seen in PSBs and likely includes contributions not only from SF, but also AGN, strong PAH features, dust-enshrouded AGB stars, and/or warm dust \citep{Alatalo_2017,Smercina_2018}. The nature of the MIR excess in PSBs is not yet understood so SED models for typical galaxies might not accurately reproduce the MIR properties of PSBs. Similar underestimations of MIR emission in PSBs by SED models have also been observed by \citet{Suess_2022}. \citet{Pereira-Santaella_2015} performed SED fitting on IC 860 with similar physical assumptions as described at the beginning of this section. Though our derived galaxy properties such as the dust mass are similar to their results, their best-fitting model underestimates the flux in both the optical and the MIR. For IC 860, we speculate that there is a source of warm dust not represented in the models, which could be an extremely obscured AGN or compact starburst \citep[also discussed in][]{Aalto_2019}. The AGN luminosity from the SED fitting is less than 10$^{-3}$ of the galaxy bolometric luminosity and is likely an underestimate given that the model underestimates MIR flux. The AGN bolometric fraction as estimated from pure MIR data, however, is also small \citep[0.06 in][]{Diaz-Santos_2017}. Thus our SED fitting results agree with the literature and we discuss more evidence for the presence of an AGN in \S5.3.

\subsubsection{The Spatially Resolved SEDs}
The spatially resolved SEDs use the \textit{HST} data described in \S2.2.2. To prepare the data for the SED fitting, we select the galaxy region from the \textit{HST} image using an elliptical mask with the major axis being the $r-$band Petrosian radius from SDSS DR16, since $r-$band is the deepest band. The axis ratio of 0.6 is from the $g-$band isophotal $b/a$ reported by SDSS following \citet{Masters_2010}. We increase the SNR of the SEDs utilizing the binning function in \texttt{piXedfit}. To achieve reasonable SNR while best preserving the spatial resolution, we set the binning target SNR to 1.5, 3, and 5 for bands F275W, F336W, and all other bands, respectively. This results in 1710 bins. We correct the UV-optical part of the binned SEDs for Milky Way extinction in the same way as we correct the entire galaxy's SED (\S2.5).

Given the large number of SEDs, we use \texttt{piXedfit}'s random densely-sampling of parameter space (RDSPS) method to achieve a faster runtime. The model library contains 800,000 random SED models. \citet{Abdurrouf_2021} has shown that the RDSPS method produces results as good as the MCMC method. Following their work, we use a Student’s t likelihood function with degree of freedom $\nu\sim 2$. Examples of spatially resolved SEDs with their best fits are shown in Figure \ref{fig:hst_sed}. 
From these SEDs at different locations of the galaxy, there appears to be a continuum color gradient from the ends to the center of the bar. The color of the SEDs at the ends of the bar is bluer. This color difference could be caused by differences in dust properties and/or stellar age. Disentangling the influences from dust and stellar age in this case requires multiwavelength data of higher spatial and spectral resolution and is beyond the scope of this paper, but could be helpful in constraining locations of recent SF. We also obtain the extinction A$_{\rm{V}}$ in each bin from fitting the SEDs. Further application of the A$_{\rm{V}}$ maps is discussed in \S4.5.

\begin{figure*}
        \centering \includegraphics[width=2\columnwidth]{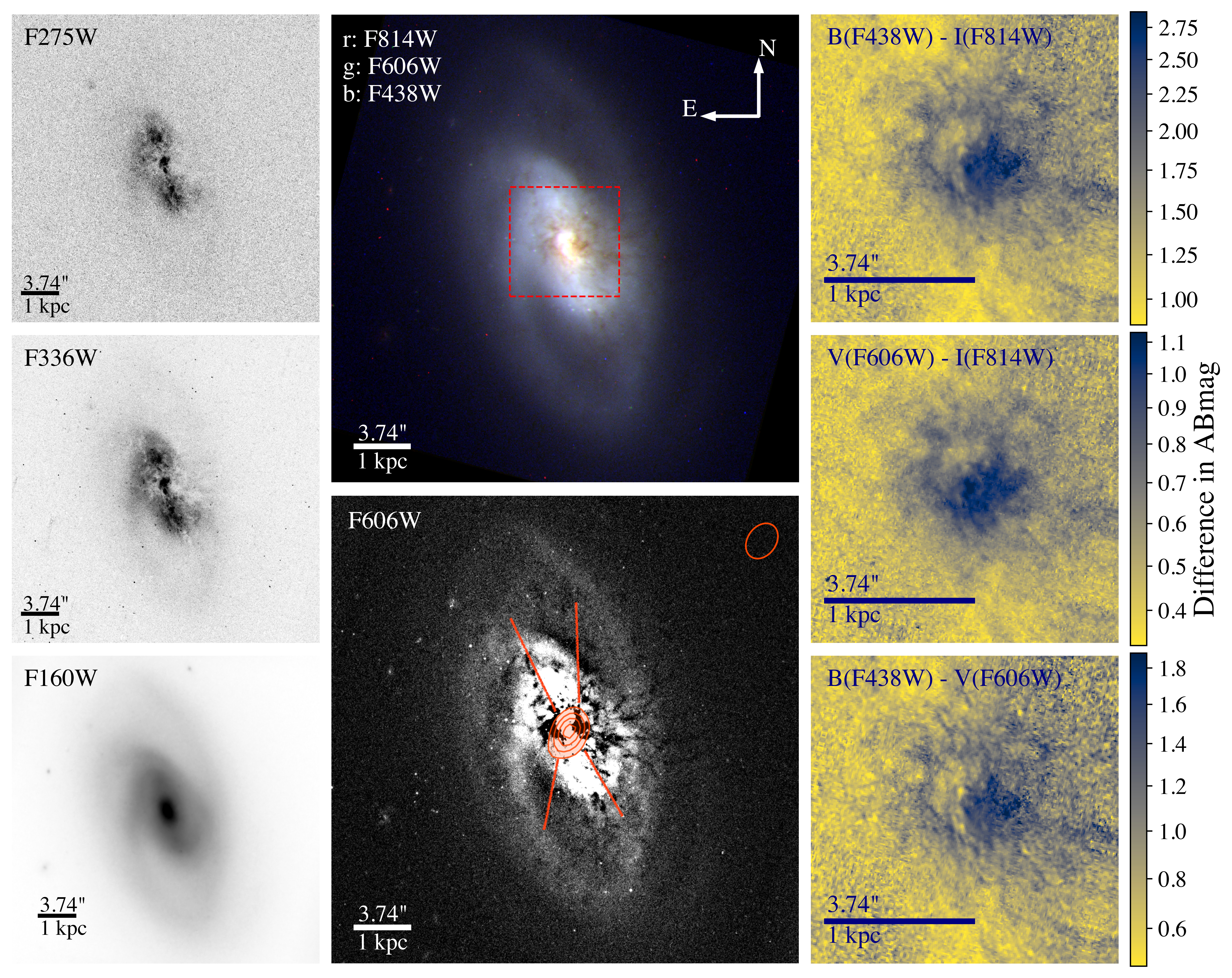}
        \caption{
                \label{fig:hst} 
                \textbf{Left:} Images from \textit{HST} WFC3 filters F275W (NUV-band), F336W (U-band), F160W (H-band). \textbf{Middle top:} A 3-color image of IC 860 constructed from \textit{HST} WFC3 filters F814W, F606W, F438W, as R, G, and B channels, respectively. This image clearly reveals that IC 860 is a barred spiral galaxy. \textbf{Middle bottom:} F606W (V-band) image after applying an unsharp mask to increase the contrast of the weak features. We can clearly see the dust cones (wide opening angle outlined by the red lines) suggesting a galaxy-wide outflow. The overlaid contours represent the 113 GHz radio continuum source from CARMA (\S2.4), with the beam size indicated at the upper right. The contour levels are 30\%, 50\%, 70\%, and 90\% of the maximum flux. \textbf{Right:} Color difference images of IC 860's central regions (marked by the red box on the 3-color image). All images follow the convention that north is up and east is left.
        }
\end{figure*}

\subsection{X-ray Luminosity with Forward Modeling}
To test for the presence of an AGN at the center of IC 860 (more discussions in \S5.3) here we present our efforts to determine the luminosity of this possible AGN using the X-ray observations.

The X-ray observations provide too few photons to reliably determine the intrinsic X-ray luminosity of IC 860's AGN, as a low count rate can be due to low luminosity, high obscuration, or both. However, we can still place constraints on the parameter space of luminosity and obscuration on the basis of the normalization set by count rate and approximate spectral shape as measured by the hardness ratio. We set these constraints using a forward modeling analysis applied to each of the three observations, following the methodology detailed in \citet{Lanz_2022}. Specifically, for each observation, we generate 1000 realizations of spectra obtainable in the observed exposure time that is consistent with 416 different combinations of intrinsic 2--10\,keV luminosity and obscuration levels (set by the column density $n_H$). These 416 power-law models have 26 different luminosities ranging from $10^{39}$\,erg/s to $10^{44}$\,erg/s and 16 obscurations between $10^{21}$\,cm$^{-2}$ and $10^{24}$\,cm$^{-2}$. We set the photon index to 1.8 \citep{Piconcelli_2005,Dadina_2008} and use \texttt{sherpa}\footnote{\url{https://cxc.cfa.harvard.edu/sherpa/index.html}} \citep{Freeman_2001} to generate a simple power-law, which is then normalized based on the desired luminosity and the 55.8 Mpc luminosity distance of IC 860. \texttt{sherpa} then creates 1000 realizations of this normalized power-law, but now with two additional absorption components: the local value for this parameter pair and a constant foreground obscuration with $n_H = 1.03\times10^{22}$\,cm$^{-2}$ \citep{HI4PI}. Absorption is applied with the \texttt{xsphabs} photoelectric absorption model. Once all of the models have been generated, we determine: 1) the likelihood that each of the 416 models is consistent with the observation by calculating the fraction of spectra (1000 per model) within the hardness ratio range of the observation as determined with the Bayesian Estimation of Hardness Ratios (BEHR; \citealt{Park_2006}) program; and 2) the likelihood for yielding a total photon number (0.5--8\,keV for \textit{Chandra}, 3--24\,keV for \textit{NuSTAR}) in the range of the observed photon number plus or minus the Gehrels uncertainty \citep{Gehrels_1986}. We then multiply these two constraints together to determine the likelihood that a particular model matches both the normalization and shape constraints imposed by the observed counts. Figure \ref{fig:xray} shows the parameter space and constraints of the models for all three datasets. Our modeling robustly rejects cases where the intrinsic luminosity is $>10^{42}$ erg/s, but tells a mixed story for lower luminosity scenarios. An intrinsic luminosity of 10$^{41}$--10$^{42}$ erg/s only works in heavily obscured cases and we cannot significantly reject the possibility of $L\sim10^{40}$--$10^{41}$ erg/s. However, the intrinsic luminosity range of $10^{39}$--10$^{40}$ erg/s, hitting the lower end of our models, has the highest number of spectra consistent with the observations from all three datasets with an average likelihood per model of 12--16\% (panels c, f, and i in Figure \ref{fig:xray}). \citet{Ricci_2021} recently investigated the X-ray luminosity of a sample of LIRGs including IC 860, and reported the intrinsic AGN luminosity of IC 860 to be $<$10$^{39.76}$ erg/s at 2--10 keV and $<$10$^{39.60}$ erg/s at 10--24 keV, which are consistent with our estimations.

\section{Results}

\begin{figure*}
        \centering \includegraphics[width=2\columnwidth]{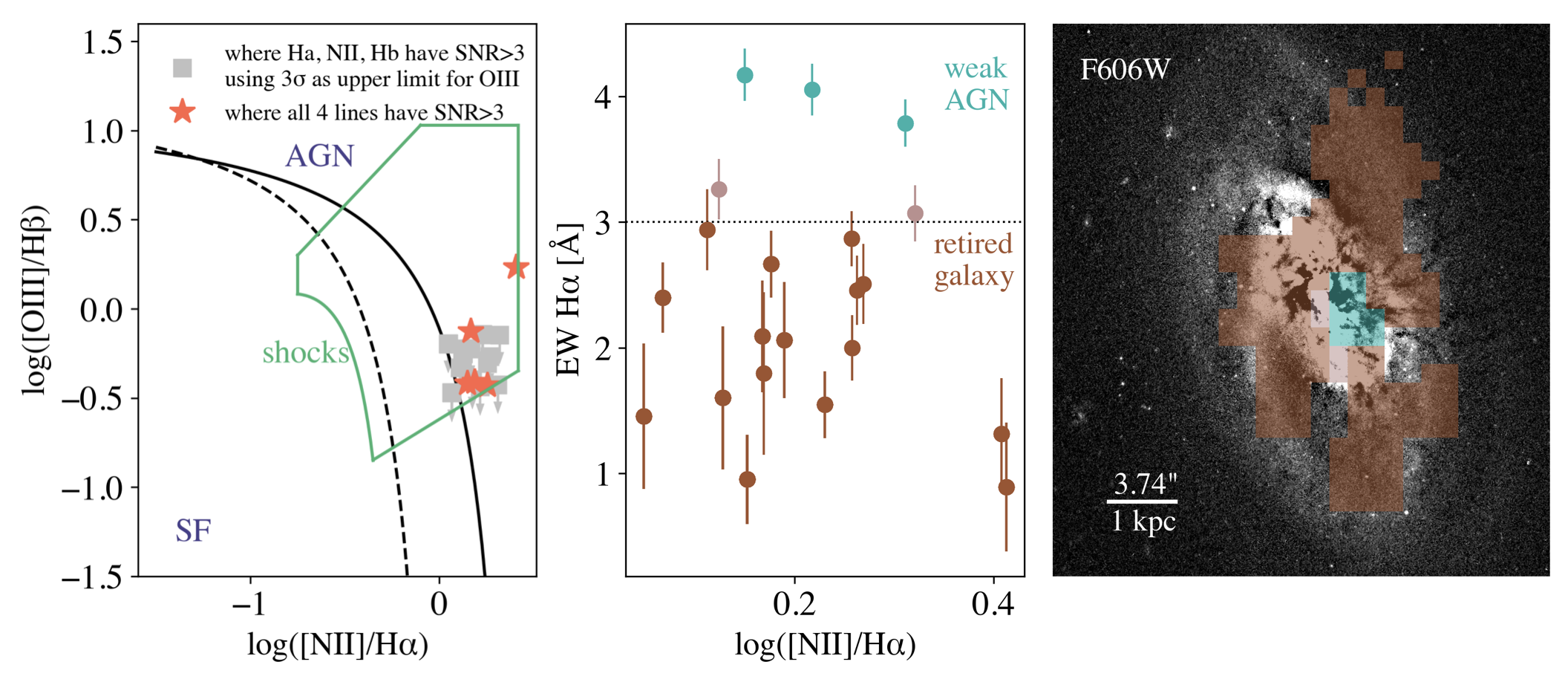}
        \caption{
                \label{fig:diagnostics} 
        \textbf{Left:} The [\ion{N}{2}]/H$\alpha$ vs [\ion{O}{3}]/H$\beta$ diagnostic diagram. The dashed and solid lines are demarcation lines for SF and AGN ionization, from \citet{Kauffmann_2003} and \citet{Kewley_2001}, respectively. The [\ion{O}{3}] line is not robustly detected ($>$3$\sigma$) in many bins, so we use the 3$\sigma$ value as the upper limit for the [\ion{O}{3}] line flux (grey squares with downward arrows). Bins where all four lines are robustly detected are shown as red stars. The green lines show the shock ionization boundaries from \citet{Alatalo_2016}. 
        \textbf{Middle:} The WHAN diagnostic diagram with classifications shown in different colors. The star forming category, which is not shown because no spaxels meet this classification, lies to the upper left of the plotted region. \textbf{Right:} The WHAN classification map on top of the \textit{HST} V-band image. The center of the galaxy is classified as weak AGN ionization while the remaining regions are consistent with the retired galaxy category (\S4.2).
        }
\end{figure*}

\subsection{Optical Morphology}
High resolution \textit{HST} images (Figure \ref{fig:hst}) reveal some interesting morphological features in IC 860. IC 860 is a barred spiral galaxy, but apart from the two spiral arms originating from the two ends of the bar, there appears to be a third ``arm" which could be a tidally-induced spiral structure, indicating that IC 860 is slightly disturbed in morphology. According to N-body simulations, minor mergers could trigger episodic SF bursts, and the most likely remnants of these minor mergers are disturbed spirals \citep{Bournaud_2007}. Thus IC 860 might have gone through a recent merger and we discuss more merger evidence in \S4.3.
In the UV images (filters F275W, F336W), the spiral arms are less prominent while the bar stands out. This means UV-emitting young stars are mostly located in the central bar region of the galaxy, and IC 860 might correspond to the outside-in quenching scenario.

One other prominent morphological feature is the dust filaments that trace the shape of a cone. Such dusty cone structures have been seen to correlate with neutral outflows traced by NaD \citep[e.g., NGC 1808, ][]{Phillips_1993}. Thus the dust filaments in IC 860 provide smoking gun evidence for a dust-rich outflow originating from the center of the galaxy. The \textit{HST} image in filter F606W (V-band) with an unsharp mask in the middle bottom panel of Figure \ref{fig:hst} not only clearly reveals the 
dust cone, but also indicates that this 
disrupted dust feature is galaxy-wide. 

The color difference images (right column of Figure. \ref{fig:hst}) show the differences in AB magnitudes between one bluer and one redder \textit{HST} band. These images reveal red, filamentary structures at the galactic center, characteristic of dust. The dust at galactic center appears very dense and the core of IC 860 is classified as a compact obscured nucleus in \citet{Aalto_2019}, where they estimated a H$_2$ column density N(H$_2$) $\gtrsim 10^{26}\ \rm{cm}^{-2}$ at the center. 

Although the orientation of the galaxy is hard to determine, we hypothesize that the right side of the galaxy is the side that is tilted away from us. This is because the dust filaments are more clearly seen on the right side of the galaxy in Figure \ref{fig:hst}. NaD absorption from the ISM also lies on the right side of the galaxy (Figure \ref{fig:nad}, middle panel).

\subsection{Ionization Mechanisms}
We use several diagnostic diagrams (Figure \ref{fig:diagnostics}) to study the ionization mechanisms in IC 860. To begin with, we plot the well-known [\ion{O}{3}]/H$\beta$ vs [\ion{N}{2}]/H$\alpha$ BPT diagram. Utilizing emission line fluxes obtained in \S3.2, we calculate the above line ratios and classify whether a bin is ionized by SF or AGN/shocks using the demarcation lines from \citet{Kauffmann_2003} and \citet{Kewley_2001}. We also plot the shock ionization boundaries from \citet{Alatalo_2016}. Only emission lines with measured fluxes greater than 3$\sigma$ are used. There are few significant detections of the [\ion{O}{3}] line, so we use the 3$\sigma$ value as the upper limit for the [\ion{O}{3}] flux in cases where the other three lines are detected above 3$\sigma$. All of our data points lie above the pure SF ionization line and fall inside the shock boundaries, indicating the presence of other ionizing sources such as AGN, shocks, and/or post-AGB stars.

We measure the EW of H$\alpha$ on the binned IFS data.  In Figure \ref{fig:diagnostics} we plot the Width of H$\alpha$ versus \ion{N}{2} \citep[WHAN diagram,][]{CidFernandes_2011}, which uses EW$_{H\alpha}$ and the [\ion{N}{2}]/H$\alpha$ flux ratio. The advantage of the WHAN diagram is that it uses only two of the strongest emission lines and can be applied to weak emission line galaxies where other lines are not detected. Accounting for the uncertainty on our EW$_{H\alpha}$ measurement, three bins are classified as weak AGN ionization at 3$\sigma$ level, and the rest are classified as retired galaxies. Retired galaxies (in our case retired “spaxels”) are systems that have stopped forming stars and whose old stellar populations make at least a substantial contribution to the ionizing field. We overlay the WHAN classification map of our bins on the unsharp-masked \textit{HST} V-band image and see that the weak AGN ionized bins are located near the galactic center.

\begin{figure}
        \centering \includegraphics[width=\columnwidth]{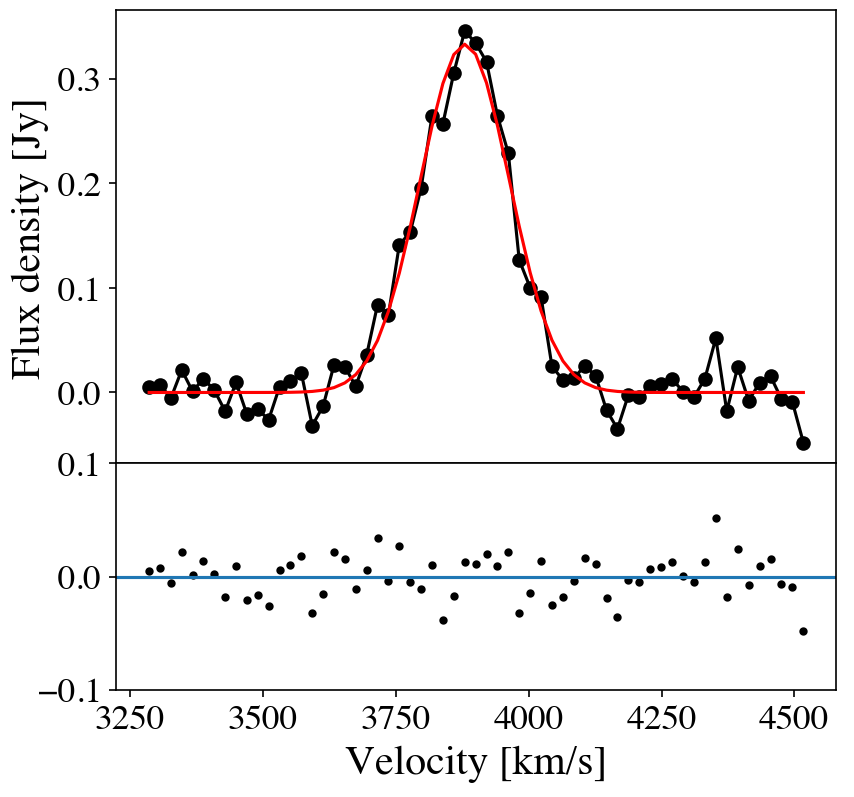}
        \caption{
                \label{fig:co_spec} 
               Integrated CO spectrum of IC 860 created using the moment0 map as a mask. The spectrum (black) is well fitted by a Gaussian function centered at the systemic velocity of the galaxy (red) and the lower panel shows the residuals. 
        }
\end{figure}

\begin{figure}
        \centering \includegraphics[width=\columnwidth]{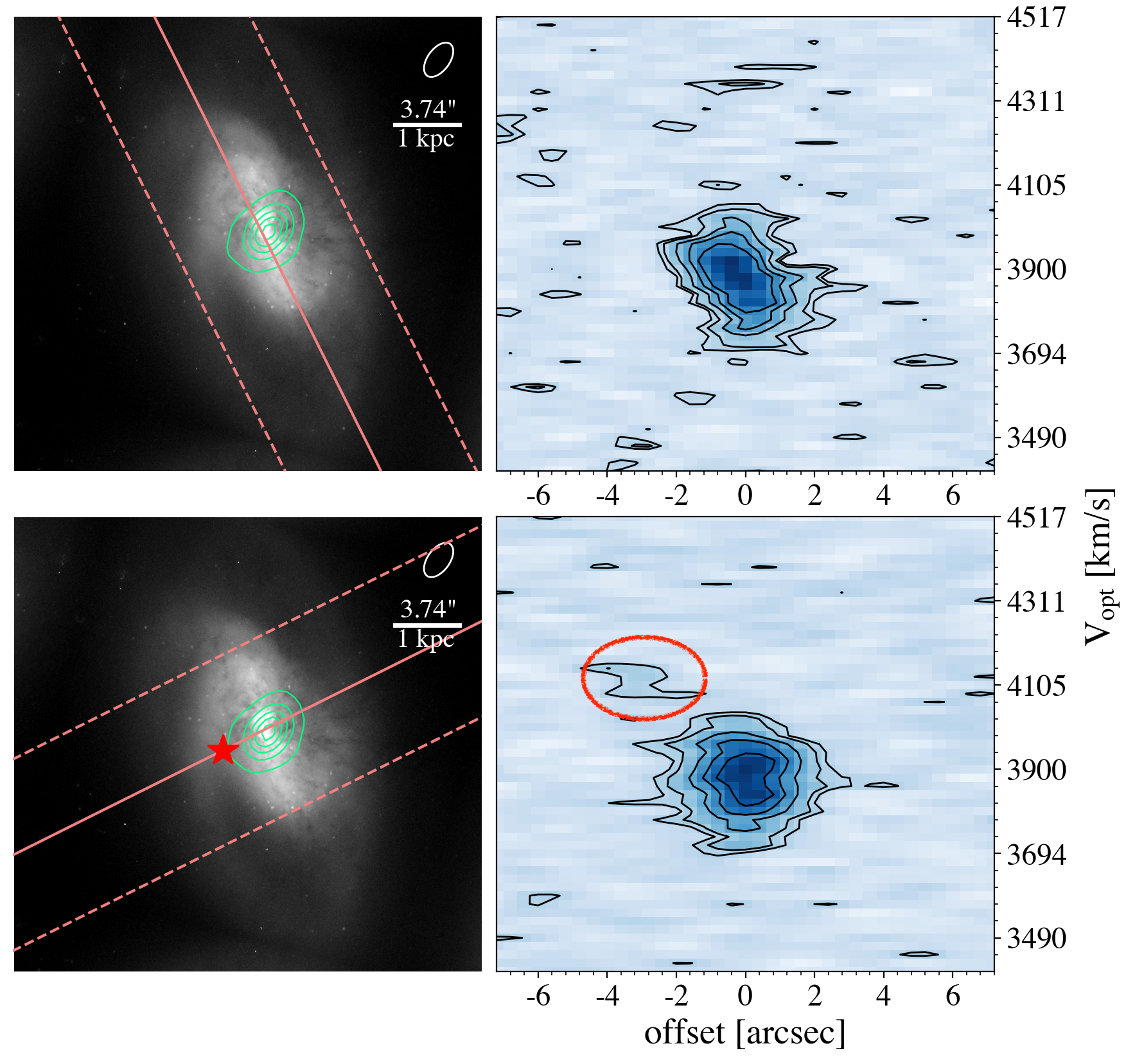}
        \caption{
                \label{fig:pv} 
        The position-velocity (PV) diagrams obtained from slicing the CO data cube along (top) and perpendicular to (bottom) the stellar kinematic PA. The black contours on the PV diagrams mark 3$\sigma$, 5$\sigma$, 10$\sigma$, 15$\sigma$ and 20$\sigma$ levels, respectively. The left columns are contours of the CO moment0 map (green) on top of the \textit{HST} F606W image. The green contour levels are logarithmically spaced from 13\% to 90\% of the peak CO moment0 flux. The small ellipse in the upper corner shows the beam and the pink lines outline the slabs for the PV diagrams. The high velocity blob marked by the red circle in the lower right PV diagram is suggestive of disturbed gas not following bulk gas motion, possibly outflowing gas moving away from us. The location of the circled blob is marked by a red star on the \textit{HST} image on the left.
        }
\end{figure}

\subsection{Kinematics of Different Components: Evidence for Merger History and Outflow}
The maps of the velocity and velocity dispersion of the stars, ionized gas, and CO molecular gas are plotted in Figure \ref{fig:vmaps}. The stars and CO show coherent rotation patterns as expected in late-type galaxies and early-type galaxies that contain gas disks \citep[e.g.,][]{Davis_2011}.

To compare the kinematics of different components in the galaxy, we fit the kinematic position angles (PA) based on their velocity fields using the \texttt{PaFit}\footnote{\url{https://pypi.org/project/pafit/}} package, which implements the method presented in Appendix C of \citet{Krajnovic_2006}. When fitting for the PA of each component, we center each velocity field by subtracting from each field the corresponding fitted velocity at the galactic center. This is to make sure that the final fitted PA accurately bisects the velocity field following the definition of PA.

The PA of the stars, the CO molecular gas, and the ionized gas are determined to be 26.5 $\pm$ 10.5$^{\circ}$, 2.5 $\pm$ 9.2$^{\circ}$, 43.0 $\pm$ 42.8$^{\circ}$, respectively\footnote{The angles quoted here are with respect to north following the convention, which is vertical up in Figure \ref{fig:vmaps}.}, and are plotted together with the velocity fields in Figure \ref{fig:vmaps}. Each velocity field is centered to the systemic velocity, which is the stellar velocity at the galactic center.

The kinematic PA of the stellar velocity and the CO velocity disagree with each other at 2$\sigma$ level and thus the stars and molecular gas could be kinematically misaligned. ISM gas could come from both stellar mass loss and external interactions such as mergers. Angular momentum conservation requires that gas from stellar mass loss to be kinematically aligned with the stars that produce them, while external gas can enter a galaxy with any angular momentum. Misalignment of the angular momenta of gas and stars have been used to determine the external origin of gas, such as a recent merger in several previous works \citep[e.g.,][]{Davis_2011,Belfiore_2017,Bryant_2019}. This misalignment in IC 860 could similarly suggest an external origin of the gas and that IC 860 could possibly have gone through a merger event. While the stellar velocity field shows a regular rotation pattern, the velocity maps of the ionized and the molecular gas show more than just simple rotation. The CO gas shows a velocity gradient along an axis that is offset from its kinematic PA. The velocity field of the ionized gas is complicated and its PA is poorly constrained. 

To look for evidence of an outflow, we plot the integrated CO spectrum in Figure~\ref{fig:co_spec}. The CO line profile is well fitted by a single Gaussian function centered at the galaxy's systemic velocity, without the need for a second Gaussian component tracing higher velocity gas. The reason for lack of outflow evidence in the CO integrated spectrum could be that the CO luminosity of the outflowing gas is small, or that the data are not sensitive enough to detect the faint wings signifying high velocity gas (e.g., as seen in NGC 1266 by \citealt{Alatalo11, Alatalo_2015}).

We further explore the CO velocity structure using the position-velocity (PV) diagram. We show in Figure \ref{fig:pv} the PV diagram of the CO spectral cube, sliced along and perpendicular to the fitted stellar kinematic PA. The uncertainty ($\sigma$) in the PV diagram is propagated using the thickness of the PV slab and the uncertainty in the CO cube, which is taken to be the robust standard deviation of several line-free slices in the cube. We observe a small redshifted blob deviating from the bulk motion of CO detected at the 3$\sigma$ level, in the slice perpendicular to the stellar PA (marked by a red circle in Figure \ref{fig:pv}). This blob is $\sim$2.5 arcsec (0.67 kpc) from the galactic center and lies on the left side of the galaxy, which is the opposite side to the location of blueshifted NaD absorption (\S3.3). Thus this redshifted emission from gas likely not in the disk might be tracing outflowing gas that is moving away from us behind the disk. A dense outflow in IC 860 was previously reported by \citet{Aalto_2019} via the detection of blueshifted CS (7–6) absorption and blueshifted HCN (3–2) emission in the central $\sim$10 pc region. They hypothesized that this nuclear outflow might be the base of a larger-scale outflow. This blob we see on kpc scales could imply a molecular outflow which might be related to the nuclear outflow, but given the distances and velocities we are uncertain about whether they belong to the same outflow structure. Comparing the velocity of the outflow blob to the systemic velocity of the CO gas, the velocity of the molecular outflow is estimated to be $v_{molecular}\sim200$ km/s. We estimate the mass contained in the blob by summing up the CO flux inside the blob and convert it to molecular mass using a conversion factor $\alpha_{\rm{CO}}=0.8$\ M$_{\odot}$ (K\ km\ s$^{-1}$\ pc$^2)^{-1}$ for outflows \citep{Bolatto_2013}. The mass of the blob is $\sim$10$^6$ M$_{\odot}$ and is 1\% of the total molecular gas mass (listed in Table \ref{tab:property}). If this feature showing disturbed gas is indeed an outflow, it 
could represent a small amount of the outflowing material, and we calculate the total hydrogen outflow mass in \S4.5. Although redshifted emission in the PV diagram is suggestive of an outflow, higher SNR CO observations with finer velocity resolution are required to confirm outflowing gas.

In addition to the PA misalignment between the stars and molecular gas, the distribution of the molecular gas also indicates a possible recent merger. On the left column of Figure \ref{fig:pv}, we plot contours of the CO(1-0) integrated intensity (moment0) on top of the \textit{HST} V-band image. These contours show that the molecular gas is concentrated within the central $\sim$1 kpc of the galaxy. Such a centrally concentrated gas distribution can happen after a merger, where the gas dissipates the angular momenta during collisions and gets funneled into the center \citep[e.g.,][]{Sanders_1996,Ueda_2014}.

\subsection{Neutral Gas Kinematics} 
In the NaD ISM profiles shown in Figure \ref{fig:nad}, the one bin where the two-component fitting performs better than one-component fitting overlaps spatially with the dust cone. The NaD profile in that bin shows a blueshifted wing in addition to the systemic component. Since NaD in the ISM traces cool, neutral gas, this blueshifted wing in the spectrum suggests the presence of a neutral outflow. Taking the outflow velocity to be the velocity difference between the systemic and outflow components, we estimate $v_{neutral} \sim$58 km/s. All bins show evidence of bulk motion of neutral gas redshifted with respect to the systemic velocity, which could indicate gas inflow.

The spectral fitting measures the velocity along our line of sight, but given the orientation of the galaxy we are likely not seeing the outflow directly face on. Thus the outflow velocity of $\sim$58 km/s is only a lower limit. To estimate the inclination angle of IC 860, we utilize the relation between the inclination angle $i$ and the axial ratio $b/a$ in \citet{Masters_2010}:
\begin{equation}
    cos^2i = \frac{(b/a)^2-q^2}{1-q^2}.
\end{equation}
We use the $g-$band isophotal $b/a$ reported by SDSS ($\sim$0.6) following \citet{Masters_2010}, and assume an intrinsic axial ratio for spirals $q=0.15$ \citep[according to][a reasonable range is 0.1 to 0.2]{Masters_2010}. The galaxy inclination angle derived this way is $\sim$54$^{\circ}$ and $v_{neutral}$ corrected for this inclination is $\sim$100 km/s\footnote{Similarly, the $v_{molecular}$ corrected for galaxy inclination is $\sim$340 km/s.}. This value is still a rough estimation due to the complexity of galaxy geometry, which the above equation is unable to account for.

In estimating the mass outflow rate, we follow \citet{Rupke_2005b}, assuming a spherically symmetric, mass conserving wind that travels at velocity $v$:
\begin{align}
    &\dot{M}_{out} = \Omega\ \mu\ m_{\rm{H}}\ N(\rm{H})\ v\ r\\
    &N(H) = \frac{N(\rm{Na\ I})}{\chi(\rm{Na\ I}))d(\rm{Na\ I}))Z(\rm{Na\ I}))}.
\end{align}
In Equation (5), $\Omega$ is the solid angle subtended by the wind at its origin, $m_{\rm{H}}$ is the mean atomic weight (with a $\mu$=1.4 correction for relative He abundance), $N(\rm{H})$ is the column density of hydrogen along the line of sight, $v$ is the central velocity of the outflow, and $r$ is the distance of the outflow to the galactic center. Equation (6) shows how the column density of hydrogen is derived from the sodium column density $N$(\ion{Na}{1}), where $\chi(\rm{Na\ I})=N(\rm{Na\ I})/N(\rm{Na})$ is the assumed ionization fraction (here we assume 0.1), d(\ion{Na}{1}) is the depletion on to dust (here we assume log d(\ion{Na}{1}) = $-$0.95, a Galactic value), and Z(\ion{Na}{1}) is the Na abundance (here we assume Z(\ion{Na}{1} = log[N(Na)/N(H)] = $-$5.69, the solar metallicity). The outflow is only detected in one spatially resolved bin, so here we only estimate the quantities of this particular bin. Using our fitted log N(\ion{Na}{1})[cm$^{-2}$] $\sim$13.97, we obtain log N(H)/cm$^{-2}$ $\sim$21.61. Following \citet{Roy_2021}, who also analyzed spatially resolved NaD absorption, we replace $\Omega vr$ in Equation (5) with $C_f A_{\rm{NaD}}/t_{out}$, where $A_{\rm{NaD}}$ is the on-sky projected area of the part comprising the NaD outflow, and $t_{out}$ is the
outflow timescale $\sim R_{out}/v$. This gives:
\begin{equation}
    \dot{M}_{out} = \mu\ m_{\rm{H}}\ C_f\ N(\rm{H})\ A_{\rm{NaD}}\ \frac{v}{R_{out}}.
\end{equation}
We take $A_{\rm{NaD}}$ to be the sky area covered by the outflow IFS bin, which is 12 arcsec$^2$, and $R_{out}$ to be the distance between the outflow bin center and the galactic center, which is $\sim$5 arcsec. The fitted $C_f$ for the outflow bin is $\sim$0.2, which is related to the clumpiness of the gas along the line of sight. Putting everything together and correcting for galaxy inclination, we estimate the neutral component of the outflow to have $\dot{M}_{out}$ $\sim$0.5 M$_{\odot}/yr$. The uncertainty associated with this number can be large since the ionization fraction and dust depletion of Na are uncertain. But since we assume typical values widely used in the literature, comparing this $\dot{M}_{out}$ to that found in other studies (\S5.2) should still be able to reflect the strength of the outflow in IC 860.

The \ion{H}{1} spectrum of IC\,860 from \citet{Mirabel_Sanders_1988} shows an absorption line profile with a blue wing, which is further evidence for a neutral gas outflow. The velocity width of the absorption line (measured as the width at 20\% of the peak absorption) is 200 km/s. The possibility of simultaneous \ion{H}{1} emission and absorption makes it challenging to estimate the column density and total mass of \ion{H}{1} in IC\,860 from spatially unresolved data, and higher spatial resolution of \ion{H}{1} are needed.

\subsection{Total Hydrogen Mass in the Outflow} 

\begin{figure*}
        \centering \includegraphics[width=2\columnwidth]{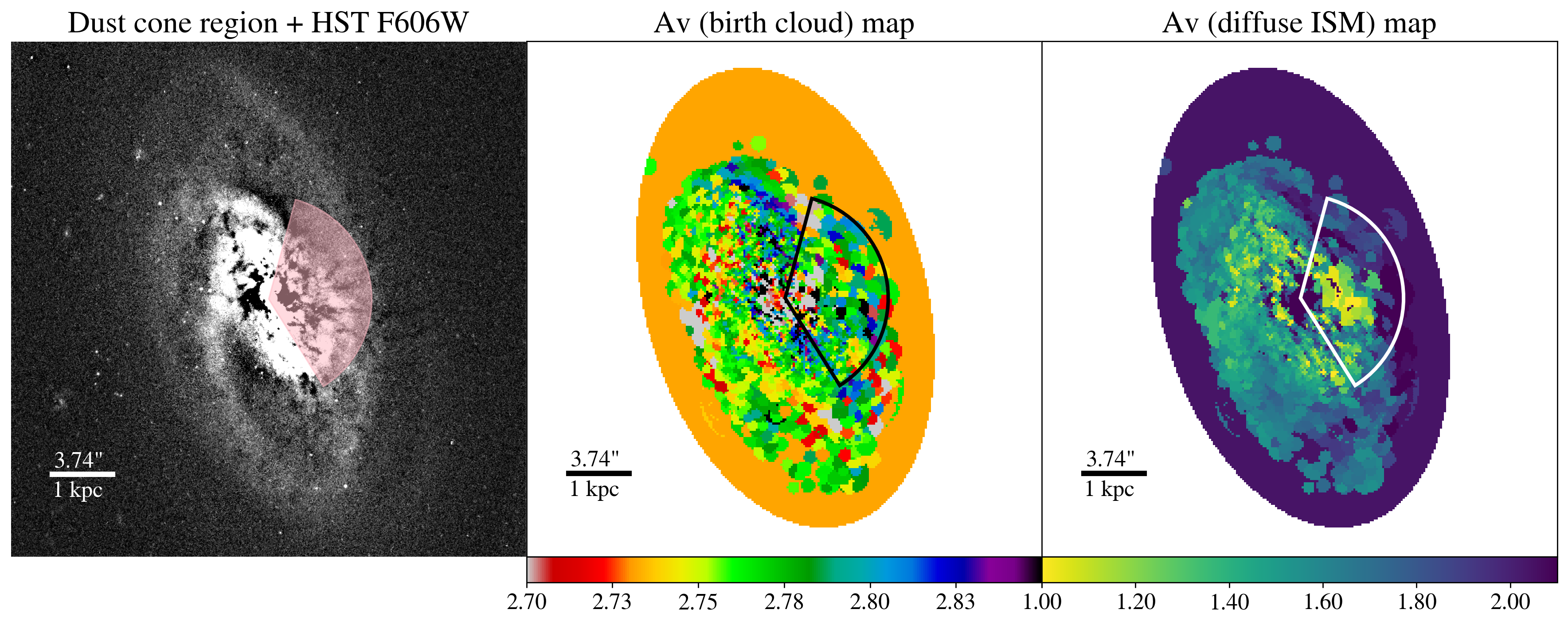}
        \caption{
                \label{fig:av} 
               The A$_V$ extinction maps from spatially resolved SED fitting with \textit{HST} data (\S3.4.2). The mean errors for A$_{\rm{V,birth\ cloud}}$ and A$_{\rm{V,diffuse\ ISM}}$ are 1.6 and 1.3, respectively. The sector marks the outflow cone region selected to constrain the hydrogen outflow mass (\S4.5).
        }
\end{figure*}

Whilst the dust cone structures in the \textit{HST} images show compelling evidence for a large-scale outflow, we find evidence for a small mass of outflowing molecular gas which may not represent the total outflow mass. We thus turn to dust to constrain the total hydrogen (\ion{H}{1} and H$_2$) outflow properties, since dust lives in the same environment as the neutral and molecular gas and has been observed to be entrained in galactic winds and outflows (see \citealt{Veilleux_2005} for a review.)

From the spatially resolved SED fitting (\S3.4.2) we obtain the A$_{\rm{V}}$ extinction maps as shown in Figure \ref{fig:av}. The two types of extinction from the two-component dust model \citep{Charlot_2000} are associated with dust in the stellar birth cloud and in the diffuse ISM. Generally, we would expect the extinction from the stellar birth cloud to dominate over that from the diffuse ISM, as the stellar birth cloud is cooler and denser and thus dustier. Our fitted range of A$_{\rm{V,birth\ cloud}}$ ($\sim$2.7--2.9) is indeed larger than that of A$_{\rm{V,diffuse\ ISM}}$ ($\sim$1.2--2). The mean errors for A$_{\rm{V,birth\ cloud}}$ and A$_{\rm{V,diffuse\ ISM}}$ are 1.6 and 1.3 from the spatially resolved SED fitting (\S3.4.2), respectively.

Using the relation between the total hydrogen column density 
and the extinction \citep{Bohlin_1978} with the assumption of a Galactic extinction law ($R_V = A_V/E(B-V)=3.1$),
\begin{equation}
    N(H)/A_V = 1.8 \times 10^{21}\ cm^{-2} mag^{-1},
\end{equation}
we can obtain the total hydrogen ($N(H)=N(H\ I) + 2N(H_2)$) 
column density in each pixel from the corresponding A$_{\rm{V}}$ value, and further convert the column density to mass in each pixel by multiplying by the pixel area and the atomic mass of hydrogen. Since we do not see high A$_{\rm{V}}$ regions delineating specific locations of the dust, we use the optical \textit{HST} image to define the region of the dust cone. We can only clearly see one side of the dust cone, so we define a sector encompassing the dust filaments on the right side of the image (see the leftmost panel in Figure \ref{fig:av}). We sum up masses in pixels within this sector to calculate the total hydrogen mass in the outflow. Finally this mass is multiplied by 2 to account for the other side of the outflow assuming a symmetric bi-conical outflow geometry. The outflow mass derived from A$_{\rm{V,birth\ cloud}}$ is (2.64 $\pm$ 0.03)$\times 10^8 \rm{M}_{\odot}$, and that from A$_{\rm{V,diffuse\ ISM}}$ is (1.69 $\pm$ 0.02)$\times 10^8 \rm{M}_{\odot}$. The errors quoted here are obtained from propagation of the A$_{\rm{V}}$ errors and are likely underestimates given the existence of other uncertainties that are difficult to quantify (e.g, the A$_{\rm{V}}$ conversion factor, the selection of the outflow region). Thus the hydrogen 
outflow mass derived from the two types of A$_{\rm{V}}$ are consistent. Although A$_{\rm{V,birth\ cloud}}$ is expected to dominate, it is not a surprise here that the A$_{\rm{V,diffuse\ ISM}}$-derived outflow mass is non-negligible, as IC 860 is not a star-forming galaxy with dense stellar birth clouds, and the large-scale outflow might have mixed the dust with the diffuse ISM, which would contribute to the extinction caused by the diffuse ISM.

To estimate the hydrogen mass outflow rate from M$_{out}/t_{out}$, we assume an outflow timescale $t_{out} \sim R_{out}/v$, where $R_{out}$ is the distance between the NaD outflow bin center and galactic center ($\sim$5 arcsec, \S4.4), and $v$ is the observed neutral gas outflow velocity ($\sim$58 km/s, \S4.4). We take M$_{out}\sim 2 \times 10^8$ M$_{\odot}$ (the average of masses derived from the two types of A$_{\rm{V}}$). After correcting for the galaxy inclination, 
the total mass of outflowing hydrogen is $\sim$12 M$_{\odot}/yr$. 
This mass outflow rate is a reasonable order of magnitude estimation, though there are several uncertainty factors. First, the conversion from A$_{\rm{V}}$ to N(H) is uncertain. The fact that part of the gas within the selected cone region could be disk gas would lead to an overestimation of the total outflowing mass. On the other hand, the A$_{\rm{V}}$ from the spatially resolved SED fitting could be an underestimation of the true extinction level in some regions where the extinction is so high that essentially the entire stellar population is obscured. In such cases the measured A$_{\rm{V}}$ values are for the visible portion of the stellar population only. These highly obscured regions likely exist near the center of IC 860 as suggested in \citet{Aalto_2019}, which also agrees with the high column density implied by the deep 9.7$\mu$m silicate absorption feature in the MIR spectrum (Figure \ref{fig:irs}). Thus our outflow mass estimation using A$_{\rm{V}}$ could miss some dense gas and underestimate the true outflowing mass.

\begin{table*}[t]
\caption{IC 860 SFR from different indicators}
\centering
\begin{tabular}{c | c | c}
\hline \hline
\textbf{SFR indicator (wavelength)} & \textbf{SFR [M$_{\odot}$/yr]} & \textbf{Uncertainty factors}\\
\hline
H$\alpha$ & 0.66 $\pm$ 0.73  & Highly extincted; Contamination by AGN/shock ionization\\
\hline
Pa$\alpha$ & 0.35 $\pm$ 0.07  &   Contamination by AGN/shock ionization \\
\hline
[\ion{Ne}{2}] + [\ion{Ne}{3}] & 0.54 $\pm$ 0.07  &   Contamination by AGN/shock ionization \\
\hline
IR (24$\mu m$) & 8.43 $\pm$ 0.68  &  \multirow{3}{10cm}{Contamination by AGN/shock ionization; dust heated by old stars} \\
\cline{1-2}
FUV (0.153$\mu m$) + Total IR (8-1000$\mu m$)  & 11.19 $\pm$ 2.24  & \\
\cline{1-2}
FUV (0.153$\mu m$) + IR (25$\mu m$) & 7.40 $\pm$ 0.60  & \\
\hline \hline
\multicolumn{3}{p{18cm}}{\textbf{Notes: } All values reported here assume a \citet{Kroupa_2001} IMF.}
\label{tab:sfr}
\end{tabular}
\end{table*}

\subsection{Star Formation Rate}
The determination of star formation rates (SFRs) for PSBs is difficult. Most commonly used SFR indicators are uncertain in PSBs either because they suffer from contamination from AGN activity and/or heating from old stellar populations, or because they are calibrated using star-forming galaxies. We calculate the SFR for IC 860 using different global SFR indicators and summarize the values in Table \ref{tab:sfr}.

We start with SFR indicators using IR fluxes or combinations of UV and IR fluxes that do not require dust extinction corrections. The IR and UV + IR SFRs are from \citet[][and references therein]{Calzetti_2013}, assuming a \citet{Kroupa_2001} IMF. We use flux densities from \textit{GALEX} (0.153$\mu m$) and \textit{Spitzer} (24$\mu m$) 
from \S2.5 and convert them to luminosities. 
The total IR luminosity (8-1000$\mu m$) is from \citet{U_2012}, and the IR luminosity at 25 $\mu m$ is approximated by that at 24 $\mu m$ (the difference is only $\sim$2\% based on \citealt{Kennicutt_2009} and \citealt{Calzetti_2010}). The uncertainties on the derived SFRs are propagated from the flux density uncertainties (assuming a 20\% uncertainty on the total IR luminosity).

The IR and UV + IR SFRs trace SF over a long timescale ($\gtrsim$100 Myr). To get a sense of recent ($\sim$10 Myr) SF activity, we use other SFR indicators based on emission line luminosity: H$\alpha$, Pa$\alpha$, and [\ion{Ne}{2}] + [\ion{Ne}{3}]. We use the measured H$\alpha$ flux from the IFS data (\S3.2), and correct for extinction using the A$_V$ maps\footnote{The H$\beta$ line is poorly detected in the IFS data as mentioned in \S4.2, and appears in absorption in the SDSS spectrum. This makes it difficult to do the extinction correction via the traditional Balmer decrement approach, so we use our measured A$_V$ values instead.} from spatially resolved SED fitting (\S3.4.2). We use the A$_V$ from the diffuse ISM as this is what affects the reddening in the stellar continuum and thus the SED, and divide the values by 0.44 to account for the differences between the reddening from the stellar continuum and from the ionized gas ($E(B-V)_{star}$ = 0.44 $E(B-V)_{gas}$ according to \citealt{Calzetti_2000}). The extinction map has finer resolution than the binned IFS flux map, so we take the median (robust standard deviation) of all A$_V$ values within a H$\alpha$ flux bin as the extinction (extinction error) of the corresponding bin. The corrected H$\alpha$ fluxes in all bins are then summed and converted to H$\alpha$ luminosity. We derive the SFR from the H$\alpha$ luminosity following equation (2) in \citet{Kennicutt_1998}, and the uncertainty of SFR$_{H\alpha}$ is estimated by propagating errors in both the A$_V$ and the measured H$\alpha$ flux. We divide this SFR$_{H\alpha}$ by 1.6 to convert from a \citet{Salpeter_1955} to \citet{Kroupa_2001} IMF following \citet{Calzetti_2013}.

The extinction-corrected Pa$\alpha$ luminosity is from \citet{Alonso-Herrero_2006} and we adjust the value according to our cosmological parameters. \citet{Alonso-Herrero_2006} measured the Pa$\alpha$ luminosity from \textit{HST} images covering the entire galaxy, so we do not further apply aperture corrections to their value. The L(Pa$\alpha$)--SFR relation is converted from the L(H$\alpha$)--SFR relation in \citet{Kennicutt_1998}, assuming case B recombination and a Kroupa IMF: SFR[M$_{\odot}$/yr] = 4.24 $\times$ 10$^{-41}$L(Pa$\alpha$) [ergs/s]. \citet{Alonso-Herrero_2006} did not provide an uncertainty for L(Pa$\alpha$) so we estimate a 20\% uncertainty in SFR$_{Pa\alpha}$. The [\ion{Ne}{2}] + [\ion{Ne}{3}] luminosity is measured in \citet{Lambrides_2019} by integrating the \textit{Spitzer} IRS spectrum. As described in \S2.8, the spectral slit covers the MIR emission of the whole galaxy. We calculate SFR$_{Ne}$ following equation (13) in \citet{Ho_2007}, assuming the fractional abundances of [\ion{Ne}{2}] and [\ion{Ne}{3}] to be 0.75 and 0.01, respectively, based on \citet{Ho_2007}, and propagate the luminosity uncertainties to obtain the uncertainty in SFR$_{Ne}$.

To put the derived SFRs into context, we compute the SFR and gas surface densities and compare them to the Kennicutt-Schmidt relation \citep[KS relation,][]{Kennicutt_1998}. This relation was revisited in \citet{Kennicutt_2021} and we use the results from this more recent study hereafter. We assume that the molecular gas and SF are co-spatial, and use the radius of the CO gas distribution (1 kpc) to calculate the area of the gas and star formation regions. The total molecular gas mass is M$(H_2)_{tot} = (1.79 \pm 0.54)\times 10^9 \rm{M}_{\odot}$ derived in \S2.4. We plot the data and the empirical relation from \citet{Kennicutt_2021} along with our results in Figure \ref{fig:ks}. The uncertainty of $\Sigma_{gas}$ includes the error in the total molecular mass, 50\% error in the gas radius, and an additional 80\% error in the total molecular gas mass due to the CO$-\rm{H}_2$ conversion factor. The uncertainty in $\Sigma_{SFR}$ includes the error in SFR and 50\% error in the SF radius. More discussion of the SFRs and KS relation are presented in \S5.1.

\section{Discussion}

\subsection{Star Formation Suppression}
\begin{figure}
        \centering \includegraphics[width=\columnwidth]{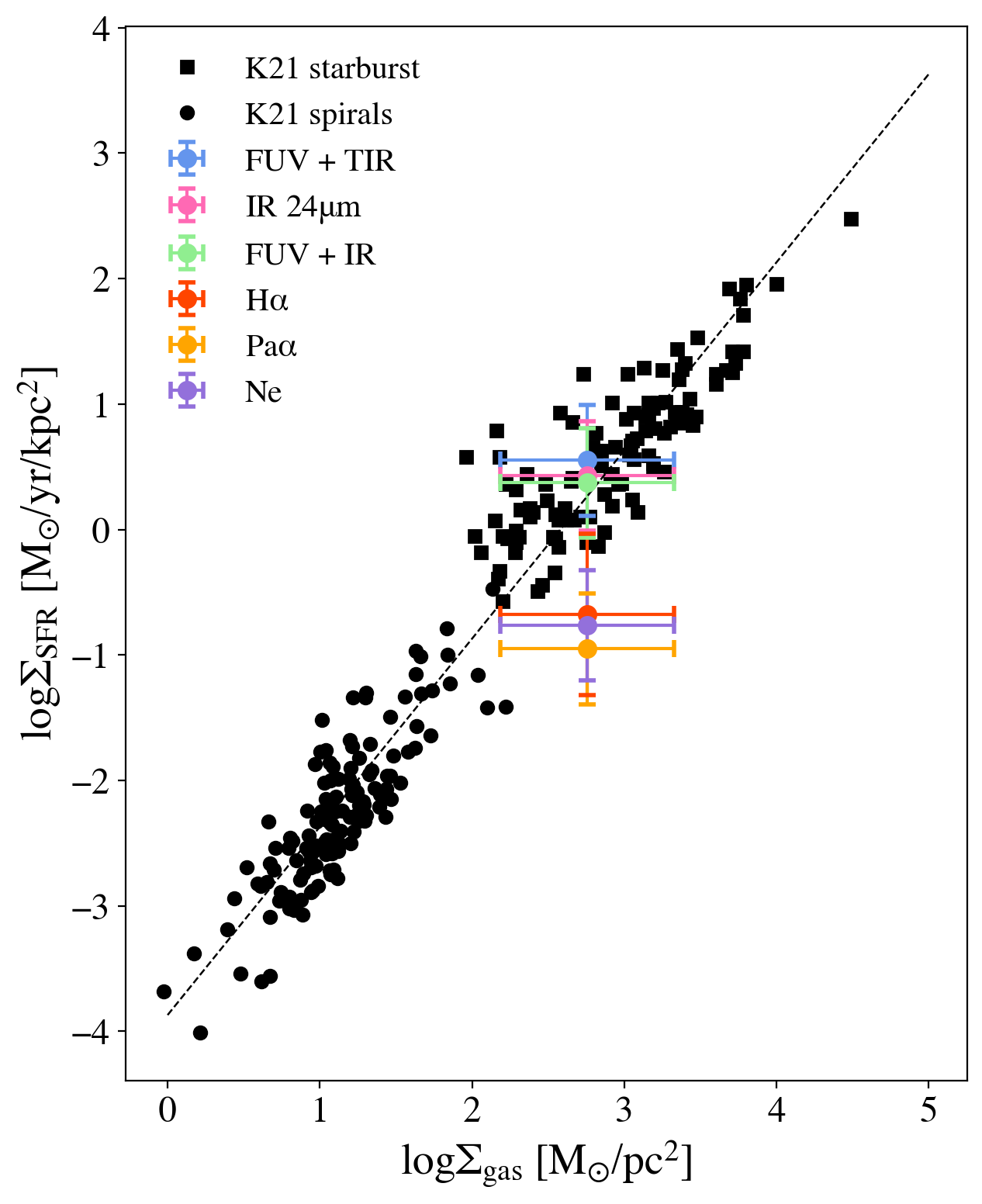}
        \caption{
                \label{fig:ks} 
               Surface densities of the molecular gas mass and star formation rate compared to the Kennicutt-Schmidt star formation relation. We assume that the molecular gas and SF are co-spatial, and use the radius of the CO gas distribution (1 kpc) to calculate the area of the gas and star formation regions. The data and the empirical relation from \citet{Kennicutt_2021} are plotted in black, while the estimations for IC 860 are plotted in color. The longer timescale SFRs (IR and UV + IR as indicators) are consistent with starburst on the relation, while the recent SFRs (H$\alpha$, Pa$\alpha$, and Ne as indicators) lie below the relation. 
        }
\end{figure}

PSBs have been observed to lie below the KS relation \citep[e.g.,][]{French_2015,Smercina_2021}. For IC 860, SFR$_{H\alpha}$, SFR$_{Pa\alpha}$, and SFR$_{Ne}$ do lie below the relation as expected (Figure \ref{fig:ks}), showing that it has already started its quenching process and is currently in a state of inefficient star formation given its molecular gas content. 

The SFRs from IR and UV + IR calibrators are $\sim$15 times larger than those from emission line calibrators and are consistent with starburst galaxies on the KS relation. The post-burst age of IC 860, defined as the time elapsed since 90\% of the stars formed in the recent burst, was estimated to be 555$^{+84}_{-83}$ Myr in \citet{French_2018}. Thus SFR indicators involving IR luminosity that are tracing a SF timescale of $\gtrsim$100 Myr may well capture the burst in the galaxy's star formation history. In this case, these SFRs being consistent with starbursts actually corroborates that IC 860 is a young PSB. 

However, dust can be heated by an AGN and re-emit the energy in IR, therefore the SFR from IR-based SF indicators could be overestimated. In addition, IR luminosity is known to overestimate the SFR in quiescent galaxies and PSBs. At longer wavelengths, the contribution to IR emission from dust heated by low-mass long-lived stars also becomes prominent. Specifically, \citet{Smercina_2018} found that the total IR flux enhanced by the A-star population could increase the estimated SFR by a factor of $\gtrsim$ 3--4, and \citet{Hayward_2014} found that the total IR flux could overestimate the true SFR by a factor of $\gtrsim$ 30 during the post-starburst phase. The indicators that combine both UV and IR fluxes take into account both direct starlight and dust-processed light, and essentially use the IR emission to ``correct" the UV emission for dust attenuation. However, they are still affected by the dust heating from the old stars and tend to overestimate SFR for galaxies that have a substantial fraction of old stars \citep[e.g.,][]{Abdurrouf_2022}. These UV + IR SFR indicators also suffer from systematic uncertainties from the different samples/simulations they are calibrated on, which involve mostly star-forming galaxies \citep{Kennicutt_2012}. 

Emission line luminosity is directly affected by the number of ionizing photons from young stars, therefore the SFRs derived from H$\alpha$, Pa$\alpha$, and Ne emission lines should be representative of the galaxy's recent SF activity. However, there are still caveats. One important uncertainty that affects all emission line SFRs is the ionization source. As discussed in \S4.2, IC 860 hosts AGN and/or shock ionization, so SF is not the only contributor to the emission line luminosity. Additionally, the high column density of dust in IC 860 makes our extinction corrections uncertain. The central region of IC 860 has such high column density \citep{Aalto_2019} that free-free emission (thus H$\alpha$ and Pa$\alpha$ luminosity) might be suppressed. Although the Pa$\alpha$ and Ne lines are less affected by extinction, the recent SFR could still be underestimated due to the extremely high dust column. \citet{Wild_2011} tested that dust extinction corrections can recover the H$\alpha$ luminosity reasonably well even in dusty ultraluminous LIRGs, and the agreement of our three emission line derived SFRs provides reassurance that they are reliable. In this case, the dominating uncertainty factor in emission line derived SFRs is AGN contamination, which indicates that the true recent SFRs are smaller than what we estimate. Thus the observed deficit relative to the KS relation should be robust and IC 860 has recently moved below the relation.

A further source of uncertainty is the assumption that the SF area is the same as the molecular gas. \citet{Aalto_2019} estimated that the infrared luminosity could come from the central $\sim$10 pc. If the area used for the SF surface density is much more compact than assumed, this would push the SFR surface density above the KS relation, consistent with extreme starbursts. However, the bulk of the molecular gas is not co-spatial with this extreme infrared knot (similar to what is seen in NGC 1266; \citealt{Alatalo_2015}). Therefore, the star formation efficiency in the majority of the molecular gas would fall even farther below the KS relation in this case. Additionally, due to the highly obscured nature of IC 860, it is unclear whether all of the infrared luminosity can be attributed to star formation \citep{Aalto_2019}.

\subsection{Strong or Weak Outflow?}
In previous sections, we have presented evidence that IC 860 could host a multiphase outflow and estimated the mass and velocity of each outflow phase. In \S4.1, we present the optical morphology of IC 860 from multiband {\em HST} observations, showing the clear presence of dust filaments that radiate outward from the nucleus. In \S4.3, the CO-traced molecular gas is shown to have non-disk component visible on the position-velocity diagram as redshifted emission, which is suggestive of outflowing gas. \S4.4 shows that there is a blueshifted component in the NaD absorption profile, and its corresponding location on the sky is spatially consistent with the dust filaments. This section aims to put these observations to context and determine the nature of the outflow present in this galaxy.

We first compare our estimated outflow properties with other results in the literature. \citet{Roberts-Borsani_2019} studied the neutral outflow properties of 300,000 local (0.025 $\leqslant z \leqslant$ 0.1) inactive and AGN-host galaxies in SDSS via NaD absorption line fitting. They reported outflow velocities in the range of 69--370 km/s with a median of 160 km/s. In higher mass galaxies (10 $\lesssim$ log(M$_*$[M$_{\odot}$]) $\lesssim$ 11.5) they found neutral mass outflow rates in the range of 0.14--1.74 M$_{\odot}/yr$ for a SFR range of $-$0.16 $\lesssim$ log SFR [M$_{\odot}/yr$] $\lesssim$ 1.23. IC 860 with log(M$_*$[M$_{\odot}$]) = 10.95 and SFR discussed in Table \ref{tab:sfr} falls within the mass and SFR range of typical local galaxies, but its neutral outflow velocity and mass outflow rate lie on the lower end of their results.

For star-forming galaxies, \citet{Chen_2010} applied a similar method using NaD absorption lines to a sample of $\sim$140,000 nearby (0.005 $\leqslant z  \leqslant$ 0.18) SF galaxies and reported an outflow velocity range of $\sim$120--160 km/s. Thus the neutral gas outflow velocity of IC 860 is lower than those of typical star-forming galaxies in the local Universe.

Outflows have been detected in PSBs as well. \citet{Tremonti_2007} reported large outflow velocities of 500--2000 km/s in a small sample of PSBs at z$\sim$0.5 using the \ion{Mg}{2} absorption lines. More recently, \citet{Baron_2022} found ionized and neutral outflows in their sample of 144 PSBs. Their ionized outflows have a mean mass outflow rate of $\sim$1 M$_{\odot}$/yr and a median outflow velocity of 766 km/s. Their neutral outflows have a mean mass outflow rate of 10 M$_{\odot}$/yr and a median outflow velocity of 633 km/s. They compared the velocities of their ionized and neutral outflows to other studies and found these are roughly similar to those outflows observed in local AGN and LIRGs. However, both the neutral outflow velocity and the neutral mass outflow rate in \citet{Baron_2022} are much larger than what is found in IC 860. While we report the neutral outflow velocity based on the line center of the blueshifted spectral component, these studies defined their outflow velocity to be the maximum velocity from the spectrum (blueward of the line center). This discrepancy would naturally lead to our outflow velocity being smaller, but should not be enough to account for the large differences we see here. 

To further investigate whether the outflow in IC 860 is able to escape the galaxy, we compare the outflow velocities in different phases to the circular velocity of the galaxy, estimated from an empirical relationship for low-redshift star-forming galaxies\footnote{Though this relationship is for star-forming galaxies, the estimated circular velocity for IC 860 from this relation is very close to that from the typical equation $v_{circ} = \sqrt{GM/r}$, where $M$ here is the total galaxy mass including stars, gas, and dust, and $r$ is the total radius from the \textit{2MASS} All-Sky Extended Source Catalog from NED. Therefore, IC 860 may have a gravitational potential similar to that of star-forming galaxies, and we adopt $v_{circ}$ from this empirical relationship in the text.} involving the stellar mass: $v_{circ} = \sqrt{2}S$, where $S$ is the kinematic parameter log $S$ = 0.29 log M$_*$ $-$ 0.93 \citep{Simons_2015,Heckman_2015}. The estimated $v_{circ}$ for IC 860 is $\sim$250 km/s, which is larger than the estimated outflow velocity in the neutral phase (58--100 km/s), and consistent with that in the molecular phase (200--340 km/s). Since the escape velocity of a galaxy is larger than the circular velocity, the multiphase outflow in IC 860 is unlikely to escape, or only marginally escape the galaxy.

Our findings imply that the outflow in IC 860 is relatively weak. The kinetic energy of the total hydrogen outflow from $\frac{1}{2}Mv^2$ is $\sim$2$\times$10$^{55}$ erg. Using SFR$_{H\alpha}$ and a timescale of $\sim$16 Myr based on the distance traveled by the neutral outflow ($\sim$1.6 kpc from the galactic center, after correcting for galaxy inclination) and the neutral outflow velocity ($\sim$100 km/s after correcting for galaxy inclination), assuming 10\% efficiency in energy transfer from supernovae to the ISM following \citet{Murray_2005}, the energy deposition from supernovae since the onset of the outflow is $\sim$1$\times$10$^{55}$ erg based on equation (34) in \citet{Murray_2005}. The outflow kinetic energy and the supernovae deposited energy are consistent considering possible uncertainties in these numbers. Thus the outflow in IC 860 could be driven by SF.

\subsection{Evidence for an AGN and Its Impact}
\begin{figure}
        \centering \includegraphics[width=\columnwidth]{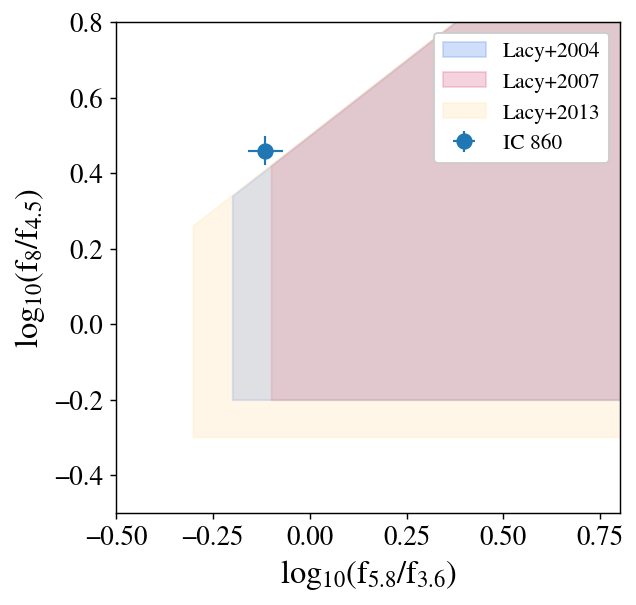}
        \caption{
                \label{fig:mir_agn} 
               MIR color plot of IC 860 with AGN selection regions shaded in colors (\S5.3). Though IC 860 does not fall in the AGN region, it is extremely close. As discussed in the text, IC 860 is actually a SF-AGN composite.
        }
\end{figure}

The study of ionization mechanisms in IC 860 in \S4.2 shows that the central region of the galaxy is consistent with being ionized by a weak AGN. However, it is known that such emission line ratio diagnostic diagrams have trouble distinguishing between shocks and AGN, and IC 860 hosts an outflow that could produce shocks. Here we discuss additional evidence from other observations that support the presence of an AGN.

\subsubsection{Mid-infrared Colors and Spectrum}
We use \textit{Spitzer} observations at 3.6$\mu m$, 4.5$\mu m$, 5.8$\mu m$, and 8.0$\mu m$ to evaluate the MIR color of IC 860, and compare it to the AGN selection criteria in \citet{Lacy_2004,Lacy_2007,Lacy_2013}. The AGN region and the position of IC 860 are plotted in Figure \ref{fig:mir_agn}. Although IC 860 does not fall in the AGN region, it is extremely close (consistent within 2$\sigma$). 
Other AGN selection criteria utilizing MIR colors based on both empirical \citep{Stern_2005,Donley_2012} and theoretical \citep{Satyapal_2018} methodology show that IC 860 does not lie close to their selected AGN regions. The reasons for an object to be missed by these criteria could be that the AGN is weak and/or too obscured, and the PAH emission from stars enters the 8.0$\mu m$ band. In fact, this is exactly the case for IC 860: its core is dusty, and its MIR spectra (Figure \ref{fig:irs}) shows PAH emission lines. The PAH emission lines in IC 860 are weaker than those in other SF galaxies in \citet{Stierwalt_2013}, and its strong rising continuum at the redder end of the MIR spectrum is strongly suggestive of an AGN. \citet{Spoon_2007} classified the MIR spectra of infrared galaxies based on the EW of the 6.2$\mu$m PAH feature and the strength of the 9.7$\mu$m silicate feature, and IC 860 falls onto the branch that could be thought of as an intermediate stage between a fully obscured galactic nucleus and an unobscured nuclear starburst. 
Furthermore, \citet{Diaz-Santos_2017} used a range of MIR indicators with the IRS spectrum, and estimated for IC 860 the AGN fractional contribution to the MIR luminosity and bolometric luminosity to be 19\% and 6\%, respectively. Thus we speculate that IC 860 may be a SF-AGN composite system from the MIR data perspective. 

The MIR emission lines of IC 860 have been investigated in previous studies \citep{petric2011, Stierwalt_2013, inami2013, petric2018}. MIR molecular H$_2$ lines probe gas with temperatures between $\sim$100 and 1000\,K, typically representing a few percent of a galaxy's total molecular gas reservoir. Galaxies hosting AGN have relatively more and warmer H$_2$ than galaxies that do not, and this excess H$_2$ appears to be connected with shocks and outflows \citep{hill2014,petric2018,lamb2019, minsley2020, riff2020}. 
For IC 860, \citet{petric2018} found a H$_2$ temperature of $\sim$400K, consistent with LIRGs whose AGN contribution is more than 50\% of the IR luminosity. IC 860 is one of the few LIRGs ($\sim$4\%) with a H$_2$ S(5) rotational line detection in \citet{petric2018}. While these MIR H$_2$ properties alone are not sufficient to determine that IC 860 hosts an AGN, they are consistent with a non-SF source heating the warm molecular gas.

\begin{figure*}
        \centering \includegraphics[width=2\columnwidth]{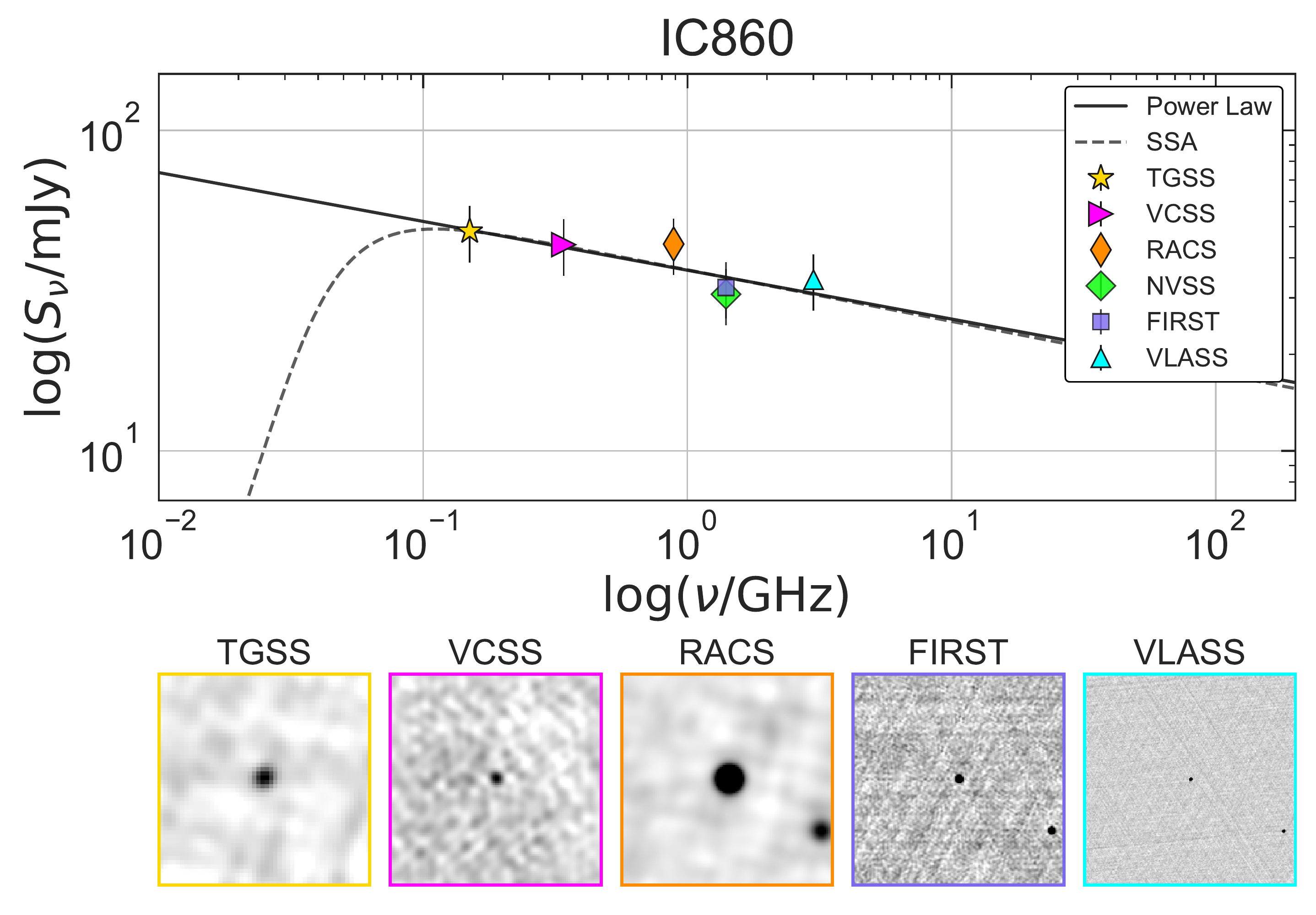}
        \caption{
                \label{fig:radio_index} 
               The radio spectrum of IC 860 based on selected, public, wide-field radio surveys.  The best-fit power-law model (with a spectral index of $\alpha=-0.15$) is shown by the solid line. The dashed line illustrates the fit to a synchrotron self-absorption (SSA) model typical of compact radio emission arising from an AGN.  The radio spectral models follow the definitions from \citet{Patil_2022}. Further details on the radio surveys used in this analysis are provided in Table~\ref{tab:radio}. The smaller panels show the selected cutout images (5$^{\prime} \times 5^{\prime}$) from the radio surveys used in the spectral modeling, which demonstrate the highly compact nature of the radio source in IC 860.
        }
\end{figure*}

\subsubsection{Radio Continuum}
Radio continuum observations provide an extinction-free tracer of emission that may arise from an AGN or compact starburst. Unambiguously determining the origin of a compact, low-luminosity radio source in the nucleus of a nearby galaxy is inherently challenging \citep[e.g.,][]{Nyland_2016}, but the radio properties (including the radio morphology, brightness temperature, and spectral shape) offer important constraints.

Utilizing data summarized in Table~\ref{tab:radio}, we measure the radio spectral shape of IC 860. In Figure \ref{fig:radio_index}, we plot the radio spectrum of IC 860 from 150~MHz to 3~GHz along with selected cutout images.  All flux values represent total fluxes, and we conservatively assume flux errors of 20\% in order to account for systematic uncertainties between the different surveys.  Since the radio source in IC 860 is known to be compact, these observations should be capturing the total emission without resolving out flux.  We used the radio spectral modeling tools\footnote{\url{https://github.com/paloween/Radio_Spectral_Fitting}} from \citet{Patil_2022} to fit the radio spectrum.  We find the radio spectrum is best fit by a power-law ($S_{\nu} \propto \nu^{\alpha}$) with $\alpha \sim -0.15$.  The flat radio spectral index is consistent with optically-thin synchrotron emission from a compact, self-absorbed radio source, such as a low-luminosity AGN jet.  Future low-frequency radio continuum observations in the 10's of MHz regime (e.g., with the Low Band Array of LOFAR) would enable a test of this hypothesis. 

Another possibility for the origin of the radio emission is SF.  Given the flat radio spectral shape, thermal emission associated with star-formation (i.e. free-free emission from \ion{H}{2} regions) is plausible.  Using the calibration from \citet{Murphy_2011}, we find that a SFR of 8.4 M$_{\odot}$/yr could lead to the observed level of radio continuum emission at 1.4~GHz. This SFR is consistent with the IR and UV + SFR estimates summarized in Table \ref{tab:sfr}.  The radio emission could therefore conceivably arise from SF. However, the radio SFR has a large uncertainty and could be biased by AGN contamination.  



Previous high-resolution imaging studies of IC 860 with the VLA have revealed a compact morphology with a deconvolved size upper limit of $\theta \lesssim0.2^{\prime \prime}$ ($\lesssim50$~pc; \citealt{Baan_2006}).  The brightness temperature of IC 860 at 1.43~GHz based on the VLA/A-configuration data from \citet{Baan_2006} is $\log(T_b/K) = 5.04$.  Since free-free emission is constrained to $\log(T_b/K) < 5.0$, this argues against a compact starburst origin \citep{Condon_1991}.  However, synchrotron emission associated with SF (e.g. radio supernova remnants) cannot be completely ruled out \citep{Nyland_2017}. Very long baseline interferometry (VLBI) may help further constrain the presence of a radio-emitting AGN in the nucleus of IC 860.  Single-baseline fringe detections with the European VLBI Network at 4.8~GHz reported in \citet{Parra_2010} support the presence of a compact, high-brightness temperature source consistent with an AGN.  A future radio imaging study on milliarcsecond scales is needed to provide more robust constraints.  Overall, we conclude that the radio properties of IC 860 are consistent with the presence of a low-luminosity radio AGN, but we cannot completely rule out a SF origin with the currently available data.

Distinguishing AGN and compact starbursts is indeed challenging. \citet{Baron_2022} found some systems selected to have post-starburst signatures based on their optical spectra were in fact obscured starbursts. \citet{Aalto_2019} studied the inner $\sim$10 pc region of IC 860 and concluded that the central energy source could be an AGN or an extremely compact starburst, or both. If the central energy source is a starburst, it would require a top-heavy IMF with a high SFR $\sim$10--20 M$_{\odot}$/yr purely in massive stars. IC 860 shows an abnormal ratio between L(Pa$\alpha$) and L(IR) compared to other local LIRGs \citep[Figrue 3, ][]{Alonso-Herrero_2006}, with L(IR) being much larger than what would be expected from  L(Pa$\alpha$). This fact again implies that IC 860 is extremely dusty and such high obscuration significantly complicates the process of determining the nuclear energy source. Nevertheless, we see
evidence pointing towards an AGN, though there could be a co-existing central compact starburst.

\subsubsection{Feedback?}
Whether AGN feedback can significantly affect the evolution of PSBs remains a mystery. \citet{Greene_2020} postulated that increased AGN activity in younger PSBs is due to the availability of gas fuel in these systems, rather than AGN feedback shutting down SF in older systems. \citet{Lanz_2022} recently analyzed \textit{Chandra} X-ray observations of 12 SPOGs and found that the AGN, if present, have high obscuration and/or low luminosity. They suggested that it is more likely that these AGN are along for the ride while the galaxies transit from blue to red rather than driving the transition, which could also be the case for IC 860.

Our results show that the AGN in IC 860, if it exists, is very weak. It is classified as a weak AGN in the WHAN diagnostic diagram (\S4.2), and has an intrinsic X-ray luminosity likely less than 10$^{40}$ erg/s (\S3.5). However, high obscuration might also contribute to the observed low luminosity. Vibrationally excited HCN emission was detected at the core of IC 860 \citep{Aalto_2019} and is thought to trace a heavily obscured stage of evolution before nuclear feedback mechanisms are fully developed \citep{Falstad_2019}. Thus the AGN feedback in IC 860 appears to be at an early stage, but we are not yet sure if it is connected to the quenching of SF.

\subsection{IC 860 as a Local Laboratory for Galaxy Evolution}
Although PSBs represent an important stage in galaxy evolution, they are rare in the local universe and more abundant at higher redshifts \citep[e.g.,][]{Wild_2016}. IC 860 is a local laboratory allowing us to study what are happening inside quenching galaxies. 

Since the discovery of large molecular gas reservoirs in PSBs \citep[e.g.,][]{Rowlands_2015,French_2015} the traditional picture of powerful AGN-driven outflows removing and/or heating gas in the galaxies \citep[e.g.,][]{Hopkins_2008} has been challenged. IC 860 is one example of the PSBs that contain large gas reservoirs. Under the current star formation rate, the gas depletion time ($\tau_{dep} = \rm{M}_{gas}/SFR$) of IC 860 is $\sim$7 Gyr. \citet{Rowlands_2015} found similar long depletion times in their sample of 11 PSBs at low redshifts. Such accumulating evidence leads to the picture where the significant decline in SFR (``quenching") and gas depletion have very different timescales. Present-day PSBs might not contribute significantly to the growth of the red sequence galaxy population even though they quenched rapidly: they are able to stay at the intermediate stage for a long time of several Gyr \citep{Rowlands_2018}. Nevertheless, there are still uncertainties in this picture of galaxy evolution. Despite the estimated long gas depletion time, we are unable to tell whether gas would stay in the galaxy, or some other events might happen to remove the gas or trigger SF again. 

What could be hindering the SF in these PSBs? The outflow in IC 860 is unlikely to escape the galaxy (\S5.2) but could disturb the gas enough that it is stable against collapsing into new stars. This quenching by turbulence scenario has been observed in another PSB, NGC 1266, where \citet{Alatalo_2015} found a molecular gas outflow with large $\dot{M}_{out}\sim$110 M$_{\odot}$/yr that was unable to escape the galaxy. More recently, \citet{Smercina_2021} used high-resolution CO data to probe the conditions of the ISM in PSBs and concluded that SF is suppressed by significant turbulent heating. Indeed, energy injection by turbulent heating is a plausible cause for the SF suppression seen in IC 860.

Mergers have been commonly found in PSBs, with evidence for $\sim$50\% of PSBs being morphologically disturbed \citep[e.g.,][]{Blake_2004,Yang_2008,Pawlik_2016}. \citet{Sazonova_2021} studied 26 PSBs in the same parent sample (SPOGs) containing IC 860 
using high resolution \textit{HST} snapshots and found that 88\% of them were structurally disturbed. We observe several signs of a recent merger in IC 860 (\S4.1, \S4.3). Our work thus complements previous findings of mergers being capable of suppressing SF \citep[e.g.,][]{van_de_Voort_2018,Davis_2019}.

Based on our findings, we summarize the possible life story of IC 860: The turning point of this galaxy's life is marked by a merger event. The merger ignited SF that became a starburst, consuming most of the gas originally inside the galaxy. Meanwhile, external gas brought by the merger dissipated its angular momentum through collisions and sank into the center of the galaxy. The central black hole of the galaxy was fueled by gas and became an AGN. The starburst episode quickly ended and the merged system finally stabilized and became a barred spiral galaxy with a 
asymmetric spiral morphology and a multiphase outflow. The gas at the center does not share the same kinematics as the stars, and is prevented from making new stars by feedback likely from the AGN. With a low level of remaining SF and a large gas reservoir, the galaxy would stay at the intermediate stage for several Gyr if no stronger outflow episodes happen, until it finally becomes a quiescent elliptical galaxy. This evolutionary picture could be applicable to other PSBs. More high resolution observations of the gas kinematics at the centers of PSBs, including IC 860, will help to understand the final fate of these quenching galaxies.

\section{Conclusions}
It has been well-established that galaxies evolve from blue and star-forming to red and quiescent. IC 860 is a nearby post-starburst galaxy (PSB) that has just experienced a starburst and is currently on its way to the red sequence. The fact that IC 860 is on the brink of transition suggests that it might contain visible evidence for the triggers of this important transformation. The exquisite multiwavelength data from X-ray to radio enable us to investigate in detail the different activities happening in this galaxy, using it as a local laboratory for studying properties of PSBs at higher redshifts. We find:
\begin{enumerate}
\item The CO molecular gas is compact (r$\sim$1 kpc) and centered on the galaxy nucleus, and is kinematically misaligned with the stars. High resolution \textit{HST} images reveal 
a disturbed, asymmetric spiral morphology. This evidence suggests that IC 860 probably went through a recent merger, which could be the trigger of its starburst and subsequent quenching. Our results complement previous findings that mergers play an important role in quenching galaxies.

\item We find evidence for a multiphase outflow from the CO position-velocity diagram (suggesting a molecular component) and in the NaD absorption spectra (suggesting a neutral component). Dust cone structures in the \textit{HST} images suggest a galaxy-wide dust-rich outflow originating from the galactic center. However, comparing our estimated outflow velocities with the circular velocity of the galaxy suggests that the outflows are unable to, or merely marginally escape the galaxy. This result supports the picture where outflows suppress star formation by disturbing rather than expelling the gas reservoir. 

\item The star formation rates (SFRs) derived from H$\alpha$, Pa$\alpha$, and Ne emission line luminosities show that the current star formation in IC 860 is suppressed given its molecular gas surface density. Under these SFRs, the depletion time of the molecular gas is $\sim$7 Gyr, meaning that the galaxy would continue to harbor a large gas reservoir, unless future mechanisms remove or heat the remaining molecular gas. Thus the timescales for significant decline in SFR (``quenching") and gas depletion are not necessarily the same. 

\item We find evidence for the existence of a weak AGN in the optical emission line ratios, MIR properties, and radio spectral shape. The AGN is probably at an early stage of developing its feedback mechanism. Our results imply that AGN very likely play a role in quenching, but it is unclear whether they are dominant factors driving the change, or merely collateral events as quenching happens. Investigations with more high resolution and deep data on a larger sample of PSBs are desirable. 
\end{enumerate}

\acknowledgments
We thank the anonymous referee for constructive comments what improve the quality of the manuscript. YL thanks Carson Adams for the preliminary work on this project, and Mack Lacy, Erini Lambrides for very useful scientific suggestions and feedback. 

YL, KA, KR, and ES have been have been partially funded by Space Telescope Science Institute Director's Discretionary Research Fund grants D0101.90241, D0101.90276, D0101.90262, D0101.90281, \textit{HST} grants GO-14715.021, GO-14649.015, and \textit{Chandra} grant GO7-18096A. AMM acknowledges support from the National Science Foundation under grant number 2009416. Parts of this research were supported by the Australian Research Council Centre of Excellence for All Sky Astrophysics in 3 Dimensions (ASTRO 3D), through project number CE170100013. LL and SS acknowledges support from \textit{Chandra} grants GO0-21107X (Cycle 22 data of IC\,860) and GO7-18093A and \textit{NuSTAR} grant 80NSSC20K0050, which helped support the development of the forward modeling methodology.

This research is based on observations made with the NASA/ESA Hubble Space Telescope obtained from the Space Telescope Science Institute, which is operated by the Association of Universities for Research in Astronomy, Inc., under NASA contract NAS 5–26555. These observations are associated with program 14715. Basic research in radio astronomy at the U.S. Naval Research Laboratory is supported by 6.1 Base Funding. This research has made use of the NASA/IPAC Extragalactic Database (NED), which is operated by the Jet Propulsion Laboratory, California Institute of Technology, under contract with the National Aeronautics and Space Administration. This research has made use of the NASA/IPAC Infrared Science Archive, which is funded by the National Aeronautics and Space Administration and operated by the California Institute of Technology. This publication makes use of data products from the Two Micron All Sky Survey, which is a joint project of the University of Massachusetts and the Infrared Processing and Analysis Center/California Institute of Technology, funded by the National Aeronautics and Space Administration and the National Science Foundation. This work is based in part on observations made with the Spitzer Space Telescope, which was operated by the Jet Propulsion Laboratory, California Institute of Technology under a contract with NASA. This publication makes use of data products from the Wide-field Infrared Survey Explorer, which is a joint project of the University of California, Los Angeles, and the Jet Propulsion Laboratory/California Institute of Technology, funded by the National Aeronautics and Space Administration. This research has made use of data from the NuSTAR mission, a project led by the California Institute of Technology, managed by the Jet Propulsion Laboratory, and funded by the National Aeronautics and Space Administration. The scientific results reported in this article are partly based on observations made by the Chandra X-Ray Observatory. This research is based on observations with AKARI, a JAXA project with the participation of ESA.

Funding for the SDSS and SDSS-II has been provided by the Alfred P. Sloan Foundation, the Participating Institutions, the National Science Foundation, the U.S. Department of Energy, the National Aeronautics and Space Administration, the Japanese Monbukagakusho, the Max Planck Society, and the Higher Education Funding Council for England. The SDSS Web Site is http://www.sdss.org/. The SDSS is managed by the Astrophysical Research Consortium for the Participating Institutions. The Participating Institutions are the American Museum of Natural History, Astrophysical Institute Potsdam, University of Basel, University of Cambridge, Case Western Reserve University, University of Chicago, Drexel University, Fermilab, the Institute for Advanced Study, the Japan Participation Group, Johns Hopkins University, the Joint Institute for Nuclear Astrophysics, the Kavli Institute for Particle Astrophysics and Cosmology, the Korean Scientist Group, the Chinese Academy of Sciences (LAMOST), Los Alamos National Laboratory, the Max-Planck-Institute for Astronomy (MPIA), the Max-Planck-Institute for Astrophysics (MPA), New Mexico State University, Ohio State University, University of Pittsburgh, University of Portsmouth, Princeton University, the United States Naval Observatory, and the University of Washington.

\software{Astropy \citep{astropy,astropy2}, photutils \citep{photutils}, SciPy \citep{scipy}, Matplotlib \citep{matplotlib}, NumPy \citep{numpy}, AstroDrizzle \citep{Gonzaga_2012}, lacosmic \citep{vanDokkum_2001}, pPXF \citep{Cappellari_2004, Cappellari_2017}, piXedfit \citep{Abdurrouf_2021}, SpectRes \citep{Carnall_2017}, mangadap \citep{Westfall_2019}, LZIFU \citep{Ho_2016}, EMCEE \citep{mcmc}, sherpa \citep{Freeman_2001}}

\appendix
\section{Photometric SED of IC 860}
We summarize in Table \ref{tab:full_sed} the observed total photometric SED data of IC 860 plotted in Figure \ref{fig:sed} (colored squares) and the 113 GHz continuum flux observed by CARMA (\S2.4). The CARMA continuum flux is not used in the SED fitting but is included here for completeness.

\begin{table*}[ht]
\caption{Photometric SED of IC 860}
\centering
\begin{tabular*}{16cm}{c|c|c|c}
\hline \hline
\textbf{Filters} & \textbf{Central wavelength [$\mu m$]} & \textbf{Flux density [Jy]} &  \textbf{Flux density error [Jy]} \\ 
\hline
GALEX FUV & 0.1516 & 2.22$\times 10^{-5}$ & 8.98$\times 10^{-6}$\\
\hline
GALEX NUV & 0.2267 & 1.91$\times 10^{-4}$ & 6.13$\times 10^{-5}$\\
\hline
SDSS $u$ & 0.3351 & 2.11$\times 10^{-3}$ & 3.78$\times 10^{-4}$\\
\hline \hline
\multicolumn{4}{p{16cm}}{\textbf{Notes: }
Fluxes with filter wavelength less than 2$\mu m$ have been corrected for Milky Way extinction (\S2.5). The full version of this table is available in the online version.}
\label{tab:full_sed}
\end{tabular*}
\end{table*}


\newpage
\bibliography{sample63}{}

\begin{thebibliography}{}
\makeatletter
\relax
\def\mn@urlcharsother{\let\do\@makeother \do\$\do\&\do\#\do\^\do\_\do\%\do\~}
\def\mn@doi{\begingroup\mn@urlcharsother \@ifnextchar [ {\mn@doi@}
  {\mn@doi@[]}}
\def\mn@doi@[#1]#2{\def\@tempa{#1}\ifx\@tempa\@empty \href
  {http://dx.doi.org/#2} {doi:#2}\else \href {http://dx.doi.org/#2} {#1}\fi
  \endgroup}
\def\mn@eprint#1#2{\mn@eprint@#1:#2::\@nil}
\def\mn@eprint@arXiv#1{\href {http://arxiv.org/abs/#1} {{\tt arXiv:#1}}}
\def\mn@eprint@dblp#1{\href {http://dblp.uni-trier.de/rec/bibtex/#1.xml}
  {dblp:#1}}
\def\mn@eprint@#1:#2:#3:#4\@nil{\def\@tempa {#1}\def\@tempb {#2}\def\@tempc
  {#3}\ifx \@tempc \@empty \let \@tempc \@tempb \let \@tempb \@tempa \fi \ifx
  \@tempb \@empty \def\@tempb {arXiv}\fi \@ifundefined
  {mn@eprint@\@tempb}{\@tempb:\@tempc}{\expandafter \expandafter \csname
  mn@eprint@\@tempb\endcsname \expandafter{\@tempc}}}

\bibitem[\protect\citeauthoryear{{AKARI Team}}{{AKARI Team}}{2020}]{akari-doi}
{AKARI Team} 2020, AKARI/FIS All-Sky Survey Bright Source Catalogue,
  \mn@doi{10.26131/IRSA180}, \url
  {https://catcopy.ipac.caltech.edu/dois/doi.php?id=10.26131/IRSA180}

\bibitem[\protect\citeauthoryear{{Aalto} et~al.,}{{Aalto}
  et~al.}{2019}]{Aalto_2019}
{Aalto} S.,  et~al., 2019, \mn@doi [\aap] {10.1051/0004-6361/201935480}, \href
  {https://ui.adsabs.harvard.edu/abs/2019A&A...627A.147A} {627, A147}

\bibitem[\protect\citeauthoryear{{Abazajian} et~al.,}{{Abazajian}
  et~al.}{2009}]{Abazajian_2009}
{Abazajian} K.~N.,  et~al., 2009, \mn@doi [\apjs]
  {10.1088/0067-0049/182/2/543}, \href
  {https://ui.adsabs.harvard.edu/abs/2009ApJS..182..543A} {182, 543}

\bibitem[\protect\citeauthoryear{{Abdurro'uf} et~al.,}{{Abdurro'uf}
  et~al.}{2021}]{Abdurrouf_2021}
{Abdurro'uf} et~al., 2021, \mn@doi [\apjs] {10.3847/1538-4365/abebe2}, \href
  {https://ui.adsabs.harvard.edu/abs/2021ApJS..254...15A} {254, 15}

\bibitem[\protect\citeauthoryear{{Abdurro'uf} et~al.,}{{Abdurro'uf}
  et~al.}{2022}]{Abdurrouf_2022}
{Abdurro'uf} et~al., 2022, \mn@doi [\apj] {10.3847/1538-4357/ac439a}, \href
  {https://ui.adsabs.harvard.edu/abs/2022ApJ...926...81A} {926, 81}

\bibitem[\protect\citeauthoryear{{Alatalo} et~al.,}{{Alatalo}
  et~al.}{2011}]{Alatalo11}
{Alatalo} K.,  et~al., 2011, \mn@doi [\apj] {10.1088/0004-637X/735/2/88}, \href
  {https://ui.adsabs.harvard.edu/abs/2011ApJ...735...88A} {735, 88}

\bibitem[\protect\citeauthoryear{{Alatalo} et~al.,}{{Alatalo}
  et~al.}{2013}]{Alatalo_2013}
{Alatalo} K.,  et~al., 2013, \mn@doi [\mnras] {10.1093/mnras/sts299}, \href
  {https://ui.adsabs.harvard.edu/abs/2013MNRAS.432.1796A} {432, 1796}

\bibitem[\protect\citeauthoryear{{Alatalo} et~al.,}{{Alatalo}
  et~al.}{2015}]{Alatalo_2015}
{Alatalo} K.,  et~al., 2015, \mn@doi [\apj] {10.1088/0004-637X/798/1/31}, \href
  {https://ui.adsabs.harvard.edu/abs/2015ApJ...798...31A} {798, 31}

\bibitem[\protect\citeauthoryear{{Alatalo} et~al.,}{{Alatalo}
  et~al.}{2016a}]{Alatalo_2016}
{Alatalo} K.,  et~al., 2016a, \mn@doi [\apjs] {10.3847/0067-0049/224/2/38},
  \href {https://ui.adsabs.harvard.edu/abs/2016ApJS..224...38A} {224, 38}

\bibitem[\protect\citeauthoryear{{Alatalo} et~al.,}{{Alatalo}
  et~al.}{2016b}]{Alatalo_2016a}
{Alatalo} K.,  et~al., 2016b, \mn@doi [\apj] {10.3847/0004-637X/827/2/106},
  \href {https://ui.adsabs.harvard.edu/abs/2016ApJ...827..106A} {827, 106}

\bibitem[\protect\citeauthoryear{{Alatalo} et~al.,}{{Alatalo}
  et~al.}{2017}]{Alatalo_2017}
{Alatalo} K.,  et~al., 2017, \mn@doi [\apj] {10.3847/1538-4357/aa72eb}, \href
  {https://ui.adsabs.harvard.edu/abs/2017ApJ...843....9A} {843, 9}

\bibitem[\protect\citeauthoryear{{Alonso-Herrero} et~al.,}{{Alonso-Herrero}
  et~al.}{2006}]{Alonso-Herrero_2006}
{Alonso-Herrero} A.,  et~al., 2006, \mn@doi [\apj] {10.1086/506958}, \href
  {https://ui.adsabs.harvard.edu/abs/2006ApJ...650..835A} {650, 835}

\bibitem[\protect\citeauthoryear{{Astropy Collaboration} et~al.,}{{Astropy
  Collaboration} et~al.}{2013}]{astropy}
{Astropy Collaboration} et~al., 2013, \mn@doi [\aap]
  {10.1051/0004-6361/201322068}, 558, A33

\bibitem[\protect\citeauthoryear{{Astropy Collaboration} et~al.,}{{Astropy
  Collaboration} et~al.}{2018}]{astropy2}
{Astropy Collaboration} et~al., 2018, \mn@doi [\aj] {10.3847/1538-3881/aabc4f},
  156, 123

\bibitem[\protect\citeauthoryear{{Baan} \& {Kl{\"o}ckner}}{{Baan} \&
  {Kl{\"o}ckner}}{2006}]{Baan_2006}
{Baan} W.~A.,  {Kl{\"o}ckner} H.~R.,  2006, \mn@doi [\aap]
  {10.1051/0004-6361:20042331}, \href
  {https://ui.adsabs.harvard.edu/abs/2006A&A...449..559B} {449, 559}

\bibitem[\protect\citeauthoryear{{Baldry} et~al.,}{{Baldry}
  et~al.}{2004}]{Baldry_2004}
{Baldry} I.~K.,  et~al., 2004, \mn@doi [\apj] {10.1086/380092}, \href
  {https://ui.adsabs.harvard.edu/abs/2004ApJ...600..681B} {600, 681}

\bibitem[\protect\citeauthoryear{Baldwin et~al.,}{Baldwin
  et~al.}{1981}]{Baldwin_1981}
Baldwin J.~A.,  et~al., 1981, \mn@doi [\pasp] {10.1086/130766}, 93, 5

\bibitem[\protect\citeauthoryear{{Baron} et~al.,}{{Baron}
  et~al.}{2022}]{Baron_2022}
{Baron} D.,  et~al., 2022, \mn@doi [\mnras] {10.1093/mnras/stab3232}, \href
  {https://ui.adsabs.harvard.edu/abs/2022MNRAS.509.4457B} {509, 4457}

\bibitem[\protect\citeauthoryear{{Becker} et~al.,}{{Becker}
  et~al.}{1995}]{Becker_1995}
{Becker} R.~H.,  et~al., 1995, \mn@doi [\apj] {10.1086/176166}, \href
  {https://ui.adsabs.harvard.edu/abs/1995ApJ...450..559B} {450, 559}

\bibitem[\protect\citeauthoryear{{Belfiore} et~al.,}{{Belfiore}
  et~al.}{2017}]{Belfiore_2017}
{Belfiore} F.,  et~al., 2017, \mn@doi [\mnras] {10.1093/mnras/stw3211}, \href
  {https://ui.adsabs.harvard.edu/abs/2017MNRAS.466.2570B} {466, 2570}

\bibitem[\protect\citeauthoryear{{Bell} et~al.,}{{Bell}
  et~al.}{2007}]{Bell_2007}
{Bell} E.~F.,  et~al., 2007, \mn@doi [\apj] {10.1086/518594}, \href
  {https://ui.adsabs.harvard.edu/abs/2007ApJ...663..834B} {663, 834}

\bibitem[\protect\citeauthoryear{{Bell} et~al.,}{{Bell}
  et~al.}{2012}]{Bell_2012}
{Bell} E.~F.,  et~al., 2012, \mn@doi [\apj] {10.1088/0004-637X/753/2/167},
  \href {https://ui.adsabs.harvard.edu/abs/2012ApJ...753..167B} {753, 167}

\bibitem[\protect\citeauthoryear{{Bertelli} et~al.,}{{Bertelli}
  et~al.}{1994}]{Bertelli_1994}
{Bertelli} G.,  et~al., 1994, \aaps, \href
  {https://ui.adsabs.harvard.edu/abs/1994A&AS..106..275B} {106, 275}

\bibitem[\protect\citeauthoryear{{Blake} et~al.,}{{Blake}
  et~al.}{2004}]{Blake_2004}
{Blake} C.,  et~al., 2004, \mn@doi [\mnras] {10.1111/j.1365-2966.2004.08351.x},
  \href {https://ui.adsabs.harvard.edu/abs/2004MNRAS.355..713B} {355, 713}

\bibitem[\protect\citeauthoryear{{Bock} et~al.,}{{Bock}
  et~al.}{2006}]{Bock_2006}
{Bock} D.~C.~J.,  et~al., 2006, in {Stepp} L.~M.,  ed.,  Society of
  Photo-Optical Instrumentation Engineers (SPIE) Conference Series Vol. 6267,
  Society of Photo-Optical Instrumentation Engineers (SPIE) Conference Series.
  p. 626713, \mn@doi{10.1117/12.674051}

\bibitem[\protect\citeauthoryear{{Bohlin} et~al.,}{{Bohlin}
  et~al.}{1978}]{Bohlin_1978}
{Bohlin} R.~C.,  et~al., 1978, \mn@doi [\apj] {10.1086/156357}, \href
  {https://ui.adsabs.harvard.edu/abs/1978ApJ...224..132B} {224, 132}

\bibitem[\protect\citeauthoryear{{Bolatto} et~al.,}{{Bolatto}
  et~al.}{2013}]{Bolatto_2013}
{Bolatto} A.~D.,  et~al., 2013, \mn@doi [\araa]
  {10.1146/annurev-astro-082812-140944}, \href
  {https://ui.adsabs.harvard.edu/abs/2013ARA&A..51..207B} {51, 207}

\bibitem[\protect\citeauthoryear{{Bournaud} et~al.,}{{Bournaud}
  et~al.}{2005}]{Bournaud_2005}
{Bournaud} F.,  et~al., 2005, \mn@doi [\aap] {10.1051/0004-6361:20042036},
  \href {https://ui.adsabs.harvard.edu/abs/2005A&A...437...69B} {437, 69}

\bibitem[\protect\citeauthoryear{{Bournaud} et~al.,}{{Bournaud}
  et~al.}{2007}]{Bournaud_2007}
{Bournaud} F.,  et~al., 2007, \mn@doi [\aap] {10.1051/0004-6361:20078010},
  \href {https://ui.adsabs.harvard.edu/abs/2007A&A...476.1179B} {476, 1179}

\bibitem[\protect\citeauthoryear{Bradley et~al.,}{Bradley
  et~al.}{2020}]{photutils}
Bradley L.,  et~al., 2020, astropy/photutils: 1.0.1,
  \mn@doi{10.5281/zenodo.596036}, \url
  {https://ui.adsabs.harvard.edu/abs/2020zndo....596036B}

\bibitem[\protect\citeauthoryear{{Brown} et~al.,}{{Brown}
  et~al.}{2014}]{Brown_2014}
{Brown} M. J.~I.,  et~al., 2014, \mn@doi [\apjs] {10.1088/0067-0049/212/2/18},
  \href {https://ui.adsabs.harvard.edu/abs/2014ApJS..212...18B} {212, 18}

\bibitem[\protect\citeauthoryear{{Bryant} et~al.,}{{Bryant}
  et~al.}{2019}]{Bryant_2019}
{Bryant} J.~J.,  et~al., 2019, \mn@doi [\mnras] {10.1093/mnras/sty3122}, \href
  {https://ui.adsabs.harvard.edu/abs/2019MNRAS.483..458B} {483, 458}

\bibitem[\protect\citeauthoryear{{Calzetti}}{{Calzetti}}{2013}]{Calzetti_2013}
{Calzetti} D.,  2013, {Star Formation Rate Indicators}.
p.~419

\bibitem[\protect\citeauthoryear{{Calzetti} et~al.,}{{Calzetti}
  et~al.}{2000}]{Calzetti_2000}
{Calzetti} D.,  et~al., 2000, \mn@doi [\apj] {10.1086/308692}, \href
  {https://ui.adsabs.harvard.edu/abs/2000ApJ...533..682C} {533, 682}

\bibitem[\protect\citeauthoryear{{Calzetti} et~al.,}{{Calzetti}
  et~al.}{2010}]{Calzetti_2010}
{Calzetti} D.,  et~al., 2010, \mn@doi [\apj] {10.1088/0004-637X/714/2/1256},
  \href {https://ui.adsabs.harvard.edu/abs/2010ApJ...714.1256C} {714, 1256}

\bibitem[\protect\citeauthoryear{{Cappellari}}{{Cappellari}}{2017}]{Cappellari_2017}
{Cappellari} M.,  2017, \mn@doi [\mnras] {10.1093/mnras/stw3020}, \href
  {https://ui.adsabs.harvard.edu/abs/2017MNRAS.466..798C} {466, 798}

\bibitem[\protect\citeauthoryear{{Cappellari} \& {Copin}}{{Cappellari} \&
  {Copin}}{2003}]{Cappellari_2003}
{Cappellari} M.,  {Copin} Y.,  2003, \mn@doi [\mnras]
  {10.1046/j.1365-8711.2003.06541.x}, \href
  {https://ui.adsabs.harvard.edu/abs/2003MNRAS.342..345C} {342, 345}

\bibitem[\protect\citeauthoryear{{Cappellari} \& {Emsellem}}{{Cappellari} \&
  {Emsellem}}{2004}]{Cappellari_2004}
{Cappellari} M.,  {Emsellem} E.,  2004, \mn@doi [\pasp] {10.1086/381875}, \href
  {https://ui.adsabs.harvard.edu/abs/2004PASP..116..138C} {116, 138}

\bibitem[\protect\citeauthoryear{{Carnall}}{{Carnall}}{2017}]{Carnall_2017}
{Carnall} A.~C.,  2017, arXiv e-prints, \href
  {https://ui.adsabs.harvard.edu/abs/2017arXiv170505165C} {p. arXiv:1705.05165}

\bibitem[\protect\citeauthoryear{{Chabrier}}{{Chabrier}}{2003}]{Chabrier_2003}
{Chabrier} G.,  2003, \mn@doi [\pasp] {10.1086/376392}, \href
  {https://ui.adsabs.harvard.edu/abs/2003PASP..115..763C} {115, 763}

\bibitem[\protect\citeauthoryear{{Charlot} \& {Fall}}{{Charlot} \&
  {Fall}}{2000}]{Charlot_2000}
{Charlot} S.,  {Fall} S.~M.,  2000, \mn@doi [\apj] {10.1086/309250}, \href
  {https://ui.adsabs.harvard.edu/abs/2000ApJ...539..718C} {539, 718}

\bibitem[\protect\citeauthoryear{{Chen} et~al.,}{{Chen}
  et~al.}{2010}]{Chen_2010}
{Chen} Y.-M.,  et~al., 2010, \mn@doi [\aj] {10.1088/0004-6256/140/2/445}, \href
  {https://ui.adsabs.harvard.edu/abs/2010AJ....140..445C} {140, 445}

\bibitem[\protect\citeauthoryear{{Childress} et~al.,}{{Childress}
  et~al.}{2014}]{Childress_2014}
{Childress} M.~J.,  et~al., 2014, \mn@doi [\apss] {10.1007/s10509-013-1682-0},
  \href {https://ui.adsabs.harvard.edu/abs/2014Ap&SS.349..617C} {349, 617}

\bibitem[\protect\citeauthoryear{{Chu} et~al.,}{{Chu} et~al.}{2017}]{Chu_2017}
{Chu} J.~K.,  et~al., 2017, \mn@doi [\apjs] {10.3847/1538-4365/aa5d15}, \href
  {https://ui.adsabs.harvard.edu/abs/2017ApJS..229...25C} {229, 25}

\bibitem[\protect\citeauthoryear{{Cid Fernandes} et~al.,}{{Cid Fernandes}
  et~al.}{2011}]{CidFernandes_2011}
{Cid Fernandes} R.,  et~al., 2011, \mn@doi [\mnras]
  {10.1111/j.1365-2966.2011.18244.x}, \href
  {https://ui.adsabs.harvard.edu/abs/2011MNRAS.413.1687C} {413, 1687}

\bibitem[\protect\citeauthoryear{{Clarke} et~al.,}{{Clarke}
  et~al.}{2016}]{Clarke_2016}
{Clarke} T.,  et~al., 2016, arXiv e-prints, \href
  {https://ui.adsabs.harvard.edu/abs/2016arXiv160303080C} {p. arXiv:1603.03080}

\bibitem[\protect\citeauthoryear{{Colombo} et~al.,}{{Colombo}
  et~al.}{2018}]{Colombo_2018}
{Colombo} D.,  et~al., 2018, \mn@doi [\mnras] {10.1093/mnras/stx3233}, \href
  {https://ui.adsabs.harvard.edu/abs/2018MNRAS.475.1791C} {475, 1791}

\bibitem[\protect\citeauthoryear{{Condon} et~al.,}{{Condon}
  et~al.}{1991}]{Condon_1991}
{Condon} J.~J.,  et~al., 1991, \mn@doi [\apj] {10.1086/170407}, \href
  {https://ui.adsabs.harvard.edu/abs/1991ApJ...378...65C} {378, 65}

\bibitem[\protect\citeauthoryear{{Condon} et~al.,}{{Condon}
  et~al.}{1998}]{Condon_1998}
{Condon} J.~J.,  et~al., 1998, \mn@doi [\aj] {10.1086/300337}, \href
  {https://ui.adsabs.harvard.edu/abs/1998AJ....115.1693C} {115, 1693}

\bibitem[\protect\citeauthoryear{{Conroy} et~al.,}{{Conroy}
  et~al.}{2009}]{Conroy_2009}
{Conroy} C.,  et~al., 2009, \mn@doi [\apj] {10.1088/0004-637X/699/1/486}, \href
  {https://ui.adsabs.harvard.edu/abs/2009ApJ...699..486C} {699, 486}

\bibitem[\protect\citeauthoryear{{Dadina}}{{Dadina}}{2008}]{Dadina_2008}
{Dadina} M.,  2008, \mn@doi [\aap] {10.1051/0004-6361:20077569}, \href
  {https://ui.adsabs.harvard.edu/abs/2008A&A...485..417D} {485, 417}

\bibitem[\protect\citeauthoryear{{Davis} et~al.,}{{Davis}
  et~al.}{2011}]{Davis_2011}
{Davis} T.~A.,  et~al., 2011, \mn@doi [\mnras]
  {10.1111/j.1365-2966.2011.19355.x}, \href
  {https://ui.adsabs.harvard.edu/abs/2011MNRAS.417..882D} {417, 882}

\bibitem[\protect\citeauthoryear{{Davis} et~al.,}{{Davis}
  et~al.}{2019}]{Davis_2019}
{Davis} T.~A.,  et~al., 2019, \mn@doi [\mnras] {10.1093/mnras/stz180}, \href
  {https://ui.adsabs.harvard.edu/abs/2019MNRAS.484.2447D} {484, 2447}

\bibitem[\protect\citeauthoryear{{D{\'\i}az-Santos} et~al.,}{{D{\'\i}az-Santos}
  et~al.}{2017}]{Diaz-Santos_2017}
{D{\'\i}az-Santos} T.,  et~al., 2017, \mn@doi [\apj]
  {10.3847/1538-4357/aa81d7}, \href
  {https://ui.adsabs.harvard.edu/abs/2017ApJ...846...32D} {846, 32}

\bibitem[\protect\citeauthoryear{{Donley} et~al.,}{{Donley}
  et~al.}{2012}]{Donley_2012}
{Donley} J.~L.,  et~al., 2012, \mn@doi [\apj] {10.1088/0004-637X/748/2/142},
  \href {https://ui.adsabs.harvard.edu/abs/2012ApJ...748..142D} {748, 142}

\bibitem[\protect\citeauthoryear{{Dopita} et~al.,}{{Dopita}
  et~al.}{2007}]{Dopita_2007}
{Dopita} M.,  et~al., 2007, \mn@doi [\apss] {10.1007/s10509-007-9510-z}, \href
  {https://ui.adsabs.harvard.edu/abs/2007Ap&SS.310..255D} {310, 255}

\bibitem[\protect\citeauthoryear{{Dopita} et~al.,}{{Dopita}
  et~al.}{2010}]{Dopita_2010}
{Dopita} M.,  et~al., 2010, \mn@doi [\apss] {10.1007/s10509-010-0335-9}, \href
  {https://ui.adsabs.harvard.edu/abs/2010Ap&SS.327..245D} {327, 245}

\bibitem[\protect\citeauthoryear{{Draine} \& {Li}}{{Draine} \&
  {Li}}{2007}]{Draine_2007}
{Draine} B.~T.,  {Li} A.,  2007, \mn@doi [\apj] {10.1086/511055}, \href
  {https://ui.adsabs.harvard.edu/abs/2007ApJ...657..810D} {657, 810}

\bibitem[\protect\citeauthoryear{{Falstad} et~al.,}{{Falstad}
  et~al.}{2019}]{Falstad_2019}
{Falstad} N.,  et~al., 2019, \mn@doi [\aap] {10.1051/0004-6361/201834586},
  \href {https://ui.adsabs.harvard.edu/abs/2019A&A...623A..29F} {623, A29}

\bibitem[\protect\citeauthoryear{{Foreman-Mackey} et~al.,}{{Foreman-Mackey}
  et~al.}{2013}]{mcmc}
{Foreman-Mackey} D.,  et~al., 2013, \mn@doi [\pasp] {10.1086/670067}, \href
  {https://ui.adsabs.harvard.edu/abs/2013PASP..125..306F} {125, 306}

\bibitem[\protect\citeauthoryear{{Freeman} et~al.,}{{Freeman}
  et~al.}{2001}]{Freeman_2001}
{Freeman} P.,  et~al., 2001, in {Starck} J.-L.,  {Murtagh} F.~D.,  eds,
  Society of Photo-Optical Instrumentation Engineers (SPIE) Conference Series
  Vol. 4477, Astronomical Data Analysis. pp 76--87 (\mn@eprint {arXiv}
  {astro-ph/0108426}), \mn@doi{10.1117/12.447161}

\bibitem[\protect\citeauthoryear{{French}}{{French}}{2021}]{French_2021}
{French} K.~D.,  2021, \mn@doi [\pasp] {10.1088/1538-3873/ac0a59}, \href
  {https://ui.adsabs.harvard.edu/abs/2021PASP..133g2001F} {133, 072001}

\bibitem[\protect\citeauthoryear{{French} et~al.,}{{French}
  et~al.}{2015}]{French_2015}
{French} K.~D.,  et~al., 2015, \mn@doi [\apj] {10.1088/0004-637X/801/1/1},
  \href {https://ui.adsabs.harvard.edu/abs/2015ApJ...801....1F} {801, 1}

\bibitem[\protect\citeauthoryear{{French} et~al.,}{{French}
  et~al.}{2018}]{French_2018}
{French} K.~D.,  et~al., 2018, \mn@doi [\apj] {10.3847/1538-4357/aacb2d}, \href
  {https://ui.adsabs.harvard.edu/abs/2018ApJ...862....2F} {862, 2}

\bibitem[\protect\citeauthoryear{{GOALS Team}}{{GOALS Team}}{2020}]{goals}
{GOALS Team} 2020, Great Observatories All-sky LIRG Survey,
  \mn@doi{10.26131/IRSA183}, \url
  {https://catcopy.ipac.caltech.edu/dois/doi.php?id=10.26131/IRSA183}

\bibitem[\protect\citeauthoryear{{Gehrels}}{{Gehrels}}{1986}]{Gehrels_1986}
{Gehrels} N.,  1986, \mn@doi [\apj] {10.1086/164079}, \href
  {https://ui.adsabs.harvard.edu/abs/1986ApJ...303..336G} {303, 336}

\bibitem[\protect\citeauthoryear{{Girardi} et~al.,}{{Girardi}
  et~al.}{2000}]{Girardi_2000}
{Girardi} L.,  et~al., 2000, \mn@doi [\aaps] {10.1051/aas:2000126}, \href
  {https://ui.adsabs.harvard.edu/abs/2000A&AS..141..371G} {141, 371}

\bibitem[\protect\citeauthoryear{{Girardi} et~al.,}{{Girardi}
  et~al.}{2002}]{Girardi_2002}
{Girardi} L.,  et~al., 2002, \mn@doi [\aap] {10.1051/0004-6361:20020612}, \href
  {https://ui.adsabs.harvard.edu/abs/2002A&A...391..195G} {391, 195}

\bibitem[\protect\citeauthoryear{{Gonzaga} \& {et al.}}{{Gonzaga} \& {et
  al.}}{2012}]{Gonzaga_2012}
{Gonzaga} S.,  {et al.} 2012, {The DrizzlePac Handbook}

\bibitem[\protect\citeauthoryear{{Gonz{\'a}lez Delgado} et~al.,}{{Gonz{\'a}lez
  Delgado} et~al.}{2005}]{Delgado_2005}
{Gonz{\'a}lez Delgado} R.~M.,  et~al., 2005, \mn@doi [\mnras]
  {10.1111/j.1365-2966.2005.08692.x}, \href
  {https://ui.adsabs.harvard.edu/abs/2005MNRAS.357..945G} {357, 945}

\bibitem[\protect\citeauthoryear{{Gordon} et~al.,}{{Gordon}
  et~al.}{2020}]{Gordon_2020}
{Gordon} Y.~A.,  et~al., 2020, \mn@doi [Research Notes of the American
  Astronomical Society] {10.3847/2515-5172/abbe23}, \href
  {https://ui.adsabs.harvard.edu/abs/2020RNAAS...4..175G} {4, 175}

\bibitem[\protect\citeauthoryear{{Goto}}{{Goto}}{2007}]{Goto_2007}
{Goto} T.,  2007, \mn@doi [\mnras] {10.1111/j.1365-2966.2007.11674.x}, \href
  {https://ui.adsabs.harvard.edu/abs/2007MNRAS.377.1222G} {377, 1222}

\bibitem[\protect\citeauthoryear{{Greene} et~al.,}{{Greene}
  et~al.}{2020}]{Greene_2020}
{Greene} J.~E.,  et~al., 2020, \mn@doi [\apjl] {10.3847/2041-8213/aba534},
  \href {https://ui.adsabs.harvard.edu/abs/2020ApJ...899L...9G} {899, L9}

\bibitem[\protect\citeauthoryear{{Gunn} \& {Gott}}{{Gunn} \&
  {Gott}}{1972}]{Gunn_1972}
{Gunn} J.~E.,  {Gott} J.~Richard I.,  1972, \mn@doi [\apj] {10.1086/151605},
  \href {https://ui.adsabs.harvard.edu/abs/1972ApJ...176....1G} {176, 1}

\bibitem[\protect\citeauthoryear{{HI4PI Collaboration} et~al.,}{{HI4PI
  Collaboration} et~al.}{2016}]{HI4PI}
{HI4PI Collaboration} et~al., 2016, \mn@doi [\aap]
  {10.1051/0004-6361/201629178}, \href
  {https://ui.adsabs.harvard.edu/abs/2016A&A...594A.116H} {594, A116}

\bibitem[\protect\citeauthoryear{Harris et~al.,}{Harris et~al.}{2020}]{numpy}
Harris C.~R.,  et~al., 2020, \mn@doi [Nature] {10.1038/s41586-020-2649-2}, 585,
  357

\bibitem[\protect\citeauthoryear{{Harrison} et~al.,}{{Harrison}
  et~al.}{2013}]{Harrison_2013}
{Harrison} F.~A.,  et~al., 2013, \mn@doi [\apj] {10.1088/0004-637X/770/2/103},
  \href {https://ui.adsabs.harvard.edu/abs/2013ApJ...770..103H} {770, 103}

\bibitem[\protect\citeauthoryear{{Hayward} et~al.,}{{Hayward}
  et~al.}{2014}]{Hayward_2014}
{Hayward} C.~C.,  et~al., 2014, \mn@doi [\mnras] {10.1093/mnras/stu1843}, \href
  {https://ui.adsabs.harvard.edu/abs/2014MNRAS.445.1598H} {445, 1598}

\bibitem[\protect\citeauthoryear{{Heckman} et~al.,}{{Heckman}
  et~al.}{2015}]{Heckman_2015}
{Heckman} T.~M.,  et~al., 2015, \mn@doi [\apj] {10.1088/0004-637X/809/2/147},
  \href {https://ui.adsabs.harvard.edu/abs/2015ApJ...809..147H} {809, 147}

\bibitem[\protect\citeauthoryear{{Hill} \& {Zakamska}}{{Hill} \&
  {Zakamska}}{2014}]{hill2014}
{Hill} M.~J.,  {Zakamska} N.~L.,  2014, \mn@doi [\mnras]
  {10.1093/mnras/stu123}, \href
  {https://ui.adsabs.harvard.edu/abs/2014MNRAS.439.2701H} {439, 2701}

\bibitem[\protect\citeauthoryear{{Ho} \& {Keto}}{{Ho} \&
  {Keto}}{2007}]{Ho_2007}
{Ho} L.~C.,  {Keto} E.,  2007, \mn@doi [\apj] {10.1086/511260}, \href
  {https://ui.adsabs.harvard.edu/abs/2007ApJ...658..314H} {658, 314}

\bibitem[\protect\citeauthoryear{{Ho} et~al.,}{{Ho} et~al.}{2016}]{Ho_2016}
{Ho} I.~T.,  et~al., 2016, \mn@doi [\apss] {10.1007/s10509-016-2865-2}, \href
  {https://ui.adsabs.harvard.edu/abs/2016Ap&SS.361..280H} {361, 280}

\bibitem[\protect\citeauthoryear{{Hopkins} et~al.,}{{Hopkins}
  et~al.}{2006}]{Hopkins_2006}
{Hopkins} P.~F.,  et~al., 2006, \mn@doi [\apjs] {10.1086/499298}, \href
  {https://ui.adsabs.harvard.edu/abs/2006ApJS..163....1H} {163, 1}

\bibitem[\protect\citeauthoryear{{Hopkins} et~al.,}{{Hopkins}
  et~al.}{2008}]{Hopkins_2008}
{Hopkins} P.~F.,  et~al., 2008, \mn@doi [\apjs] {10.1086/524363}, \href
  {https://ui.adsabs.harvard.edu/abs/2008ApJS..175..390H} {175, 390}

\bibitem[\protect\citeauthoryear{{Hubble}}{{Hubble}}{1926}]{Hubble_1926}
{Hubble} E.~P.,  1926, \mn@doi [\apj] {10.1086/143018}, \href
  {https://ui.adsabs.harvard.edu/abs/1926ApJ....64..321H} {64, 321}

\bibitem[\protect\citeauthoryear{Hunter}{Hunter}{2007}]{matplotlib}
Hunter J.~D.,  2007, \mn@doi [Computing in Science {\&} Engineering]
  {10.1109/mcse.2007.55}, 9, 90

\bibitem[\protect\citeauthoryear{{Inami} et~al.,}{{Inami}
  et~al.}{2013}]{inami2013}
{Inami} H.,  et~al., 2013, \mn@doi [\apj] {10.1088/0004-637X/777/2/156}, \href
  {https://ui.adsabs.harvard.edu/abs/2013ApJ...777..156I} {777, 156}

\bibitem[\protect\citeauthoryear{{Intema} et~al.,}{{Intema}
  et~al.}{2017}]{Intema_2017}
{Intema} H.~T.,  et~al., 2017, \mn@doi [\aap] {10.1051/0004-6361/201628536},
  \href {https://ui.adsabs.harvard.edu/abs/2017A&A...598A..78I} {598, A78}

\bibitem[\protect\citeauthoryear{{Jarrett} et~al.,}{{Jarrett}
  et~al.}{2012}]{Jarrett_2012}
{Jarrett} T.~H.,  et~al., 2012, \mn@doi [\aj] {10.1088/0004-6256/144/2/68},
  \href {https://ui.adsabs.harvard.edu/abs/2012AJ....144...68J} {144, 68}

\bibitem[\protect\citeauthoryear{{Kauffmann} et~al.,}{{Kauffmann}
  et~al.}{2003}]{Kauffmann_2003}
{Kauffmann} G.,  et~al., 2003, \mn@doi [\mnras]
  {10.1111/j.1365-2966.2003.07154.x}, \href
  {https://ui.adsabs.harvard.edu/abs/2003MNRAS.346.1055K} {346, 1055}

\bibitem[\protect\citeauthoryear{{Kawada} et~al.,}{{Kawada}
  et~al.}{2007}]{akari_fis}
{Kawada} M.,  et~al., 2007, \mn@doi [\pasj] {10.1093/pasj/59.sp2.S389}, \href
  {https://ui.adsabs.harvard.edu/abs/2007PASJ...59S.389K} {59, S389}

\bibitem[\protect\citeauthoryear{{Kennicutt}}{{Kennicutt}}{1998}]{Kennicutt_1998}
{Kennicutt} Robert~C. J.,  1998, \mn@doi [\apj] {10.1086/305588}, \href
  {https://ui.adsabs.harvard.edu/abs/1998ApJ...498..541K} {498, 541}

\bibitem[\protect\citeauthoryear{{Kennicutt} \& {De Los Reyes}}{{Kennicutt} \&
  {De Los Reyes}}{2021}]{Kennicutt_2021}
{Kennicutt} Robert~C. J.,  {De Los Reyes} M. A.~C.,  2021, \mn@doi [\apj]
  {10.3847/1538-4357/abd3a2}, \href
  {https://ui.adsabs.harvard.edu/abs/2021ApJ...908...61K} {908, 61}

\bibitem[\protect\citeauthoryear{{Kennicutt} \& {Evans}}{{Kennicutt} \&
  {Evans}}{2012}]{Kennicutt_2012}
{Kennicutt} R.~C.,  {Evans} N.~J.,  2012, \mn@doi [\araa]
  {10.1146/annurev-astro-081811-125610}, \href
  {https://ui.adsabs.harvard.edu/abs/2012ARA&A..50..531K} {50, 531}

\bibitem[\protect\citeauthoryear{{Kennicutt} Robert~C. et~al.,}{{Kennicutt}
  et~al.}{2009}]{Kennicutt_2009}
{Kennicutt} Robert~C. J.,  et~al., 2009, \mn@doi [\apj]
  {10.1088/0004-637X/703/2/1672}, \href
  {https://ui.adsabs.harvard.edu/abs/2009ApJ...703.1672K} {703, 1672}

\bibitem[\protect\citeauthoryear{{Kewley} et~al.,}{{Kewley}
  et~al.}{2001}]{Kewley_2001}
{Kewley} L.~J.,  et~al., 2001, \mn@doi [\apj] {10.1086/321545}, \href
  {https://ui.adsabs.harvard.edu/abs/2001ApJ...556..121K} {556, 121}

\bibitem[\protect\citeauthoryear{{Krajnovi{\'c}} et~al.,}{{Krajnovi{\'c}}
  et~al.}{2006}]{Krajnovic_2006}
{Krajnovi{\'c}} D.,  et~al., 2006, \mn@doi [\mnras]
  {10.1111/j.1365-2966.2005.09902.x}, \href
  {https://ui.adsabs.harvard.edu/abs/2006MNRAS.366..787K} {366, 787}

\bibitem[\protect\citeauthoryear{{Kroupa}}{{Kroupa}}{2001}]{Kroupa_2001}
{Kroupa} P.,  2001, \mn@doi [\mnras] {10.1046/j.1365-8711.2001.04022.x}, \href
  {https://ui.adsabs.harvard.edu/abs/2001MNRAS.322..231K} {322, 231}

\bibitem[\protect\citeauthoryear{{Lacy} et~al.,}{{Lacy}
  et~al.}{2004}]{Lacy_2004}
{Lacy} M.,  et~al., 2004, \mn@doi [\apjs] {10.1086/422816}, \href
  {https://ui.adsabs.harvard.edu/abs/2004ApJS..154..166L} {154, 166}

\bibitem[\protect\citeauthoryear{{Lacy} et~al.,}{{Lacy}
  et~al.}{2007}]{Lacy_2007}
{Lacy} M.,  et~al., 2007, \mn@doi [\aj] {10.1086/509617}, \href
  {https://ui.adsabs.harvard.edu/abs/2007AJ....133..186L} {133, 186}

\bibitem[\protect\citeauthoryear{{Lacy} et~al.,}{{Lacy}
  et~al.}{2013}]{Lacy_2013}
{Lacy} M.,  et~al., 2013, \mn@doi [\apjs] {10.1088/0067-0049/208/2/24}, \href
  {https://ui.adsabs.harvard.edu/abs/2013ApJS..208...24L} {208, 24}

\bibitem[\protect\citeauthoryear{{Lacy} et~al.,}{{Lacy}
  et~al.}{2020}]{Lacy_2020}
{Lacy} M.,  et~al., 2020, \mn@doi [\pasp] {10.1088/1538-3873/ab63eb}, \href
  {https://ui.adsabs.harvard.edu/abs/2020PASP..132c5001L} {132, 035001}

\bibitem[\protect\citeauthoryear{{Lambrides} et~al.,}{{Lambrides}
  et~al.}{2019a}]{Lambrides_2019}
{Lambrides} E.~L.,  et~al., 2019a, \mn@doi [\mnras] {10.1093/mnras/stz1316},
  \href {https://ui.adsabs.harvard.edu/abs/2019MNRAS.487.1823L} {487, 1823}

\bibitem[\protect\citeauthoryear{{Lambrides} et~al.,}{{Lambrides}
  et~al.}{2019b}]{lamb2019}
{Lambrides} E.~L.,  et~al., 2019b, \mn@doi [\mnras] {10.1093/mnras/stz1316},
  \href {https://ui.adsabs.harvard.edu/abs/2019MNRAS.487.1823L} {487, 1823}

\bibitem[\protect\citeauthoryear{{Lanz} et~al.,}{{Lanz}
  et~al.}{2022}]{Lanz_2022}
{Lanz} L.,  et~al., 2022, \mn@doi [\apj] {10.3847/1538-4357/ac7d56}, \href
  {https://ui.adsabs.harvard.edu/abs/2022ApJ...935...29L} {935, 29}

\bibitem[\protect\citeauthoryear{{Lin} et~al.,}{{Lin} et~al.}{2019}]{Lin_2019}
{Lin} L.,  et~al., 2019, \mn@doi [\apj] {10.3847/1538-4357/aafa84}, \href
  {https://ui.adsabs.harvard.edu/abs/2019ApJ...872...50L} {872, 50}

\bibitem[\protect\citeauthoryear{{Marshall} et~al.,}{{Marshall}
  et~al.}{2004}]{Marshall_2004}
{Marshall} H.~L.,  et~al., 2004, in {Flanagan} K.~A.,  {Siegmund} O. H.~W.,
  eds,  Society of Photo-Optical Instrumentation Engineers (SPIE) Conference
  Series Vol. 5165, X-Ray and Gamma-Ray Instrumentation for Astronomy XIII. pp
  497--508 (\mn@eprint {arXiv} {astro-ph/0308332}), \mn@doi{10.1117/12.508310}

\bibitem[\protect\citeauthoryear{{Martig} et~al.,}{{Martig}
  et~al.}{2009}]{Martig_2009}
{Martig} M.,  et~al., 2009, \mn@doi [\apj] {10.1088/0004-637X/707/1/250}, \href
  {https://ui.adsabs.harvard.edu/abs/2009ApJ...707..250M} {707, 250}

\bibitem[\protect\citeauthoryear{{Martin} et~al.,}{{Martin}
  et~al.}{2005}]{Martin_2005}
{Martin} D.~C.,  et~al., 2005, \mn@doi [\apjl] {10.1086/426387}, \href
  {https://ui.adsabs.harvard.edu/abs/2005ApJ...619L...1M} {619, L1}

\bibitem[\protect\citeauthoryear{{Masters} et~al.,}{{Masters}
  et~al.}{2010}]{Masters_2010}
{Masters} K.~L.,  et~al., 2010, \mn@doi [\mnras]
  {10.1111/j.1365-2966.2010.16335.x}, \href
  {https://ui.adsabs.harvard.edu/abs/2010MNRAS.404..792M} {404, 792}

\bibitem[\protect\citeauthoryear{{McBride} et~al.,}{{McBride}
  et~al.}{2014}]{McBride_2014}
{McBride} J.,  et~al., 2014, \mn@doi [\apj] {10.1088/0004-637X/780/2/182},
  \href {https://ui.adsabs.harvard.edu/abs/2014ApJ...780..182M} {780, 182}

\bibitem[\protect\citeauthoryear{{McConnell} et~al.,}{{McConnell}
  et~al.}{2020}]{McConnell_2020}
{McConnell} D.,  et~al., 2020, \mn@doi [\pasa] {10.1017/pasa.2020.41}, \href
  {https://ui.adsabs.harvard.edu/abs/2020PASA...37...48M} {37, e048}

\bibitem[\protect\citeauthoryear{{M{\'e}ndez-Abreu} et~al.,}{{M{\'e}ndez-Abreu}
  et~al.}{2019}]{Mendez-Abreu_2019}
{M{\'e}ndez-Abreu} J.,  et~al., 2019, \mn@doi [\mnras] {10.1093/mnrasl/slz103},
  \href {https://ui.adsabs.harvard.edu/abs/2019MNRAS.488L..80M} {488, L80}

\bibitem[\protect\citeauthoryear{{Minsley} et~al.,}{{Minsley}
  et~al.}{2020}]{minsley2020}
{Minsley} R.,  et~al., 2020, \mn@doi [\apj] {10.3847/1538-4357/ab86a1}, \href
  {https://ui.adsabs.harvard.edu/abs/2020ApJ...894..157M} {894, 157}

\bibitem[\protect\citeauthoryear{{Mirabel} \& {Sanders}}{{Mirabel} \&
  {Sanders}}{1988}]{Mirabel_Sanders_1988}
{Mirabel} I.~F.,  {Sanders} D.~B.,  1988, \mn@doi [\apj] {10.1086/166909},
  \href {https://ui.adsabs.harvard.edu/abs/1988ApJ...335..104M} {335, 104}

\bibitem[\protect\citeauthoryear{{Morton}}{{Morton}}{1991}]{Morton_1991}
{Morton} D.~C.,  1991, \mn@doi [\apjs] {10.1086/191601}, \href
  {https://ui.adsabs.harvard.edu/abs/1991ApJS...77..119M} {77, 119}

\bibitem[\protect\citeauthoryear{{Murakami} et~al.,}{{Murakami}
  et~al.}{2007}]{akari_mission}
{Murakami} H.,  et~al., 2007, \mn@doi [\pasj] {10.1093/pasj/59.sp2.S369}, \href
  {https://ui.adsabs.harvard.edu/abs/2007PASJ...59S.369M} {59, S369}

\bibitem[\protect\citeauthoryear{{Murphy} et~al.,}{{Murphy}
  et~al.}{2011}]{Murphy_2011}
{Murphy} E.~J.,  et~al., 2011, \mn@doi [\apj] {10.1088/0004-637X/737/2/67},
  \href {https://ui.adsabs.harvard.edu/abs/2011ApJ...737...67M} {737, 67}

\bibitem[\protect\citeauthoryear{{Murray} et~al.,}{{Murray}
  et~al.}{2005}]{Murray_2005}
{Murray} N.,  et~al., 2005, \mn@doi [\apj] {10.1086/426067}, \href
  {https://ui.adsabs.harvard.edu/abs/2005ApJ...618..569M} {618, 569}

\bibitem[\protect\citeauthoryear{{Naab} \& {Burkert}}{{Naab} \&
  {Burkert}}{2003}]{Naab_2003}
{Naab} T.,  {Burkert} A.,  2003, \mn@doi [\apj] {10.1086/378581}, \href
  {https://ui.adsabs.harvard.edu/abs/2003ApJ...597..893N} {597, 893}

\bibitem[\protect\citeauthoryear{{Nenkova} et~al.,}{{Nenkova}
  et~al.}{2008a}]{Nenkova_2008a}
{Nenkova} M.,  et~al., 2008a, \mn@doi [\apj] {10.1086/590482}, \href
  {https://ui.adsabs.harvard.edu/abs/2008ApJ...685..147N} {685, 147}

\bibitem[\protect\citeauthoryear{{Nenkova} et~al.,}{{Nenkova}
  et~al.}{2008b}]{Nenkova_2008b}
{Nenkova} M.,  et~al., 2008b, \mn@doi [\apj] {10.1086/590483}, \href
  {https://ui.adsabs.harvard.edu/abs/2008ApJ...685..160N} {685, 160}

\bibitem[\protect\citeauthoryear{{Nyland} et~al.,}{{Nyland}
  et~al.}{2016}]{Nyland_2016}
{Nyland} K.,  et~al., 2016, \mn@doi [\mnras] {10.1093/mnras/stw391}, \href
  {https://ui.adsabs.harvard.edu/abs/2016MNRAS.458.2221N} {458, 2221}

\bibitem[\protect\citeauthoryear{{Nyland} et~al.,}{{Nyland}
  et~al.}{2017}]{Nyland_2017}
{Nyland} K.,  et~al., 2017, \mn@doi [\apj] {10.3847/1538-4357/aa7ecf}, \href
  {https://ui.adsabs.harvard.edu/abs/2017ApJ...845...50N} {845, 50}

\bibitem[\protect\citeauthoryear{{Oh} et~al.,}{{Oh} et~al.}{2011}]{Oh_2011}
{Oh} K.,  et~al., 2011, \mn@doi [\apjs] {10.1088/0067-0049/195/2/13}, \href
  {https://ui.adsabs.harvard.edu/abs/2011ApJS..195...13O} {195, 13}

\bibitem[\protect\citeauthoryear{{Park} et~al.,}{{Park}
  et~al.}{2006}]{Park_2006}
{Park} T.,  et~al., 2006, \mn@doi [\apj] {10.1086/507406}, \href
  {https://ui.adsabs.harvard.edu/abs/2006ApJ...652..610P} {652, 610}

\bibitem[\protect\citeauthoryear{{Parra} et~al.,}{{Parra}
  et~al.}{2010}]{Parra_2010}
{Parra} R.,  et~al., 2010, \mn@doi [\apj] {10.1088/0004-637X/720/1/555}, \href
  {https://ui.adsabs.harvard.edu/abs/2010ApJ...720..555P} {720, 555}

\bibitem[\protect\citeauthoryear{{Patil} et~al.,}{{Patil}
  et~al.}{2022}]{Patil_2022}
{Patil} P.,  et~al., 2022, arXiv e-prints, \href
  {https://ui.adsabs.harvard.edu/abs/2022arXiv220107349P} {p. arXiv:2201.07349}

\bibitem[\protect\citeauthoryear{{Pawlik} et~al.,}{{Pawlik}
  et~al.}{2016}]{Pawlik_2016}
{Pawlik} M.~M.,  et~al., 2016, \mn@doi [\mnras] {10.1093/mnras/stv2878}, \href
  {https://ui.adsabs.harvard.edu/abs/2016MNRAS.456.3032P} {456, 3032}

\bibitem[\protect\citeauthoryear{{Pereira-Santaella}
  et~al.,}{{Pereira-Santaella} et~al.}{2015}]{Pereira-Santaella_2015}
{Pereira-Santaella} M.,  et~al., 2015, \mn@doi [\aap]
  {10.1051/0004-6361/201425359}, \href
  {https://ui.adsabs.harvard.edu/abs/2015A&A...577A..78P} {577, A78}

\bibitem[\protect\citeauthoryear{{Peters} et~al.,}{{Peters}
  et~al.}{2021}]{Peters_2021}
{Peters} W.,  et~al., 2021, in American Astronomical Society Meeting Abstracts.
  p. 211.06

\bibitem[\protect\citeauthoryear{{Petric} et~al.,}{{Petric}
  et~al.}{2011}]{petric2011}
{Petric} A.~O.,  et~al., 2011, \mn@doi [\apj] {10.1088/0004-637X/730/1/28},
  \href {https://ui.adsabs.harvard.edu/abs/2011ApJ...730...28P} {730, 28}

\bibitem[\protect\citeauthoryear{{Petric} et~al.,}{{Petric}
  et~al.}{2018}]{petric2018}
{Petric} A.~O.,  et~al., 2018, \mn@doi [\aj] {10.3847/1538-3881/aaca35}, \href
  {https://ui.adsabs.harvard.edu/abs/2018AJ....156..295P} {156, 295}

\bibitem[\protect\citeauthoryear{{Phillips}}{{Phillips}}{1993}]{Phillips_1993}
{Phillips} A.~C.,  1993, \mn@doi [\aj] {10.1086/116447}, \href
  {https://ui.adsabs.harvard.edu/abs/1993AJ....105..486P} {105, 486}

\bibitem[\protect\citeauthoryear{{Piconcelli} et~al.,}{{Piconcelli}
  et~al.}{2005}]{Piconcelli_2005}
{Piconcelli} E.,  et~al., 2005, \mn@doi [\aap] {10.1051/0004-6361:20041621},
  \href {https://ui.adsabs.harvard.edu/abs/2005A&A...432...15P} {432, 15}

\bibitem[\protect\citeauthoryear{{Pilbratt} et~al.,}{{Pilbratt}
  et~al.}{2010}]{Pilbratt_2010}
{Pilbratt} G.~L.,  et~al., 2010, \mn@doi [\aap] {10.1051/0004-6361/201014759},
  \href {https://ui.adsabs.harvard.edu/abs/2010A&A...518L...1P} {518, L1}

\bibitem[\protect\citeauthoryear{{Polisensky} et~al.,}{{Polisensky}
  et~al.}{2016}]{Polisensky_2016}
{Polisensky} E.,  et~al., 2016, \mn@doi [\apj] {10.3847/0004-637X/832/1/60},
  \href {https://ui.adsabs.harvard.edu/abs/2016ApJ...832...60P} {832, 60}

\bibitem[\protect\citeauthoryear{{Ricci} et~al.,}{{Ricci}
  et~al.}{2021}]{Ricci_2021}
{Ricci} C.,  et~al., 2021, \mn@doi [\mnras] {10.1093/mnras/stab2052}, \href
  {https://ui.adsabs.harvard.edu/abs/2021MNRAS.506.5935R} {506, 5935}

\bibitem[\protect\citeauthoryear{{Riffel} et~al.,}{{Riffel}
  et~al.}{2020}]{riff2020}
{Riffel} R.~A.,  et~al., 2020, \mn@doi [\mnras] {10.1093/mnras/stz3137}, \href
  {https://ui.adsabs.harvard.edu/abs/2020MNRAS.491.1518R} {491, 1518}

\bibitem[\protect\citeauthoryear{{Roberts-Borsani} \&
  {Saintonge}}{{Roberts-Borsani} \& {Saintonge}}{2019}]{Roberts-Borsani_2019}
{Roberts-Borsani} G.~W.,  {Saintonge} A.,  2019, \mn@doi [\mnras]
  {10.1093/mnras/sty2824}, \href
  {https://ui.adsabs.harvard.edu/abs/2019MNRAS.482.4111R} {482, 4111}

\bibitem[\protect\citeauthoryear{{Rowlands} et~al.,}{{Rowlands}
  et~al.}{2015}]{Rowlands_2015}
{Rowlands} K.,  et~al., 2015, \mn@doi [\mnras] {10.1093/mnras/stu2714}, \href
  {https://ui.adsabs.harvard.edu/abs/2015MNRAS.448..258R} {448, 258}

\bibitem[\protect\citeauthoryear{{Rowlands} et~al.,}{{Rowlands}
  et~al.}{2018}]{Rowlands_2018}
{Rowlands} K.,  et~al., 2018, \mn@doi [\mnras] {10.1093/mnras/stx1903}, \href
  {https://ui.adsabs.harvard.edu/abs/2018MNRAS.473.1168R} {473, 1168}

\bibitem[\protect\citeauthoryear{{Roy} et~al.,}{{Roy} et~al.}{2021}]{Roy_2021}
{Roy} N.,  et~al., 2021, arXiv e-prints, \href
  {https://ui.adsabs.harvard.edu/abs/2021arXiv210614901R} {p. arXiv:2106.14901}

\bibitem[\protect\citeauthoryear{{Rupke} et~al.,}{{Rupke}
  et~al.}{2005a}]{Rupke_2005a}
{Rupke} D.~S.,  et~al., 2005a, \mn@doi [\apjs] {10.1086/432886}, \href
  {https://ui.adsabs.harvard.edu/abs/2005ApJS..160...87R} {160, 87}

\bibitem[\protect\citeauthoryear{{Rupke} et~al.,}{{Rupke}
  et~al.}{2005b}]{Rupke_2005b}
{Rupke} D.~S.,  et~al., 2005b, \mn@doi [\apjs] {10.1086/432889}, \href
  {https://ui.adsabs.harvard.edu/abs/2005ApJS..160..115R} {160, 115}

\bibitem[\protect\citeauthoryear{{Ryan} R.~E. et~al.,}{{Ryan}
  et~al.}{2016}]{Ryan_2016}
{Ryan} R.~E. J.,  et~al., 2016, {The Updated Calibration Pipeline for
  WFC3/UVIS: a Reference Guide to calwf3 (version 3.3)}, Space Telescope WFC
  Instrument Science Report

\bibitem[\protect\citeauthoryear{{Salpeter}}{{Salpeter}}{1955}]{Salpeter_1955}
{Salpeter} E.~E.,  1955, \mn@doi [\apj] {10.1086/145971}, \href
  {https://ui.adsabs.harvard.edu/abs/1955ApJ...121..161S} {121, 161}

\bibitem[\protect\citeauthoryear{{Sanders} \& {Mirabel}}{{Sanders} \&
  {Mirabel}}{1996}]{Sanders_1996}
{Sanders} D.~B.,  {Mirabel} I.~F.,  1996, \mn@doi [\araa]
  {10.1146/annurev.astro.34.1.749}, \href
  {https://ui.adsabs.harvard.edu/abs/1996ARA&A..34..749S} {34, 749}

\bibitem[\protect\citeauthoryear{{Satyapal} et~al.,}{{Satyapal}
  et~al.}{2018}]{Satyapal_2018}
{Satyapal} S.,  et~al., 2018, \mn@doi [\apj] {10.3847/1538-4357/aab7f8}, \href
  {https://ui.adsabs.harvard.edu/abs/2018ApJ...858...38S} {858, 38}

\bibitem[\protect\citeauthoryear{{Sault} et~al.,}{{Sault}
  et~al.}{1995}]{Sault_1995}
{Sault} R.~J.,  et~al., 1995, in {Shaw} R.~A.,  et~al., eds,  Astronomical
  Society of the Pacific Conference Series Vol. 77, Astronomical Data Analysis
  Software and Systems IV. p.~433 (\mn@eprint {arXiv} {astro-ph/0612759})

\bibitem[\protect\citeauthoryear{{Sazonova} et~al.,}{{Sazonova}
  et~al.}{2021}]{Sazonova_2021}
{Sazonova} E.,  et~al., 2021, \mn@doi [\apj] {10.3847/1538-4357/ac0f7f}, \href
  {https://ui.adsabs.harvard.edu/abs/2021ApJ...919..134S} {919, 134}

\bibitem[\protect\citeauthoryear{{Schawinski} et~al.,}{{Schawinski}
  et~al.}{2014}]{Schawinski_2014}
{Schawinski} K.,  et~al., 2014, \mn@doi [\mnras] {10.1093/mnras/stu327}, \href
  {https://ui.adsabs.harvard.edu/abs/2014MNRAS.440..889S} {440, 889}

\bibitem[\protect\citeauthoryear{{Schlafly} \& {Finkbeiner}}{{Schlafly} \&
  {Finkbeiner}}{2011}]{Schlafly_2011}
{Schlafly} E.~F.,  {Finkbeiner} D.~P.,  2011, \mn@doi [\apj]
  {10.1088/0004-637X/737/2/103}, \href
  {https://ui.adsabs.harvard.edu/abs/2011ApJ...737..103S} {737, 103}

\bibitem[\protect\citeauthoryear{{Simons} et~al.,}{{Simons}
  et~al.}{2015}]{Simons_2015}
{Simons} R.~C.,  et~al., 2015, \mn@doi [\mnras] {10.1093/mnras/stv1298}, \href
  {https://ui.adsabs.harvard.edu/abs/2015MNRAS.452..986S} {452, 986}

\bibitem[\protect\citeauthoryear{{Skrutskie} et~al.,}{{Skrutskie}
  et~al.}{2006}]{Skrutskie_2006}
{Skrutskie} M.~F.,  et~al., 2006, \mn@doi [\aj] {10.1086/498708}, \href
  {https://ui.adsabs.harvard.edu/abs/2006AJ....131.1163S} {131, 1163}

\bibitem[\protect\citeauthoryear{{Smercina} et~al.,}{{Smercina}
  et~al.}{2018}]{Smercina_2018}
{Smercina} A.,  et~al., 2018, \mn@doi [\apj] {10.3847/1538-4357/aaafcd}, \href
  {https://ui.adsabs.harvard.edu/abs/2018ApJ...855...51S} {855, 51}

\bibitem[\protect\citeauthoryear{{Smercina} et~al.,}{{Smercina}
  et~al.}{2021}]{Smercina_2021}
{Smercina} A.,  et~al., 2021, arXiv e-prints, \href
  {https://ui.adsabs.harvard.edu/abs/2021arXiv210803231S} {p. arXiv:2108.03231}

\bibitem[\protect\citeauthoryear{{Smethurst} et~al.,}{{Smethurst}
  et~al.}{2017}]{Smethurst_2017}
{Smethurst} R.~J.,  et~al., 2017, \mn@doi [\mnras] {10.1093/mnras/stx973},
  \href {https://ui.adsabs.harvard.edu/abs/2017MNRAS.469.3670S} {469, 3670}

\bibitem[\protect\citeauthoryear{{Snyder} et~al.,}{{Snyder}
  et~al.}{2011}]{Snyder_2011}
{Snyder} G.~F.,  et~al., 2011, \mn@doi [\apj] {10.1088/0004-637X/741/2/77},
  \href {https://ui.adsabs.harvard.edu/abs/2011ApJ...741...77S} {741, 77}

\bibitem[\protect\citeauthoryear{{Sparre} \& {Springel}}{{Sparre} \&
  {Springel}}{2016}]{Sparre_2016}
{Sparre} M.,  {Springel} V.,  2016, \mn@doi [\mnras] {10.1093/mnras/stw1793},
  \href {https://ui.adsabs.harvard.edu/abs/2016MNRAS.462.2418S} {462, 2418}

\bibitem[\protect\citeauthoryear{{Spoon} et~al.,}{{Spoon}
  et~al.}{2007}]{Spoon_2007}
{Spoon} H.~W.~W.,  et~al., 2007, \mn@doi [\apjl] {10.1086/511268}, \href
  {https://ui.adsabs.harvard.edu/abs/2007ApJ...654L..49S} {654, L49}

\bibitem[\protect\citeauthoryear{{Stern} et~al.,}{{Stern}
  et~al.}{2005}]{Stern_2005}
{Stern} D.,  et~al., 2005, \mn@doi [\apj] {10.1086/432523}, \href
  {https://ui.adsabs.harvard.edu/abs/2005ApJ...631..163S} {631, 163}

\bibitem[\protect\citeauthoryear{{Stierwalt} et~al.,}{{Stierwalt}
  et~al.}{2013}]{Stierwalt_2013}
{Stierwalt} S.,  et~al., 2013, \mn@doi [\apjs] {10.1088/0067-0049/206/1/1},
  \href {https://ui.adsabs.harvard.edu/abs/2013ApJS..206....1S} {206, 1}

\bibitem[\protect\citeauthoryear{{Stierwalt} et~al.,}{{Stierwalt}
  et~al.}{2014}]{Stierwalt_2014}
{Stierwalt} S.,  et~al., 2014, \mn@doi [\apj] {10.1088/0004-637X/790/2/124},
  \href {https://ui.adsabs.harvard.edu/abs/2014ApJ...790..124S} {790, 124}

\bibitem[\protect\citeauthoryear{{Suess} et~al.,}{{Suess}
  et~al.}{2022}]{Suess_2022}
{Suess} K.~A.,  et~al., 2022, \mn@doi [\apj] {10.3847/1538-4357/ac404a}, \href
  {https://ui.adsabs.harvard.edu/abs/2022ApJ...926...89S} {926, 89}

\bibitem[\protect\citeauthoryear{{Terrazas} et~al.,}{{Terrazas}
  et~al.}{2016}]{Terrazas_2016}
{Terrazas} B.~A.,  et~al., 2016, \mn@doi [\apjl] {10.3847/2041-8205/830/1/L12},
  \href {https://ui.adsabs.harvard.edu/abs/2016ApJ...830L..12T} {830, L12}

\bibitem[\protect\citeauthoryear{{Terrazas} et~al.,}{{Terrazas}
  et~al.}{2017}]{Terrazas_2017}
{Terrazas} B.~A.,  et~al., 2017, \mn@doi [\apj] {10.3847/1538-4357/aa7d07},
  \href {https://ui.adsabs.harvard.edu/abs/2017ApJ...844..170T} {844, 170}

\bibitem[\protect\citeauthoryear{{Tremonti} et~al.,}{{Tremonti}
  et~al.}{2007}]{Tremonti_2007}
{Tremonti} C.~A.,  et~al., 2007, \mn@doi [\apjl] {10.1086/520083}, \href
  {https://ui.adsabs.harvard.edu/abs/2007ApJ...663L..77T} {663, L77}

\bibitem[\protect\citeauthoryear{{U} et~al.,}{{U} et~al.}{2012}]{U_2012}
{U} V.,  et~al., 2012, \mn@doi [\apjs] {10.1088/0067-0049/203/1/9}, \href
  {https://ui.adsabs.harvard.edu/abs/2012ApJS..203....9U} {203, 9}

\bibitem[\protect\citeauthoryear{{Ueda} et~al.,}{{Ueda}
  et~al.}{2014}]{Ueda_2014}
{Ueda} J.,  et~al., 2014, \mn@doi [\apjs] {10.1088/0067-0049/214/1/1}, \href
  {https://ui.adsabs.harvard.edu/abs/2014ApJS..214....1U} {214, 1}

\bibitem[\protect\citeauthoryear{Veilleux \& Osterbrock}{Veilleux \&
  Osterbrock}{1987}]{Veilleux_1987}
Veilleux S.,  Osterbrock D.~E.,  1987, \mn@doi [\apjs] {10.1086/191166}, 63,
  295

\bibitem[\protect\citeauthoryear{{Veilleux} et~al.,}{{Veilleux}
  et~al.}{2005}]{Veilleux_2005}
{Veilleux} S.,  et~al., 2005, \mn@doi [\araa]
  {10.1146/annurev.astro.43.072103.150610}, \href
  {https://ui.adsabs.harvard.edu/abs/2005ARA&A..43..769V} {43, 769}

\bibitem[\protect\citeauthoryear{Virtanen et~al.,}{Virtanen
  et~al.}{2020}]{scipy}
Virtanen P.,  et~al., 2020, \mn@doi [Nature Methods]
  {10.1038/s41592-019-0686-2}, 17, 261

\bibitem[\protect\citeauthoryear{{Weisskopf} et~al.,}{{Weisskopf}
  et~al.}{2000}]{Weisskopf_2000}
{Weisskopf} M.~C.,  et~al., 2000, in {Truemper} J.~E.,  {Aschenbach} B.,  eds,
  Society of Photo-Optical Instrumentation Engineers (SPIE) Conference Series
  Vol. 4012, X-Ray Optics, Instruments, and Missions III. pp 2--16 (\mn@eprint
  {arXiv} {astro-ph/0004127}), \mn@doi{10.1117/12.391545}

\bibitem[\protect\citeauthoryear{{Werner} et~al.,}{{Werner}
  et~al.}{2004}]{Werner_2004}
{Werner} M.~W.,  et~al., 2004, \mn@doi [\apjs] {10.1086/422992}, \href
  {https://ui.adsabs.harvard.edu/abs/2004ApJS..154....1W} {154, 1}

\bibitem[\protect\citeauthoryear{{Westfall} et~al.,}{{Westfall}
  et~al.}{2019}]{Westfall_2019}
{Westfall} K.~B.,  et~al., 2019, \mn@doi [\aj] {10.3847/1538-3881/ab44a2},
  \href {https://ui.adsabs.harvard.edu/abs/2019AJ....158..231W} {158, 231}

\bibitem[\protect\citeauthoryear{{Wild} et~al.,}{{Wild}
  et~al.}{2011}]{Wild_2011}
{Wild} V.,  et~al., 2011, \mn@doi [\mnras] {10.1111/j.1365-2966.2010.17536.x},
  \href {https://ui.adsabs.harvard.edu/abs/2011MNRAS.410.1593W} {410, 1593}

\bibitem[\protect\citeauthoryear{{Wild} et~al.,}{{Wild}
  et~al.}{2016}]{Wild_2016}
{Wild} V.,  et~al., 2016, \mn@doi [\mnras] {10.1093/mnras/stw1996}, \href
  {https://ui.adsabs.harvard.edu/abs/2016MNRAS.463..832W} {463, 832}

\bibitem[\protect\citeauthoryear{{Wright} et~al.,}{{Wright}
  et~al.}{2010}]{Wright_2010}
{Wright} E.~L.,  et~al., 2010, \mn@doi [\aj] {10.1088/0004-6256/140/6/1868},
  \href {https://ui.adsabs.harvard.edu/abs/2010AJ....140.1868W} {140, 1868}

\bibitem[\protect\citeauthoryear{Wright et~al.,}{Wright et~al.}{2019}]{allwise}
Wright E.~L.,  et~al., 2019, AllWISE Source Catalog, \mn@doi{10.26131/IRSA1},
  \url {https://catcopy.ipac.caltech.edu/dois/doi.php?id=10.26131/IRSA1}

\bibitem[\protect\citeauthoryear{{Yang} et~al.,}{{Yang}
  et~al.}{2008}]{Yang_2008}
{Yang} Y.,  et~al., 2008, \mn@doi [\apj] {10.1086/591656}, \href
  {https://ui.adsabs.harvard.edu/abs/2008ApJ...688..945Y} {688, 945}

\bibitem[\protect\citeauthoryear{{Zabludoff} et~al.,}{{Zabludoff}
  et~al.}{1996}]{Zabludoff_1996}
{Zabludoff} A.~I.,  et~al., 1996, \mn@doi [\apj] {10.1086/177495}, \href
  {https://ui.adsabs.harvard.edu/abs/1996ApJ...466..104Z} {466, 104}

\bibitem[\protect\citeauthoryear{{Zheng} et~al.,}{{Zheng}
  et~al.}{2020}]{Zheng_2020}
{Zheng} Y.,  et~al., 2020, \mn@doi [\mnras] {10.1093/mnras/staa2358}, \href
  {https://ui.adsabs.harvard.edu/abs/2020MNRAS.498.1259Z} {498, 1259}

\bibitem[\protect\citeauthoryear{{da Cunha} et~al.,}{{da Cunha}
  et~al.}{2008}]{daCunha_2008}
{da Cunha} E.,  et~al., 2008, \mn@doi [\mnras]
  {10.1111/j.1365-2966.2008.13535.x}, \href
  {https://ui.adsabs.harvard.edu/abs/2008MNRAS.388.1595D} {388, 1595}

\bibitem[\protect\citeauthoryear{{van Dokkum}}{{van
  Dokkum}}{2001}]{vanDokkum_2001}
{van Dokkum} P.~G.,  2001, \mn@doi [\pasp] {10.1086/323894}, \href
  {https://ui.adsabs.harvard.edu/abs/2001PASP..113.1420V} {113, 1420}

\bibitem[\protect\citeauthoryear{{van de Voort} et~al.,}{{van de Voort}
  et~al.}{2018}]{van_de_Voort_2018}
{van de Voort} F.,  et~al., 2018, \mn@doi [\mnras] {10.1093/mnras/sty228},
  \href {https://ui.adsabs.harvard.edu/abs/2018MNRAS.476..122V} {476, 122}

\makeatother
\end{thebibliography}
\bibliographystyle{mnras_custom}



\end{document}